\documentclass[11pt]{article}
\usepackage{amsmath,amssymb,amsthm,amsfonts,mathtools,bm}
\usepackage{mathrsfs}
\usepackage{algorithm}
\usepackage{algpseudocode}
\usepackage{geometry}
\usepackage{setspace}
\usepackage{pdfpages}
\usepackage{graphicx}
\graphicspath{{figures/}}
\usepackage{booktabs,multirow,float}
\usepackage{tabularx}
\usepackage{longtable}
\usepackage[section]{placeins}
\usepackage{microtype}
\usepackage{caption}
\usepackage[hidelinks]{hyperref}
\hypersetup{bookmarksnumbered=true}
\newcommand{\safeincludegraphics}[2][]{\includegraphics[#1]{#2}}
\newtheorem{theorem}{Theorem}[section]
\newtheorem{proposition}[theorem]{Proposition}
\newtheorem{lemma}[theorem]{Lemma}
\newtheorem{corollary}[theorem]{Corollary}

\newtheorem{remark}[theorem]{Remark}
\newtheorem{assumption}[theorem]{Assumption}

\newcommand{\E}{\mathbb E}
\newcommand{\Q}{\mathbb Q}
\newcommand{\R}{\mathbb R}
\newcommand{\cL}{\mathcal L}
\newcommand{\cJ}{\mathcal J}

\newcommand{\tablenote}[1]{\par\smallskip
  {\footnotesize\raggedright\noindent\textit{Notes:}~#1\par}}

\title{Fourier-Block CTMC Pricing of American Futures Options under Jump-Diffusion Models}
\ifdefined\AnonymousSubmission
\author{}
\hypersetup{pdfauthor={}}
\else
\author{%
Shunyi Wang\thanks{Department of Applied Mathematics, University of
Washington, Seattle, USA. Email: \href{mailto:shunyi25@uw.edu}{\nolinkurl{shunyi25@uw.edu}}.}
\and
Junyi Zhou\thanks{Department of Financial Mathematics, Xi'an Jiaotong-Liverpool University, Suzhou, China.
Email: \href{mailto:JunyiZhou20040716@gmail.com}{\nolinkurl{JunyiZhou20040716@gmail.com}}.}
\and
Qiang Niu\thanks{Department of Applied Mathematics, Xi'an Jiaotong-Liverpool
University, Suzhou, China. Email: \href{mailto:Qiang.Niu@xjtlu.edu.cn}{\nolinkurl{Qiang.Niu@xjtlu.edu.cn}}.}
\and
Yi Hong\thanks{Department of Financial Mathematics, Xi'an Jiaotong-Liverpool
University, Suzhou, China. Email: \href{mailto:Yi.Hong@xjtlu.edu.cn}{\nolinkurl{Yi.Hong@xjtlu.edu.cn}}.}}
\hypersetup{pdfauthor={Shunyi Wang, Junyi Zhou, Qiang Niu, Yi Hong}}
\fi
\date{}
\begin{document}
\maketitle
\begin{abstract}
American futures-option pricing under stochastic volatility and jumps
requires repeated propagation of option values over a large joint
price--variance grid. We develop a Fourier-block continuous-time
Markov chain (CTMC) framework that exploits conditional translation
invariance in the transformed price coordinate. For the constructed
periodic CTMC, a discrete Fourier transform exactly reduces the
$(N_xN_v)\times(N_xN_v)$ matrix-exponential problem to $N_x$
independent $N_v\times N_v$ problems, retaining the full discrete
variance coupling without directional splitting. The conservative
generator provides a structural basis for positivity, contractivity,
and stability of the Bermudan stopping recursion. Cached block
propagators and strike-normalized value surfaces further enable
computational reuse across exercise dates and compatible strikes.
We establish sequential convergence to the continuous-model
American put value under the stated Feller, mesh, expanding-domain,
jump-winding, and exercise-mesh conditions. Numerical experiments
verify the Fourier-block representation against direct
matrix-exponential propagation of the identical periodic generator
to machine precision and demonstrate favorable cost--discrepancy
performance relative to endpoint-CTMC, operator-splitting, and
PROJ implementations under the tested configurations. Spatial
refinement, exercise-mesh refinement, and domain-expansion
experiments further assess the numerical robustness of the proposed
method. Empirical calibration and cross-date pricing evaluations
using Shanghai Futures Exchange gold options provide complementary
evidence of its practical applicability.
\end{abstract}
\noindent\textbf{Keywords:} American futures options; Heston jump-diffusion;
Continuous-time Markov chains; Fourier block diagonalization; Optimal stopping.

\noindent\textbf{JEL classification:} C63; G13.
\clearpage
\section{Introduction}
\label{sec:introduction}
American futures-option pricing under stochastic volatility and jumps requires
tracking both the futures price and its variance while allowing early exercise.
A joint CTMC discretization with $N_x$ price nodes and $N_v$ variance nodes
produces a generator $Q$ of dimension $N_xN_v$. Between exercise dates, pricing
requires applying $e^{\Delta t Q}$ to the value surface, discounting
the result, and then comparing continuation with immediate exercise.
Efficient evaluation of this large coupled propagation problem is central to
repeated pricing across exercise dates, strikes, and calibration candidates.

Several established numerical approaches exploit structure in this problem.
Fourier space time-stepping (FST) applies the Fourier transform directly to the
pricing PIDE, so that continuation is propagated frequency by frequency before
the obstacle is imposed in real space; see Jackson, Jaimungal and
Surkov~\cite{JacksonJaimungalSurkov2008}. CTMC--Fourier and PROJ methods instead
combine finite-state approximations of stochastic volatility with Fourier
representations of transition kernels and projected densities; see
Kirkby~\cite{Kirkby2015}, Kirkby, Nguyen and Cui~\cite{KirkbyNguyenCui2017},
and the related CTMC--Heston/SWIFT construction of Leitao, Kirkby and
Ortiz-Gracia~\cite{LeitaoKirkbyOrtiz2021}. A third line of work uses
price--variance operator splitting, replacing the multidimensional propagator
by successive lower-dimensional actions; related constructions appear in the
Heston--jump/FST literature~\cite{ZhangWangLi2026}.

Viewed from the joint-CTMC perspective, the central computational bottleneck is
the repeated action of the large semigroup $e^{\Delta t Q}$. Directional
splitting addresses this bottleneck by decomposing the joint generator into
successive lower-dimensional actions. We pursue an alternative that
combines the frequency-domain continuation principle of FST with the
finite-state volatility--Fourier structure underlying CTMC--PROJ, but moves
the Fourier reduction directly to the level of a conservative joint CTMC
generator.

We discretize the transformed price and variance jointly, including jumps in
the same generator. On a uniform periodic price grid, conditional translation
invariance allows the large joint matrix-exponential problem to be replaced
by independent smaller problems, one for each price frequency. Each smaller
matrix retains all variance states and their coupling. The computational advantage
comes from reducing the dimension of the matrices being exponentiated from
$N_xN_v$ to $N_v$, while preserving the joint discrete dynamics.

This construction therefore provides an unsplit alternative to
price--variance operator splitting for the same joint propagation problem.
Splitting approximates the joint semigroup through successive directional
actions, whereas the Fourier-block method retains the full variance coupling
within each frequency block and evaluates the semigroup of the specified
periodic generator without a splitting approximation. Relative to PIDE-based
Fourier time-stepping, the frequency symbols arise from a conservative CTMC
generator, linking continuation to positivity, probability conservation, and
contraction of the exact discrete semigroup. Relative to CTMC--PROJ, the
method discretizes the price state as well as the variance state and propagates
the value surface directly, without constructing an intermediate projected
transition density. In this way, the Fourier reduction is tied directly to the
Markov structure of the joint discretization.

After a decorrelating transformation, the transformed price coordinate is
spatially homogeneous conditional on variance. We approximate the continuous
variance factor by a CTMC and discretize the transformed price direction on a
uniform periodic grid, obtaining
\[
 Q_{h,\mathrm{per}}=Q_V^h\otimes I_{N_x}
  +\operatorname{diag}_j(C_j)+I_{N_v}\otimes C_J.
\]
The conditional price matrices $C_j$ and the jump matrix $C_J$ are circulant,
whereas $Q_V^h$ retains the full variance-state coupling. A DFT in the price
coordinate yields the exact finite-state similarity reduction
\[
 \mathcal F_XQ_{h,\mathrm{per}}\mathcal F_X^{-1}
   =\bigoplus_k \widehat Q_k,
 \qquad
 \mathcal F_Xe^{\Delta t Q_{h,\mathrm{per}}}\mathcal F_X^{-1}
   =\bigoplus_k e^{\Delta t\widehat Q_k}.
\]
Thus one $(N_xN_v)\times(N_xN_v)$ matrix-exponential problem becomes $N_x$
independent $N_v\times N_v$ problems. This is an exact computational
dimensionality reduction for the periodic CTMC; the CTMC discretization
provides the approximation to the continuous stochastic model. After block
propagation, an inverse Fourier transform recovers the value surface for the
early-exercise comparison. The block propagators can be cached by exercise interval,
and strike normalization permits compatible contracts to reuse the same
propagated value surface.

The construction requires conditional translation invariance in the transformed
price coordinate and a uniform periodic price grid. General local--stochastic-
volatility specifications with explicit price dependence do not inherit this
block structure directly, and the periodic closure introduces a domain choice
that requires numerical validation. These scope conditions are examined through
the domain-sensitivity diagnostics reported below.

Our contributions are threefold. First, we develop a unified computational
framework for American futures-option pricing that connects multidimensional
stochastic modeling, efficient numerical valuation, and empirical calibration.
Second, we develop a Fourier-block representation of the joint two-dimensional
CTMC semigroup: the large price--variance matrix exponential is reduced to
independent variance-sized matrix exponentials while the full discrete variance
coupling is retained. This provides an unsplit alternative to directional
propagation and substantially simplifies repeated pricing, with stability and
convergence established under the stated assumptions. Third, we provide an
empirical CTMC study of SHFE gold futures options under a stochastic
volatility jump-diffusion model: a multi-day alternating
calibration on a ten-day estimation window and a separate ten-day stress
window, with futures-side initialization of the daily variance
states, loss-profile identification diagnostics, a two-measure
state-space panel estimation, and a 15-date
cross-date pricing evaluation with daily-state re-estimation.
The empirical implementation also
demonstrates reusable cross-strike pricing. These results contribute to
the literature on CTMC-based futures-option valuation.

Section~\ref{sec:theory} defines the model and the endpoint reference chain.
Section~\ref{sec:algorithm} constructs and diagonalizes the periodic chain.
Section~\ref{sec:architectures} develops stopping, error, representation, and
cost analysis. Section~\ref{sec:numerics} reports the numerical benchmarks,
algebraic verification, and refinement studies. Section~\ref{sec:data}
describes the SHFE gold-option data and calibration sample.
Section~\ref{sec:empirical} presents a
multi-day alternating calibration with all structural parameters freely
estimated on a ten-day window spanning three maturity chains
(September--October 2023),
together with identification diagnostics, a two-measure state-space
estimation on a 120-day panel, and cross-date pricing
evaluation over 15 selected trading dates.
Technical proofs and auxiliary numerical and market-data results are collected
in the appendices.

\section{Model transformation and CTMC approximation}
\label{sec:theory}
\subsection{Risk-neutral dynamics and decorrelating transformation}
\label{subsec:model}

Let $(\Omega,\mathcal F,\mathbb F,\Q)$ satisfy the usual conditions and let
$\mathcal N(dt,dz)$ be a Poisson random measure with compensator
$\lambda \nu_Z(dz)dt$. Let $F_t$ denote the price of a futures contract whose
delivery date is no earlier than the option maturity. Under the risk-neutral
measure, $F$ is a martingale and the futures price and instantaneous variance
$V$ obey
\begin{equation}
\left\{
\begin{aligned}
\frac{dF_t}{F_{t-}}
 &= -\lambda\kappa_J\,dt+\sqrt{V_t}\,dW_t^{(1)}
    +\int_{\R}(e^z-1)\mathcal N(dt,dz),\\
dV_t
 &= \kappa(\theta-V_t)\,dt+\sigma\sqrt{V_t}\,dW_t^{(2)},
\end{aligned}
\right.
\label{eq:jump_heston}
\end{equation}
with $F_0>0$ and $V_0>0$. Here $r\ge0$ is the short rate,
$\kappa,\theta,\sigma>0$, and
$d\langle W^{(1)},W^{(2)}\rangle_t=\rho\,dt$ for $\rho\in(-1,1)$. The
total jump-arrival rate is $\lambda\ge0$, and the Poisson random measure is
independent of the Brownian motion. Conditional on an arrival, the log-jump
has a general probability law $\nu_Z$ on $\R$. We assume
\begin{equation}
 \int_{\R}e^z\nu_Z(dz)<\infty,
 \qquad \int_{\R}z^2\nu_Z(dz)<\infty,
 \qquad
 \kappa_J=\int_{\R}(e^z-1)\nu_Z(dz).
\label{eq:jump_compensator}
\end{equation}
Thus jumps occur in the futures price only; the variance path is continuous.
Neither symmetry nor a particular parametric form is imposed in the theory.

Define
\begin{equation}
 X_t=\log(F_t/F_0)-\frac{\rho}{\sigma}(V_t-V_0).
\label{eq:decorrelation}
\end{equation}
Writing $W^{(1)}=\rho W^{(2)}+\sqrt{1-\rho^2}\,B$, where $B$ is independent
of $W^{(2)}$, It\^o's formula for jump processes gives
\begin{equation}
\left\{
\begin{aligned}
dX_t
 &= b(V_t)\,dt+a(V_t)\,dB_t+\int_{\R}z\,\mathcal N(dt,dz),\\
dV_t
 &= \mu(V_t)\,dt+\eta(V_t)\,dW_t^{(2)},
\end{aligned}
\right.
\label{eq:transformed_system}
\end{equation}
where
\begin{equation}
\begin{aligned}
b(v)&=-\lambda\kappa_J-\frac12v
      -\frac{\rho\kappa(\theta-v)}{\sigma},
&a^2(v)&=(1-\rho^2)v,\\
\mu(v)&=\kappa(\theta-v),
&\eta^2(v)&=\sigma^2v.
\end{aligned}
\label{eq:coefficients}
\end{equation}
The transformation removes the diffusion mixed derivative, while a futures-price
jump remains a translation $x\mapsto x+z$ in the first coordinate.
This is a statement about the diffusion covariance and generator, not
independence of $X$ and $V$: both the drift and diffusion coefficient of $X$
still depend on $V$.

For a strike $K>0$, the American put payoff written in the transformed state is
the bounded continuous function
\begin{equation}
 \Phi_K(x,v):=
 \left(K-F_0\exp\left(x+\frac{\rho}{\sigma}(v-V_0)\right)\right)^+,
 \qquad 0\le \Phi_K\le K.
\label{eq:put_payoff_state}
\end{equation}

For $f\in C_c^2(\R\times(0,\infty))$, the infinitesimal generator is
\begin{equation}
\begin{aligned}
(\cL f)(x,v)
={}&b(v)f_x(x,v)+\frac12a^2(v)f_{xx}(x,v)
  +\mu(v)f_v(x,v)+\frac12\eta^2(v)f_{vv}(x,v)\\
 &+\lambda\int_{\R}\bigl[f(x+z,v)-f(x,v)\bigr]\nu_Z(dz).
\end{aligned}
\label{eq:generator}
\end{equation}
The last term is denoted by $\cJ f$.

\begin{assumption}[Model and payoff conventions]
\label{ass:wellposed}
The strict Feller condition $2\kappa\theta>\sigma^2$ holds, and the
jump-diffusion has its nonexplosive solution from each initial state in
$E=\mathbb R\times(0,\infty)$. The pricing model is specified under a
measure for which $F$ is a true martingale. The interest rate is deterministic
and constant, and the option uses the discounted cash-payoff convention in
\eqref{eq:put_payoff_state}.
\end{assumption}
For semigroup approximation it is convenient to use the canonical affine
state space $D=\mathbb R\times\mathbb R_+$, including the CIR entrance boundary
$v=0$. To apply the canonical affine-process result, reorder the state as
$(v,x)\in\mathbb R_+\times\mathbb R$. Its drift is $(\mu(v),b(v))$ and its
diffusion covariance is $v\,\operatorname{diag}(\sigma^2,1-\rho^2)$.
At $v=0$ the variance drift is inward, $\mu(0)=\kappa\theta>0$, and the
diffusion covariance vanishes. The state-independent finite-activity jumps
act only in the real coordinate and cannot leave the nonnegative variance
half-space. These coefficients are admissible in the sense of
\cite[Definition~2.6 and Theorem~2.7]{DuffieFilipovicSchachermayer2003};
writing the finite jump integral in that theorem's truncated form only
adjusts the constant price-coordinate drift. The CIR equation determines
$V$ on $\mathbb R_+$, and the integrals in \eqref{eq:transformed_system}
then determine $X$, giving the nonexplosive law corresponding to this
conservative regular affine process. Its transition semigroup is therefore
Feller on $C_0(D)$ and $C_c^\infty(D)$ is a core of its generator.
Under the strict Feller condition, a process started from $v>0$ does not hit
zero, so this Feller extension agrees with the pricing process on the
interior. The cash-payoff convention must also be distinguished from a
futures-style margined option contract.
\subsection{Reference CTMC for consistency and convergence analysis}
\label{subsec:construction}

This subsection constructs the nonperiodic finite-domain CTMC used as the
reference chain for consistency and convergence analysis; the Fourier-block
pricing algorithm propagates a different chain. We introduce
it separately to approximate the continuous transformed process with local
boundary closures before imposing the periodic price representation needed for
the DFT. Section~\ref{sec:algorithm} then changes the price-boundary and
wrapped-jump representation to obtain $Q_{h,\mathrm{per}}^{[W]}$. Keeping the
reference and periodic chains distinct separates convergence to the continuous
model from the algebraic Fourier representation of the numerical solver.

For a refinement index $h$, let
\[
 \Pi_V^h=\{v_1^h<\cdots<v_{N_v(h)}^h\},\qquad
 \Pi_X^h=\{x_1^h<\cdots<x_{N_x(h)}^h\},
\]
and write $\delta_{j,-}^v=v_j^h-v_{j-1}^h$,
$\delta_{j,+}^v=v_{j+1}^h-v_j^h$, with analogous notation for $x$.
Superscripts $h$ are suppressed when no ambiguity arises.

For any scalar $z$, write $z^+=\max\{z,0\}$ and
$z^-=\max\{-z,0\}$. At interior nodes define the effective local variances
\begin{equation}
\begin{aligned}
 \widehat\eta_{h,j}^{\,2}
 &:=\max\left\{\eta^2(v_j),
 \mu^+(v_j)\delta_{j,+}^v+\mu^-(v_j)\delta_{j,-}^v\right\},\\
 \widehat a_{h,ij}^{\,2}
 &:=\max\left\{a^2(v_j),
 b^+(v_j)\delta_{i,+}^x+b^-(v_j)\delta_{i,-}^x\right\}.
\end{aligned}
\label{eq:effective_local_variances}
\end{equation}
These are the smallest second moments compatible with the prescribed drift
and nonnegative nearest-neighbour rates whenever the unmodified centered
rates are not admissible.

The first layer is a birth--death chain $V^h$ with generator
$Q_V^h=(q^v_{j\ell})$. At an interior node,
\begin{equation}
\left\{
\begin{aligned}
 q^v_{j,j-1}
 &=\frac{\widehat\eta_{h,j}^{\,2}-\delta_{j,+}^v\mu(v_j)}
 {\delta_{j,-}^v(\delta_{j,-}^v+\delta_{j,+}^v)},\\
 q^v_{j,j+1}
 &=\frac{\widehat\eta_{h,j}^{\,2}+\delta_{j,-}^v\mu(v_j)}
 {\delta_{j,+}^v(\delta_{j,-}^v+\delta_{j,+}^v)},\\
q^v_{j,j}&=-q^v_{j,j-1}-q^v_{j,j+1}.
\end{aligned}
\right.
\label{eq:v_rates_new}
\end{equation}
Conditional on $V^h_t=v_j$, the local part of the transformed price is a
birth--death chain with generator $Q_{X,D}^{h,(j)}=(q^{x,j}_{i\ell})$ and
\begin{equation}
\left\{
\begin{aligned}
 q^{x,j}_{i,i-1}
 &=\frac{\widehat a_{h,ij}^{\,2}-\delta_{i,+}^x b(v_j)}
 {\delta_{i,-}^x(\delta_{i,-}^x+\delta_{i,+}^x)},\\
 q^{x,j}_{i,i+1}
 &=\frac{\widehat a_{h,ij}^{\,2}+\delta_{i,-}^x b(v_j)}
 {\delta_{i,+}^x(\delta_{i,-}^x+\delta_{i,+}^x)},\\
q^{x,j}_{i,i}&=-q^{x,j}_{i,i-1}-q^{x,j}_{i,i+1}.
\end{aligned}
\right.
\label{eq:x_rates_new}
\end{equation}
These rates match the drifts exactly and their second moments equal the
effective local variances:
\begin{equation}
\begin{aligned}
 \sum_{\ell}q^v_{j\ell}(v_\ell-v_j)&=\mu(v_j),
 &\sum_{\ell}q^v_{j\ell}(v_\ell-v_j)^2&=\widehat\eta_{h,j}^{\,2},\\
 \sum_{\ell}q^{x,j}_{i\ell}(x_\ell-x_i)&=b(v_j),
 &\sum_{\ell}q^{x,j}_{i\ell}(x_\ell-x_i)^2&=\widehat a_{h,ij}^{\,2}.
\end{aligned}
\label{eq:moment_matching}
\end{equation}
When the original centered rates are nonnegative, the hats in
\eqref{eq:effective_local_variances} equal the model variances and exact
two-moment matching is recovered. Otherwise, one candidate rate is set to
zero and the remaining rate is rebalanced to retain the exact drift. In
either case,
\begin{equation}
\begin{aligned}
 0&\le \widehat\eta_{h,j}^{\,2}-\eta^2(v_j)
 \le |\mu(v_j)|\max\{\delta_{j,-}^v,\delta_{j,+}^v\},\\
 0&\le \widehat a_{h,ij}^{\,2}-a^2(v_j)
 \le |b(v_j)|\max\{\delta_{i,-}^x,\delta_{i,+}^x\}.
\end{aligned}
\label{eq:rebalancing_error}
\end{equation}
Thus the second-moment correction vanishes locally as the grid is refined.

To discretize jumps, let the price cells be
\begin{equation}
C_m^h=
\begin{cases}
(-\infty,(x_1+x_2)/2],&m=1,\\
((x_{m-1}+x_m)/2,(x_m+x_{m+1})/2],&1<m<N_x,\\
((x_{N_x-1}+x_{N_x})/2,\infty),&m=N_x,
\end{cases}
\label{eq:cells}
\end{equation}
and define the nearest-cell projection $\pi_h(y)=x_m$ for $y\in C_m^h$.
The nonlocal transition rates from $x_i$ are
\begin{equation}
q^J_{im}=\lambda \nu_Z(C_m^h-x_i),\quad m\ne i,
\qquad
q^J_{ii}=-\sum_{m\ne i}q^J_{im}.
\label{eq:jump_rates_new}
\end{equation}
Self-jumps falling in $C_i^h-x_i$ do not change the chain state and therefore
do not appear in its generator. Let $Q_J^h=(q^J_{im})$ and, on the product
grid, let $Q_{h,J}:=I_{N_v}\otimes Q_J^h$ denote the corresponding global
jump generator. Define
\begin{equation}
 Q_X^{h,(j)}=Q_{X,D}^{h,(j)}+Q_J^h.
\label{eq:conditional_generator}
\end{equation}

The directional components of the joint-state generator can now be displayed
as a single operator system. For a grid function $U=(U_{j,i})$, define
\begin{equation}
\left\{
\begin{aligned}
(\mathscr Q_V^hU)_{j,i}
  &=\sum_{\ell=1}^{N_v}q^v_{j\ell}U_{\ell,i},
  &&\text{variance transitions},\\
(\mathscr Q_{X,D}^{h}U)_{j,i}
  &=\sum_{m=1}^{N_x}q^{x,j}_{im}U_{j,m},
  &&\text{conditional price diffusion},\\
(\mathscr Q_J^hU)_{j,i}
  &=\sum_{m=1}^{N_x}q^J_{im}U_{j,m},
  &&\text{futures-price jumps},\\
\mathscr Q_h^JU
  &=\mathscr Q_V^hU+\mathscr Q_{X,D}^hU+\mathscr Q_J^hU.
  &&
\end{aligned}
\right.
\label{eq:two_layer_operator_system}
\end{equation}
These are three directional actions of one conservative joint-state
generator; no propagation split is introduced by this decomposition.
Under row-major vectorization, the coupled generator is
\begin{equation}
Q_h^J=Q_V^h\otimes I_{N_x}
+\operatorname{diag}\bigl(Q_X^{h,(1)},\ldots,Q_X^{h,(N_v)}\bigr).
\label{eq:global_jump_generator}
\end{equation}
Equivalently, its explicit $N_v\times N_v$ block representation is
\begin{equation}
Q_h^J=
\left(
\begin{array}{cccc}
q^v_{11}I_{N_x}+Q_X^{h,(1)} & q^v_{12}I_{N_x} & \cdots
    & q^v_{1N_v}I_{N_x}\\
q^v_{21}I_{N_x} & q^v_{22}I_{N_x}+Q_X^{h,(2)} & \cdots
    & q^v_{2N_v}I_{N_x}\\
\vdots & \vdots & \ddots & \vdots\\
q^v_{N_v1}I_{N_x} & q^v_{N_v2}I_{N_x} & \cdots
    & q^v_{N_vN_v}I_{N_x}+Q_X^{h,(N_v)}
\end{array}
\right),
\label{eq:global_jump_generator_blocks}
\end{equation}
where each diagonal conditional block contains the complete price generator
$Q_X^{h,(j)}=Q_{X,D}^{h,(j)}+Q_J^h$, while the off-diagonal blocks describe
transitions between variance states.

At the endpoints, we use one-sided local closures. For a scalar grid
$z_1<\cdots<z_n$, drift $d(z)$ and local variance $s^2(z)$, write
$\Delta_1=z_2-z_1$ and $\Delta_n=z_n-z_{n-1}$ and set
\begin{equation}
 q_{1,2}=\left(\frac{s^2(z_1)}{\Delta_1^2}
                 +\frac{d(z_1)}{\Delta_1}\right)^+,
 \qquad
 q_{n,n-1}=\left(\frac{s^2(z_n)}{\Delta_n^2}
                 -\frac{d(z_n)}{\Delta_n}\right)^+.
 \label{eq:endpoint_closures}
\end{equation}
All other endpoint off-diagonal entries of the local operator are zero, and
diagonal entries are negative row sums. Apply this construction to variance
with $(d,s^2)=(\mu,\eta^2)$ and to conditional price diffusion with
$(d,s^2)=(b,a^2)$. These are conservative finite-domain closures, not exact
moment matches at artificial boundaries. Jumps retain the cell projection in
\eqref{eq:jump_rates_new}. The periodic solver replaces only the price closure
and projected price-jump allocation as specified in Section~\ref{sec:algorithm}.

\begin{assumption}[Admissible refining grids]
\label{ass:grid}
The endpoint-chain grids and closures satisfy the following conditions.
\begin{enumerate}
\item The price endpoints tend to $-\infty,+\infty$, the upper variance
endpoint tends to $+\infty$, and $v_1^h\to0$. Grids become dense on every
compact subset of $D=\R\times[0,\infty)$.
\item For every finite $R,H>0$, the largest adjacent price spacing of any
stencil meeting $[-R,R]$, and the largest adjacent variance spacing of any
stencil meeting $[0,H]$, tend to zero. In particular, this condition includes
the variance layer approaching zero and the neighbours of compact sets.
\item Local endpoint closures are nearest-neighbour, conservative and
nonnegative, as in \eqref{eq:endpoint_closures}, and do not alter interior
rates. The immediate inward neighbours of both price endpoints and of the
upper variance endpoint also escape every fixed compact set in their
respective coordinates. No remote reset to an interior state is permitted.
\item The lower variance closure is either zero-inclusive,
$v_1^h=0$, or uses a positive lower endpoint with
\begin{equation}
 \frac{v_1^h}{v_2^h-v_1^h}\longrightarrow0.
 \label{eq:lower_mesh_condition}
\end{equation}
More general lower closures are admissible if their first two local moments
converge uniformly to $\kappa\theta$ and zero, respectively, at $v_1^h\to0$.
\end{enumerate}
\end{assumption}

The rebalancing in \eqref{eq:effective_local_variances} is inactive exactly
when the original centered rates satisfy
\begin{equation}
-\frac{a^2(v_j)}{\delta_{i,-}^x}\le b(v_j)
\le\frac{a^2(v_j)}{\delta_{i,+}^x},
\qquad
-\frac{\eta^2(v_j)}{\delta_{j,-}^v}\le \mu(v_j)
\le\frac{\eta^2(v_j)}{\delta_{j,+}^v}.
\label{eq:positivity}
\end{equation}

\begin{proposition}[Conservative generator]
\label{prop:generator}
Under Assumption~\ref{ass:grid}, $Q_h^J$ has nonnegative off-diagonal entries
and zero row sums. Consequently, it is the generator of a conservative
finite-state CTMC $Y^h=(X^h,V^h)$ and $e^{tQ_h^J}$ is stochastic.
\end{proposition}

\begin{proof}
The effective variances in \eqref{eq:effective_local_variances} make the
interior off-diagonal rates in \eqref{eq:v_rates_new} and
\eqref{eq:x_rates_new} nonnegative. Assumption~\ref{ass:grid} imposes the same
property on the boundary closures, and all diagonal entries are defined as
negative off-diagonal row sums. Equation~\eqref{eq:jump_rates_new} has these
properties because $\nu_Z$ is a probability measure. Sums, block-diagonal
embeddings, and $Q_V^h\otimes I$ preserve them.
\end{proof}

\begin{remark}[Probability conservation and the futures martingale]
The identity $Q_h^J\mathbf1=0$ preserves constants. It is distinct from
$Q_h^JF_h=0$, where $(F_h)_{j,i}=F_0\exp(x_i+\rho(v_j-V_0)/\sigma)$.
Polynomial local moment matching and endpoint jump projection do not
ensure the latter identity. Thus the finite-state model is not claimed
to preserve the futures martingale exactly. The computable defect
$d_h=Q_h^JF_h$ satisfies
\[
 \|e^{tQ_h^J}F_h-F_h\|_\infty\le t\|d_h\|_\infty,
\]
by integrating the semigroup derivative and using contractivity. This is
a finite-grid diagnostic, not a bound on its limit on expanding domains.
\end{remark}
\subsection{Consistency of the discrete generator}
\label{subsec:consistency}

For a continuous function $f$, write $f_h$ for its restriction to the grid.
Split $Q_h^J=Q_{h,D}+Q_{h,J}$ into its local and nonlocal parts. We regard
the finite grids as subsets of the common state space
$D=\R\times[0,\infty)$; the discrete generators act on continuous test
functions by restriction to the grid.

\begin{lemma}[Local consistency]
\label{lem:local}
Let $Q_{h,D}$ be constructed from the rebalanced local rates
\eqref{eq:v_rates_new} and \eqref{eq:x_rates_new}. For every
$f\in C_c^3(\R\times(0,\infty))$ and compact $K$,
\begin{equation}
\sup_{(x_i,v_j)\in K\cap(\Pi_X^h\times\Pi_V^h)}
\left|(Q_{h,D}f_h)(x_i,v_j)-(\cL_Df)(x_i,v_j)\right|\longrightarrow0,
\label{eq:local_consistency}
\end{equation}
where $\cL_D=\cL-\cJ$.
\end{lemma}

\begin{proof}
See Appendix~\ref{app:local-proof}. The moment expansion is the local
consistency mechanism used in CTMC approximation, including
Ding and Ning~\cite{DingNing2021}; related CTMC constructions in financial
applications include MacKay, Vachon and Cui~\cite{MacKayVachonCui2022}.
\end{proof}

\begin{lemma}[Finite-jump consistency]
\label{lem:jump}
For every $f\in C_c(\R\times(0,\infty))$ and compact $K$,
\begin{equation}
\sup_{(x_i,v_j)\in K\cap(\Pi_X^h\times\Pi_V^h)}
\left|(Q_{h,J}f_h)(x_i,v_j)-(\cJ f)(x_i,v_j)\right|\longrightarrow0.
\label{eq:jump_consistency_new}
\end{equation}
\end{lemma}

Lemma~\ref{lem:jump} is the finite-activity counterpart of the jump-generator
construction in Mijatovi\'c and Pistorius
\cite[Appendix~A.1, case O]{MijatovicPistorius2009}. Their appendix treats the
more general bounded-variation kernel; in the present compound-Poisson case,
the jump measure is finite and nearest-cell projection reduces the proof to
uniform continuity plus control of the jump-measure tails.

\begin{proof}
See Appendix~\ref{app:jump-proof}. The proof specializes the case-O
finite-measure construction of Mijatovi\'c and Pistorius
\cite[Appendix~A.1]{MijatovicPistorius2009} to compound-Poisson jumps.
\end{proof}

\begin{corollary}[Full-generator consistency]
\label{cor:generator_consistency}
For every $f\in C_c^3(\R\times(0,\infty))$ and compact $K$,
\begin{equation}
\sup_{y\in K\cap(\Pi_X^h\times\Pi_V^h)}
|(Q_h^Jf_h)(y)-(\cL f)(y)|\longrightarrow0.
\label{eq:full_consistency}
\end{equation}
\end{corollary}
\subsection{From CTMC convergence to Bermudan and American values}
\label{subsec:weak}
We use the zero-inclusive state space to control the entire lower variance
layer, not only nodes separated from zero. At $v_1^h=0$,
\eqref{eq:endpoint_closures} reduces to
\begin{equation}
 q^v_{1,2}=\frac{\kappa\theta}{v_2^h},\qquad
 q^v_{1,1}=-q^v_{1,2}.
 \label{eq:cir_entrance_closure}
\end{equation}
For a positive lower endpoint, put $\Delta_h=v_2^h-v_1^h$. For sufficiently
small $h$, the lower rate has local moments
\begin{equation}
 q^v_{1,2}\Delta_h=\mu(v_1^h)+\frac{\sigma^2v_1^h}{\Delta_h},
 \qquad
 q^v_{1,2}\Delta_h^2=\mu(v_1^h)\Delta_h+\sigma^2v_1^h.
 \label{eq:lower_boundary_moments}
\end{equation}
Assumption~\ref{ass:grid} therefore makes these moments converge to
$\kappa\theta$ and zero. A fixed positive lower endpoint with ever-smaller
$\Delta_h$ does not meet this condition. The production choice
$v_{\min}=10^{-4}$ is a finite-domain approximation whose sensitivity is
distinct from the expanding-domain limit assumed in the convergence results.

Let $D=\mathbb R\times\mathbb R_+$, let $T(t)$ be the Feller semigroup of the
transformed jump--Heston process on $C_0(D)$, and let
$T_h(t)=e^{tQ_h^J}$ act on grid functions on $E_h$. Denote by
$\pi_h:C_0(D)\to\mathbb R^{E_h}$ the restriction map.

\begin{lemma}[Global consistency on a Feller core]
\label{lem:global_core_consistency}
Under Assumption~\ref{ass:grid}, for every $f\in C_c^\infty(D)$,
\begin{equation}
 \|Q_h^J\pi_h f-\pi_h\cL f\|_{\infty,E_h}\longrightarrow0.
 \label{eq:global_core_consistency}
\end{equation}
Here $C_c^\infty(D)$ denotes restrictions of smooth functions to the closed
half-space, and derivatives at zero are right derivatives.
\end{lemma}
The proof is given in Appendix~\ref{app:proof-global_core_consistency}.

\begin{theorem}[Kurtz semigroup convergence]
\label{thm:kurtz_semigroup}
Under Assumptions~\ref{ass:wellposed} and \ref{ass:grid}, for every $f\in C_0(D)$ and
$T_0<\infty$,
\begin{equation}
 \sup_{0\le t\le T_0}
 \bigl\|T_h(t)\pi_h f-\pi_hT(t)f\bigr\|_{\infty,E_h}
 \longrightarrow0.
 \label{eq:kurtz_semigroup}
\end{equation}
Consequently, if $y_h\in E_h$ and $y_h\to y\in D$, then for every fixed
$t>0$ the transition laws satisfy
\begin{equation}
 P_t^h(y_h,\cdot)\Rightarrow P_t(y,\cdot),
 \label{eq:transition_weak}
\end{equation}
and hence
\[
 P_t^h g(y_h)\longrightarrow P_tg(y)
 \qquad\text{for every }g\in C_b(D).
\]
\end{theorem}
The proof is given in Appendix~\ref{app:proof-kurtz_semigroup}.

\begin{corollary}[Convergence of fixed-schedule Bermudan puts]
\label{cor:values}
Let $\mathcal E=\{0=t_0<\cdots<t_M=T\}$ be fixed. Under the
conditions of Theorem~\ref{thm:kurtz_semigroup}, the exact finite-state Bermudan values
satisfy, for every $y_h\to y$,
\[
 V_K^{h,\mathcal E}(y_h)\longrightarrow V_K^{\mathcal E}(y).
\]
\end{corollary}
The proof is given in Appendix~\ref{app:proof-values}.

\begin{corollary}[Sequential American put limit]
\label{cor:american}
Let $\mathcal E_m$ be nested deterministic schedules containing $0,T$, with
maximum interval $|\mathcal E_m|\to0$. Define
\begin{equation}
 V_K^A(y)=\sup_{\tau\in\mathfrak T_{[0,T]}}
 \mathbb E_y[e^{-r\tau}\Phi_K(Y_\tau)].
 \label{eq:american_value}
\end{equation}
Under the conditions of Theorem~\ref{thm:kurtz_semigroup},
\[
 \lim_{m\to\infty}\lim_{h\to0}V_K^{h,\mathcal E_m}(y_h)=V_K^A(y).
\]
\end{corollary}
The proof is given in Appendix~\ref{app:proof-american}.
\section{Fourier-block CTMC method}
\label{sec:algorithm}

Building on the joint CTMC of Section~\ref{sec:theory}, this section
constructs the periodic price representation, derives its Fourier blocks,
and uses them to propagate continuation values between exercise dates.

\subsection{Periodic CTMC representation of the transformed-price direction}
\label{subsec:periodic-construction}

\paragraph{Uniform price grid and periodic local dynamics.}
Let $N=N_x\ge3$, $m=N_v$, and use the distinct price nodes
\[
 x_i=x_0+i\delta,\qquad i=0,\ldots,N-1,\qquad L=N\delta.
\]
Indices are taken modulo $N$. If the nodes are generated on a closed interval
$[x_{\min},x_{\max}]$ with both endpoints included, then
$\delta=(x_{\max}-x_{\min})/(N-1)$ and the ring length is $N\delta$,
not $x_{\max}-x_{\min}$. Variance nodes need not be uniform; their generator
and conservative endpoint closure are those in Section~\ref{sec:theory}.

We retain the price-direction CTMC rates defined in
\eqref{eq:x_rates_new}. On the uniform transformed-price grid these interior
rates are translation invariant for each fixed variance state $v_j$. The
periodic operator used by the Fourier method changes only the boundary closure:
the price index is interpreted modulo $N$. Thus, for a price vector $w$,
\begin{equation}
 (C_jw)_i=
 q^{x,j}_{i,i-1}w_{(i-1)\bmod N}
 +q^{x,j}_{i,i}w_i
 +q^{x,j}_{i,i+1}w_{(i+1)\bmod N}.
 \label{eq:circulant_diffusion}
\end{equation}
The three rates on the right-hand side are independent of $i$ on the uniform
grid. Hence $C_j$ is circulant. The local CTMC stencil is exactly the one
already constructed for the reference chain; only the transformed-price
boundary closure is replaced by periodic wrapping. At interior nodes this is
the same moment-matching approximation as before. At the seam, the wrapped
transition is a periodic boundary realization rather than an ordinary
coordinate increment on $\mathbb R$.

\paragraph{Wrapped jump operator, Kou weights, and the meaning of $W$.}
Let $I_\ell=((\ell-1/2)\delta,(\ell+1/2)\delta]$.
For a general finite-activity jump law, define
\begin{equation}
 p_\ell=\sum_{z\in\mathbb Z}\nu_Z(I_\ell+zL),\qquad
 \ell=0,\ldots,N-1.
 \label{eq:wrapped_weights}
\end{equation}
These intervals partition the real line and $\sum_\ell p_\ell=1$.
A fixed translation-equivariant tie convention is used for cell endpoints.
Define
\begin{equation}
 c_\ell=\lambda p_\ell\ (\ell\ne0),\qquad
 c_0=-\lambda\sum_{\ell=1}^{N-1}p_\ell,\qquad
 (C_Jw)_i=\sum_{\ell=0}^{N-1}c_\ell w_{i+\ell}.
 \label{eq:periodic_jump}
\end{equation}
Jumps returning to the same ring cell have no effect. Endpoint-tail
allocation in \eqref{eq:cells} is replaced by wrapping, not by zero-padding
the endpoint-closed transition matrix.

The numerical experiments specialize to the Kou law~\cite{Kou2002},
\begin{align}
 f_Z(z)&=p\eta_+e^{-\eta_+z}\mathbf1_{z\ge0}
 +(1-p)\eta_-e^{\eta_-z}\mathbf1_{z<0},\\
 F_Z(z)&=\begin{cases}(1-p)e^{\eta_-z},&z<0,\\
 1-pe^{-\eta_+z},&z\ge0,\end{cases}\\
 \kappa_J&=\frac{p\eta_+}{\eta_+-1}
 +\frac{(1-p)\eta_-}{\eta_-+1}-1,
 \label{eq:kou_cdf}
\end{align}
with $0<p<1$, $\eta_+>1$ and $\eta_->0$.
The infinite sum appears because a jump distribution with unbounded support
has infinitely many periodic images on a ring of length $L$: jump sizes that
differ by an integer multiple of $L$ land in the same periodic price cell.
Equivalently, the periodic chain replaces the real-line jump destination
$x+Z$ by its location modulo $L$, so a jump crossing one end of the
computational interval re-enters from the opposite end. This wrapping is
part of the periodic approximation; the separate winding error below arises
only from truncating the infinite image sum at finite $W$.
We therefore introduce a \emph{winding radius} $W$. It is the number of
periodic image intervals retained on each side of the fundamental price
interval. It is not the number of Fourier modes, an FFT tolerance, or a
frequency cutoff; it controls only the truncation of the wrapped jump law.

For numerical evaluation the infinite winding sum is truncated at a finite
radius:
\begin{equation}
 p_\ell^{[W]}=\sum_{z=-W}^{W}
 \left[F_Z((\ell+1/2)\delta+zL)
       -F_Z((\ell-1/2)\delta+zL)\right].
 \label{eq:finite_winding}
\end{equation}
Replace $p_\ell$ by $p_\ell^{[W]}$ in \eqref{eq:periodic_jump} and reset the
diagonal by the off-diagonal row sum to obtain $C_J^{[W]}$.
No renormalization of retained jump probabilities is required: omitted jumps
are suppressed, and the resulting generator remains conservative.
Before self-cell exclusion, the omitted probability is
\begin{equation}
 \varepsilon_W=1-\sum_{\ell=0}^{N-1}p_\ell^{[W]}
 =F_Z(-WL-\delta/2)+1-F_Z((W+1)L-\delta/2).
 \label{eq:winding_tail}
\end{equation}
For $W\ge1$ it is explicitly
$(1-p)e^{-\eta_-(WL+\delta/2)}
 +p e^{-\eta_+((W+1)L-\delta/2)}$.
The winding radius $W$ is initialized using the slowest exponential tail and a
requested alias tolerance (set to $10^{-14}$ in the numerical experiments),
with a safety increment and a cap of $10^4$. The actual tail in
\eqref{eq:winding_tail}, rather than that initialization rule, determines
whether the requested tolerance is met. An active cap must be reported. CDF
differencing, nonnegative clipping of roundoff-level weights, and diagonal
summation add floating-point errors; materially negative weights require
diagnosis, not silent clipping.
\subsection{Exact Fourier reduction of the joint CTMC}
\label{subsec:fourier-reduction}

The periodic price construction makes the joint finite-state generator
translation invariant in the price index.  With $m=N_v$ variance states and
$N=N_x$ price states, the generator is
\begin{equation}
 Q_{h,\mathrm{per}}^{[W]}
 =Q_V^h\otimes I_N
 +\operatorname{diag}(C_1,\ldots,C_m)
 +I_m\otimes C_J^{[W]}.
 \label{eq:periodic_full}
\end{equation}
The matrices $C_j$ describe the conditional price drift--diffusion at variance
state $v_j$, and $C_J^{[W]}$ is the wrapped jump generator.  All of them are
circulant in the price coordinate.  The pricing semigroup is
$e^{t(Q_{h,\mathrm{per}}^{[W]}-rI)}$.

\paragraph{Fourier transform in the price direction.}
For a value surface $U=(U_{j,i})\in\mathbb R^{m\times N}$, take the discrete
Fourier transform only across the price index:
\begin{equation}
 \widehat U_{j,k}
 =\sum_{i=0}^{N-1}U_{j,i}e^{-\mathrm i\omega_k i},
 \qquad
 \omega_k=\frac{2\pi k}{N},
 \qquad k=0,\ldots,N-1.
 \label{eq:dft_convention}
\end{equation}
For each frequency, collect the variance components into the vector
\[
 \widehat U_k
 =\bigl(\widehat U_{1,k},\ldots,\widehat U_{m,k}\bigr)^\top
 \in\mathbb C^m.
\]
Thus the Fourier transform removes the price index but leaves all $m$ variance
states coupled.

On the uniform price grid, write $q_{j,-},q_{j,0},q_{j,+}$ for the three
nearest-neighbour rates in $C_j$.  Since $C_j$ is circulant,
\begin{equation}
 \widehat{(C_jU_j)}_k
 =\left(q_{j,-}e^{-\mathrm i\omega_k}
       +q_{j,0}
       +q_{j,+}e^{\mathrm i\omega_k}\right)\widehat U_{j,k}.
 \label{eq:local_fourier_action}
\end{equation}
Likewise, from $(C_J^{[W]}w)_i=\sum_{\ell=0}^{N-1}c_\ell^{[W]}w_{i+\ell}$,
\begin{equation}
 \widehat{(C_J^{[W]}U_j)}_k
 =\left(\sum_{\ell=0}^{N-1}c_\ell^{[W]}e^{\mathrm i\omega_k\ell}\right)
   \widehat U_{j,k}.
 \label{eq:jump_fourier_action}
\end{equation}
The opposite signs in \eqref{eq:local_fourier_action}--\eqref{eq:jump_fourier_action}
come directly from the shifts $i-1$ and $i+\ell$ under the convention
\eqref{eq:dft_convention}.  In an FFT library this means that the jump
multiplier is obtained from $N\operatorname{IFFT}(c^{[W]})$ under this
convention.

\paragraph{The variance block at one Fourier frequency.}
Equations~\eqref{eq:local_fourier_action} and
\eqref{eq:jump_fourier_action} show that, at frequency $\omega_k$, the full
joint generator acts on the $m$-vector $\widehat U_k$ through
\begin{equation}
 \widehat Q_k^{[W]}
 =Q_V^h
 +\operatorname{diag}_{j=1,\ldots,m}\!\left(
      q_{j,-}e^{-\mathrm i\omega_k}
      +q_{j,0}
      +q_{j,+}e^{\mathrm i\omega_k}
      +\sum_{\ell=0}^{N-1}c_\ell^{[W]}e^{\mathrm i\omega_k\ell}
   \right).
 \label{eq:fourier_block_generator}
\end{equation}
The key structural point is that $Q_V^h$ remains inside every frequency
block.  The Fourier transform diagonalizes only the homogeneous price
direction; it does not separate or discard the variance dynamics.

\begin{theorem}[Fourier-block semigroup]
\label{thm:fft_block}
For every frequency $k$ and every surface $U$,
\begin{equation}
 \widehat{\bigl(Q_{h,\mathrm{per}}^{[W]}U\bigr)}_k
 =\widehat Q_k^{[W]}\widehat U_k.
 \label{eq:fourier_generator_action}
\end{equation}
Consequently, for $t\ge0$,
\begin{equation}
 \widehat{\left(e^{t(Q_{h,\mathrm{per}}^{[W]}-rI)}U\right)}_k
 =e^{t(\widehat Q_k^{[W]}-rI_m)}\widehat U_k.
 \label{eq:fourier_mode_propagation}
\end{equation}
Equivalently, the DFT block-diagonalizes the specified periodic finite-state
generator into $N$ independent $m\times m$ blocks.
\end{theorem}

\begin{proof}
The variance term $Q_V^h\otimes I_N$ acts only on the variance index and is
therefore unchanged by a Fourier transform in price.  Each $C_j$ and
$C_J^{[W]}$ is circulant, so their actions at frequency $\omega_k$ are the
scalar multipliers in \eqref{eq:local_fourier_action} and
\eqref{eq:jump_fourier_action}.  Combining these terms gives
\eqref{eq:fourier_generator_action}.  The semigroup identity
\eqref{eq:fourier_mode_propagation} follows because matrix exponentiation
preserves similarity transformations.
\end{proof}

This reduction is exact for the periodic finite-state generator; it is not a
low-rank approximation.  The total state dimension remains $mN$, but the
representation changes from one coupled $mN$-state operator to $N$ independent
$m\times m$ frequency blocks.  For real value surfaces,
$\widehat Q_{N-k}^{[W]}=\overline{\widehat Q_k^{[W]}}$ and
$\widehat U_{N-k}=\overline{\widehat U_k}$, so rFFT/irFFT needs only
$N_f=\lfloor N/2\rfloor+1$ blocks.  This use of conjugate symmetry is lossless;
it is not a frequency cutoff.  At $k=0$, all price-generator row sums vanish,
so $\widehat Q_0^{[W]}=Q_V^h$, which provides a simple structural check.
\subsection{Bermudan continuation with reusable Fourier-block propagators}
\label{subsec:bermudan-fourier}

For a Bermudan step of length $\Delta t$, the continuation calculation is
therefore
\[
 U^{n+1}
 \xrightarrow{\mathrm{rFFT}_X}
 \widehat U^{n+1}_k
 \xrightarrow{\ e^{\Delta t(\widehat Q_k^{[W]}-rI_m)}\ }
 \widehat C^n_k
 \xrightarrow{\mathrm{irFFT}_X}
 C^n.
\]
The obstacle is then applied in real space,
$U^n=\max\{\Phi,C^n\}$.  Because the obstacle is nonlinear, Fourier modes
cannot be exercised independently; each exercise date requires a return to
real space.  For a fixed model, grid, winding radius, rate, and interval
length, the block exponentials can be precomputed and reused across all
exercise dates with that interval length.

\begin{remark}[Relation to FST, PROJ, and splitting]
\label{rem:fst-proj}
The propagation formula \eqref{eq:fourier_mode_propagation} has the same
operator architecture as regime-switching FST: at each Fourier frequency, a
finite-state coupling matrix is exponentiated and applied directly to the
Fourier coefficients of the value function.  The difference is how the
finite states and the Fourier symbol arise.  In regime-switching FST the
finite states are part of the model and the price symbol comes from a
continuous homogeneous process.  Here the variance states approximate a
continuous diffusion, while the price multiplier comes from a discretized
periodic CTMC. The resulting real-space semigroup inherits positivity and
probability conservation from the joint generator. Fourier transformation
preserves this semigroup exactly; these Markov properties concern the
assembled real-space operator, not individual complex-valued Fourier blocks.

PROJ also begins with a finite-state approximation of stochastic volatility
and obtains a frequency-dependent matrix transform. It then uses frame
projection to construct a transition-density/kernel representation and applies
the continuation operator through convolution. The extra density/projection
representation gives PROJ substantial modeling flexibility, while introducing
a separate resolution choice whose numerical effect is examined in
Section~\ref{sec:numerics}. The present method moves the Fourier step to the
generator level: it keeps the fully discrete joint CTMC and applies
$e^{\Delta t(\widehat Q_k^{[W]}-rI_m)}$ directly to the Fourier-transformed
value surface. Thus both approaches combine finite volatility states with
Fourier structure, but the object being propagated is different: a projected
transition representation in PROJ versus the exact semigroup of the specified
discrete generator here.

In this sense, the Fourier-block construction brings the frequency-domain
continuation idea of FST and the finite-state Fourier coupling of CTMC--PROJ
into the joint CTMC semigroup itself. For the same continuation task,
directional splitting composes price and variance propagators and introduces a
propagation approximation when the directional generators do not commute.
Fourier-block propagation instead retains their joint action within each
frequency block. This makes splitting the principal architectural comparator
in our numerical study. The comparison implementations use the respective
boundary conventions stated below, so their observed price differences are
not pure splitting errors.
\end{remark}
\section{Algorithm, error analysis, and computational complexity}
\label{sec:architectures}

This section presents the backward pricing algorithm, its error analysis,
and the setup, storage, and computational costs of propagator reuse.

\subsection{Backward pricing algorithm and propagator reuse}
\label{subsec:pricing-algorithm}

For $\mathcal E=\{0=t_0<\cdots<t_M=T\}$, let $\phi$ be the sampled payoff
and $\Delta t_n=t_{n+1}-t_n$. The exact periodic-chain Bermudan values are
\begin{equation}
 u^M=\phi,\qquad u^n=\max\left\{\phi,
     e^{\Delta t_n(Q_{h,\mathrm{per}}^{[W]}-rI)}u^{n+1}\right\}.
 \label{eq:fft_stopping}
\end{equation}
By backward induction and the Markov property, this is the optimal stopping
value of the specified periodic chain on $\mathcal E$. It takes values in
$[0,K]$, is monotone in terminal/obstacle values, and is exact before
floating-point propagation. It is a Bermudan approximation to the
continuous-exercise problem, not an exact continuous-time American solver.

\begin{algorithm}[htbp]
\caption{FFT-block CTMC pricing with reusable half-spectrum propagators.}
\label{alg:fft}
\begin{algorithmic}[1]
\Require Uniform price nodes, variance generator, model parameters, rate $r$,
exercise dates, payoff $\Phi$, winding radius $W$
\State Construct the periodic price blocks $C_j$ from the rates in \eqref{eq:x_rates_new} and construct $c^{[W]}$; record $\varepsilon_W$.
\State Form the Fourier blocks $\widehat Q_k^{[W]}$ from
\eqref{eq:fourier_block_generator}, $0\le k\le\lfloor N/2\rfloor$.
\For{each distinct interval $\Delta$ in the exercise schedule}
\State Cache $\exp\!\left(\Delta(\widehat Q_k^{[W]}-rI_m)\right)$ for every retained $k$.
\EndFor
\State $U\gets\Phi$.
\For{$n=M-1,\ldots,0$}
\State $\widehat U\gets\operatorname{rFFT}_X(U)$.
\State $\widehat C_k\gets
\exp\!\left(\Delta t_n(\widehat Q_k^{[W]}-rI_m)\right)\widehat U_k$ for every $k$.
\State $C\gets\operatorname{irFFT}_X(\widehat C;N)$.
\State $U\gets\max\{\Phi,C\}$ componentwise.
\EndFor
\State \Return interpolation of $U$ at the required initial state(s).
\end{algorithmic}
\end{algorithm}

Discounting is already inside the cached block exponentials and must not be applied a second time.
A cached propagator is valid only for the same model parameters, grid and
boundary choices, winding radius, rate, and interval length. Changing the
payoff or strike does not require rebuilding it. For equally spaced dates
there is a single family of cached matrices. The numerical implementation
evaluates frequency products using batched matrix multiplication and reuses
work arrays.

\paragraph{Strike normalization and off-grid evaluation.}
For multiplicative futures dynamics,
\[
 P(t,F,v;K)=Kp(t,F/K,v;1).
\]
Using $y=\log(F/K)-\rho(v-V_0)/\sigma$ gives the strike-independent payoff
\[
 \phi_1(y,v)=\left[1-e^{y+\rho(v-V_0)/\sigma}\right]^+.
\]
One normalized surface can be sampled at $y_0(K)=\log(F_0/K)$ and multiplied
by $K$ for each strike. All requested initial states must lie in a validated
interior region of the same normalized grid.
Off-grid initial values are obtained with tensor-product interpolation, cubic
in each direction when at least four nodes are available and of degree one less
than the number of available nodes otherwise. The final normalized value is
floored at $\max\{1-F_0/K,0\}$ and multiplied by $K$. Cubic interpolation need
not preserve monotonicity or the upper price bound even though the on-grid
recursion does. No interpolation is used between propagation and application
of the obstacle.
\subsection{Error decomposition and stability}
\label{subsec:error-analysis}

The Fourier block diagonalization itself is exact for the specified periodic
finite-state generator.  The pricing error therefore comes from the model
discretization and numerical layers surrounding that identity.  The following
results distinguish propagation error, winding
truncation, periodization, spatial approximation, exercise refinement, and
off-grid interpolation.

\paragraph{Numerical propagation error through exercise.}
Let
$P_n=e^{\Delta t_n(Q_{h,\mathrm{per}}^{[W]}-rI)}$ and suppose the computed
continuation on its actual input $\widetilde u^{n+1}$ satisfies
\[
 \|\widetilde c^n-P_n\widetilde u^{n+1}\|_\infty\le\epsilon_n.
\]
This error includes small-matrix exponential errors, FFT and multiplication
roundoff, and any explicitly introduced numerical approximation.
\begin{proposition}[Stopping stability]
\label{prop:fft_error}
With a common terminal payoff,
\begin{equation}
 \|u^n-\widetilde u^n\|_\infty
 \le\sum_{\ell=n}^{M-1}e^{-r(t_\ell-t_n)}\epsilon_\ell.
 \label{eq:fft_error_budget}
\end{equation}
\end{proposition}
\begin{proof}
The obstacle map is nonexpansive in the infinity norm and
$\|P_n\|_\infty=e^{-r\Delta t_n}$. Consequently
$e_n\le e^{-r\Delta t_n}e_{n+1}+\epsilon_n$.
Backward substitution proves the assertion.
\end{proof}
The bound applies to the actual continuation error, including numerical
block propagation and floating-point arithmetic.

\paragraph{Winding truncation and periodization are different errors.}
The exact infinite-winding and finite-winding matrices satisfy
\begin{equation}
 \|Q_{h,\mathrm{per}}-Q_{h,\mathrm{per}}^{[W]}\|_\infty
 \le2\lambda\varepsilon_W.
 \label{eq:winding_generator_error}
\end{equation}
Indeed, the total missing off-diagonal rate is at most
$\lambda\varepsilon_W$ and diagonal correction contributes the same amount.
Duhamel's formula and contractivity give
\[
 \|e^{t(Q_{h,\mathrm{per}}-rI)}
      -e^{t(Q_{h,\mathrm{per}}^{[W]}-rI)}\|_\infty
 \le2\lambda t e^{-rt}\varepsilon_W.
\]
Applying this estimate to the bounded value vectors in the stopping recursion
bounds the winding-induced value error by $2K\lambda T\varepsilon_W$.
This controls omitted winding probabilities, not the error from identifying
the two ends of the price domain.

Price-domain expansion at nearly fixed spacing is used below to assess
periodization sensitivity at the tested initial state.

\paragraph{Spatial approximation and the American limit.}
\begin{proposition}[Convergence of the expanding periodic family]
\label{prop:periodic-convergence}
Under Assumption~\ref{ass:wellposed}, suppose the variance grids and closures
satisfy Assumption~\ref{ass:grid},
the price grids are uniform with $\delta_h\to0$, and
$x_{\min}^h\to-\infty$, $x_{\max}^h\to+\infty$. Their adjacent seam
nodes likewise escape every fixed price compact set. If
$\varepsilon_{W_h}\to0$, then for every $f\in C_c^\infty(D)$,
\[
 \|Q_{h,\mathrm{per}}^{[W_h]}\pi_h f-\pi_h\cL f\|_{\infty,E_h}\to0.
\]
The conclusions of Theorem~\ref{thm:kurtz_semigroup},
Corollary~\ref{cor:values}, and the sequential limit in
Corollary~\ref{cor:american} hold for this family.
\end{proposition}
The proof is given in Appendix~\ref{app:proof-periodic-convergence}.

This proposition requires an expanding price domain and a consistent lower
variance closure. Refinement on the fixed production ring and at a fixed
positive variance lower endpoint does not by itself invoke this theorem.
No joint spatial/exercise convergence rate is asserted.

For a common initial grid point, a useful bookkeeping inequality is
\begin{align}
 |V^A-\widetilde V_{\mathrm{per},W}^{h,\mathcal E}|
 \le{}&|V^A-V^{\mathcal E}|
 +|V^{\mathcal E}-V_{\mathrm{end}}^{h,\mathcal E}|\nonumber\\
 &+|V_{\mathrm{end}}^{h,\mathcal E}-V_{\mathrm{per}}^{h,\mathcal E}|
 +|V_{\mathrm{per}}^{h,\mathcal E}-V_{\mathrm{per},W}^{h,\mathcal E}|
 +\epsilon_{\mathrm{num}}.
 \label{eq:error_decomposition}
\end{align}
The five terms correspond to exercise dates, reference-chain approximation,
periodization, winding truncation, and numerical propagation. Interpolation
at an off-grid initial state adds its own error. Refining only $N_x,N_v$ on
a fixed ring does not remove periodization error. Uniform-norm convergence
of the nonperiodic payoff's Fourier series is not required for the finite
DFT identity and is not claimed.
\subsection{Computational complexity, storage, and reuse}
\label{subsec:complexity}

Let $N_f=\lfloor N/2\rfloor+1$ and let $D$ be the number of distinct
exercise intervals. Constructing finite-winding weights costs $O(WN)$;
computing their symbol costs $O(N\log N)$. Local rates and grid surfaces cost
at most $O(mN)$. Dense small-matrix exponentiation gives the conventional
setup estimate $O(DN_fm^3)$, with constants depending on the exponential
algorithm and matrix norms. A full cache stores $O(DN_fm^2)$ complex numbers;
forming all block generators in a batch temporarily adds $O(N_fm^2)$.
Streaming their construction can reduce that temporary allocation.

One dense-block backward step costs
\begin{equation}
 O(mN\log N)+O(N_fm^2)+O(mN)
 =O(mN\log N+N_fm^2).
 \label{eq:fft_complexity}
\end{equation}
For one equally spaced schedule the total leading dense-block cost is
\[
 O(WN+N\log N+mN+N_fm^3
       +M[mN\log N+N_fm^2]).
\]
Memory is $O(N_fm^2+mN)$ with a single dense propagator family.

The method avoids forming an $(mN)\times(mN)$ matrix exponential. No exponential or Krylov
basis is rebuilt online while the cached parameters remain fixed. Nonuniform
dates may require multiple caches; changed parameters usually require fresh
block exponentials. No automatic calibration speedup is inferred from
warm-cache timing.

\begin{table}[htbp]
\centering\small
\caption{Leading costs under the stated implementations; $d=mN$ and one interval length.}
\label{tab:architecture}
\begin{tabular}{llll}
\toprule
Method & Propagator setup & Per interval & Main storage\\
\midrule
FFT & $O(N_fm^3)$ & $O(mN\log N+N_fm^2)$ & $O(N_fm^2+d)$\\
Layer split & $O(mN^3+m^3)$ & $O(mN^2+Nm^2)$ & $O(mN^2+m^2)$\\
\bottomrule
\end{tabular}
\tablenote{The layer-split row describes stored dense price-layer and variance
transitions used in the benchmark implementation.}
\end{table}

The benchmarks below compare setup, online recursion, and total computational
cost for the same pricing tasks. The stated grids, exercise schedules, and reuse
settings specify the computational workload for each method.

\section{Numerical validation and benchmark comparisons}
\label{sec:numerics}
\subsection{Benchmark models and numerical settings}
\label{subsec:benchmark-settings}
The common-grid benchmarks compare computational behavior under matched
spatial grids and exercise intervals. They provide a controlled view of
pricing behavior and setup/recursion cost under the stated configurations;
matching these inputs does not imply equal pricing accuracy across methods.
In particular, projection resolution and splitting substeps are not optimized
separately for each method in these tables. The dedicated FFT--splitting
exercise-mesh study in Section~\ref{subsec:spatial-exercise-sensitivity}
complements this comparison by examining price differences and computational
cost as the time discretization is refined.

The synthetic benchmark uses $F_0=40$, $V_0=0.04$, $r=0.05$, $T=5$ years,
and strikes $K\in\{35,40,45\}$. Exercise schedules use
$M\in\{500,1000,2000\}$ equally spaced intervals, so
$\Delta t=T/M$. Two parameter sets are considered:
\begin{table}[htbp]\centering\small
\caption{Synthetic benchmark parameters.}\label{tab:scenario-parameters}
\begin{tabular}{lrrrrrrrr}\toprule
Scenario &$\kappa$&$\theta$&$\sigma$&$\rho$&$\lambda$&$p$&$\eta_+$&$\eta_-$\\\midrule
A&3.00&0.04&0.30&$-0.60$&0.25&0.35&25&15\\
B&2.50&0.06&0.45&$-0.70$&0.50&0.30&20&10\\\bottomrule
\end{tabular}\end{table}

The transformed-price/variance grids contain $700\times120$ states---the fixed
benchmark configuration used throughout this section. It is distinct from the
$700\times112$ and $980\times157$ configurations appearing in the common-grid
cost study below and in the empirical refinement diagnostics, which form a
separate refinement family that refines both directions proportionally.
Scenario A
uses $x\in[-1.5,1.5]$ and $v\in[10^{-4},0.60]$; Scenario B uses
$x\in[-2,2]$ and $v\in[10^{-4},0.80]$. We use the short method labels
\emph{Poly}, \emph{FFT}, \emph{Layer split}, \emph{PROJ}, and
\emph{PROJ-HiRes}. Poly is the unsplit endpoint-closed CTMC propagated by the
polynomial semigroup approximation and provides a nonperiodic CTMC reference.
FFT is Algorithm~\ref{alg:fft}. Layer split follows the published
price--variance product/splitting architecture represented by
\cite{ZhangWangLi2026}, with the endpoint closure of the CTMC reference. PROJ
follows the density-projection/FFT-convolution framework of
\cite{Kirkby2015,KirkbyNguyenCui2017}. These comparators follow numerical
architectures from the cited literature, with the CTMC grids and boundary
closures specified here. They are evaluated on the common model parameters,
contracts, and exercise schedules, using the computational domains stated
above; method-specific propagation and boundary treatments are described below.
PROJ-HiRes doubles the complete price/projection grid. LSM provides an
independent stochastic cross-check. Because Poly and FFT use different price
closures, their difference includes boundary effects.

The layer-split implementation applies the conditional price transition first
to a backward value surface and then the variance transition, once per
exercise interval and without substeps. In operator notation its continuation
action is $P_V(P_XU)$, with discounting included once. The order stated here
is the order of actions on values, not a claim that the two operators commute.
The PROJ implementation uses a linear projection basis and
$N_{\mathrm{dens}}=2N_x$ density coefficients, with frequency spacing
$2\pi/(N_{\mathrm{dens}}\Delta x)$. Its transformed-process
regime-switching characteristic exponent uses the same variance generator
$Q_V$ as the CTMC comparators, but does not discretize the price generator
with the CTMC price stencil. The density padding and the value-grid resolution
are therefore distinct settings. Supplementary real-space kernel and
finite-iteration continuation diagnostics are reported in
Appendix~\ref{subsec:proj-operator-diagnostic}. Their scope is the four
tested Scenario A grids at $\Delta t=0.005$, not all reported PROJ runs.

The LSM implementation simulates log futures and variance with Euler
substeps, projects variance to nonnegative values at each substep, and uses
correlated Gaussian increments and compensated Kou jumps. Jump counts are
Poisson and the summed upward and downward jump sizes are sampled through
Gamma distributions. Each of 300 replications uses 60,000 paths and four
simulation substeps per exercise interval. Regression uses in-the-money
paths only and the six-term basis $(1,f,f^2,v,v^2,fv)$, where $f$ is the
simulated futures price itself, not its logarithm, and $v$ is simulated
variance. With fewer
than six in-the-money paths, that step permits no early exercise. The same
path panel is used for regression and valuation within a replication and
is shared across the three strikes. Independent replication streams are
generated by \texttt{SeedSequence(23).spawn(300)}. For replication estimates
$\widehat V_b$, the reported uncertainty is
\[
 \operatorname{SE}_{\mathrm{rep}}
 =\frac{\operatorname{sd}(\widehat V_1,\ldots,\widehat V_{300})}{\sqrt{300}},
\]
using the sample standard deviation with denominator 299. Sharing paths
across strikes correlates the strike estimates within a replication; it
does not change this marginal standard-error definition.

Both state grids in the benchmark are uniform. The local endpoints use
\eqref{eq:endpoint_closures}, while the periodic price stencil uses the same
conditional interior rates with periodic wrapping. Fourier blocks are
exponentiated with SciPy matrix exponentials, and the endpoint CTMC is
propagated by uniformization. For a uniformization rate
$q\ge\max_i(-Q_{ii})$,
\[
e^{\Delta Q}=e^{-q\Delta}\sum_{k=0}^{\infty}
       \frac{(q\Delta)^k}{k!}(I+Q/q)^k.
\]
For an unrenormalized truncation after degree $L$, let
$\epsilon_{\mathrm{Pois}}=\mathbb P\{\operatorname{Pois}(q\Delta)>L\}$.
Because $I+Q/q$ is stochastic, the omitted series has infinity norm
$\epsilon_{\mathrm{Pois}}$; its discounted action on a value vector bounded
by $K$ is therefore at most $Ke^{-r\Delta}\epsilon_{\mathrm{Pois}}$.
This propagation tolerance is separate from the spatial grid and the
jump-winding tolerance, and its accumulation through exercise is governed
by Proposition~\ref{prop:fft_error}.
The implementation chooses
$q=(1+32\epsilon_{\mathrm{mach}})\max_i(-Q_{ii})$ from the full endpoint
generator, including variance, local-price and jump escape rates. It uses
the first degree whose Poisson tail is at most $2\times10^{-8}$, with a
maximum degree of 512; exceeding that cap raises an error. There are no
propagation substeps or weight renormalizations. For $r\ge0$, the resulting
on-grid stopping-value error due only to this series truncation is at most
$MK(2\times10^{-8})$. At $M=1000$ and the largest benchmark strike $K=45$,
this conservative bound is $9\times10^{-4}$; it excludes spatial,
boundary, exercise and final interpolation errors.

The production benchmark uses $M=1000$, corresponding to 200 exercise dates
per year over the five-year horizon; the market application of
Sections~\ref{sec:data}--\ref{sec:empirical} uses 100 per year. The
synthetic benchmarks exercise more frequently to stress the repeated
propagation and obstacle recursion, while the market setting aligns with
common conventions for exchange-traded options. The $M=2000$ rows probe
finer exercise meshes.
Runtime results below use a repeated single-thread protocol: one warm-up followed
by five measured repetitions for setup and online recursion, with medians reported.
The recorded environment is Python 3.13.9, NumPy 2.3.5 and SciPy 1.16.3
on an AMD Ryzen processor (family 25, model 116). The repeated protocol
uses \texttt{JH\_BLAS\_THREADS=1} with \texttt{threadpoolctl}.
The exp10 single-shot runs instead set the OMP, OpenBLAS, MKL and NumExpr
thread-count environment variables to one before importing NumPy. These
thread settings do not make the two timing protocols interchangeable.
LSM is used only as an independent price benchmark and is not included in the
deterministic runtime comparison.
\subsection{Pricing and runtime benchmarks}
\begin{table}[htbp]\centering\scriptsize
\caption{Scenario A: five-year Bermudan put values on the $700\times120$ grid.}
\label{tab:prices-A}
\begin{tabular}{rlrrr}\toprule
$M$&Method&$K=35$&$K=40$&$K=45$\\\midrule
1000 & Poly & 3.834043 & 6.073339 & 8.864995\\
 & FFT & 3.836746 & 6.073954 & 8.865182\\
 & LayerSplit & 3.838159 & 6.079192 & 8.872647\\
 & PROJ & 3.834113 & 6.073435 & 8.865109\\
 & PROJ-HiRes & 3.834114 & 6.073436 & 8.865111\\
 & LSM benchmark & 3.822691 & 6.053149 & 8.835820\\
2000 & Poly & 3.834147 & 6.073508 & 8.865248\\
 & FFT & 3.836928 & 6.074166 & 8.865476\\
 & LayerSplit & 3.836226 & 6.076465 & 8.869113\\
 & PROJ (unstable) & 0.000000 & 0.000000 & 5.000000\\
 & PROJ-HiRes & 3.834240 & 6.073636 & 8.865402\\
\bottomrule\end{tabular}
\par\smallskip\footnotesize At $M=1000$, the LSM benchmark uses 60,000 paths per replication, 300 replications, and four simulation substeps per exercise interval. The replication standard errors for $K=35,40,45$ are $1.124\times10^{-3}$, $1.450\times10^{-3}$, and $1.734\times10^{-3}$, respectively.
\end{table}

\begin{table}[htbp]\centering\scriptsize
\caption{Scenario B: five-year Bermudan put values on the $700\times120$ grid. ``PROJ-HiRes'' doubles the price direction only; it is not a projection-only refinement.}
\label{tab:prices-B}
\begin{tabular}{rlrrr}\toprule
$M$&Method&$K=35$&$K=40$&$K=45$\\\midrule
1000 & Poly & 5.143006 & 7.474694 & 10.245749\\
 & FFT & 5.143109 & 7.474754 & 10.245809\\
 & LayerSplit & 5.151393 & 7.485732 & 10.259420\\
 & PROJ & 7.436070 & 9.389909 & 11.592346\\
 & PROJ-HiRes & 5.142424 & 7.473988 & 10.244964\\
 & LSM benchmark & 5.126478 & 7.447195 & 10.208793\\
2000 & Poly & 5.143104 & 7.474848 & 10.245977\\
 & FFT & 5.143287 & 7.475013 & 10.246164\\
 & LayerSplit & 5.147376 & 7.480472 & 10.252943\\
 & PROJ (unstable) & $7.24\times10^13$ & $2.02\times10^13$ & $4.35\times10^13$\\
 & PROJ-HiRes (unstable) & $8.79\times10^{8}$ & $4.99\times10^{8}$ & $1.75\times10^{8}$\\
\bottomrule\end{tabular}
\par\smallskip\footnotesize At $M=1000$, the LSM benchmark uses the same Monte Carlo design as in Scenario A. The replication standard errors for $K=35,40,45$ are $1.455\times10^{-3}$, $1.814\times10^{-3}$, and $2.073\times10^{-3}$, respectively.
\end{table}

Several points are important for interpretation. First, the stable Scenario A
difference between FFT and Poly is about $2.6$--$2.8\times10^{-3}$ in the most
sensitive strike and should not be labeled a Fourier propagation error,
because the two methods use different price-boundary conventions. Scenario B
shows a much smaller boundary discrepancy on the tested domain.

LSM is included as an auxiliary simulation-based reference rather than an
accuracy standard for Fourier-block propagation. LSM combines path simulation
with an estimated stopping policy, whereas the proposed method propagates the
semigroup of a discretized joint-state generator. Their price differences can
therefore reflect the different numerical approximations to continuation and
stopping, including simulation and regression effects, CTMC discretization,
boundary treatment, and exercise approximation. At $M=1000$, LSM lies below
the deterministic cluster; the FFT--LSM gaps are approximately 11--18 reported
standard errors across the six scenario--strike pairs. Sampling uncertainty
alone does not explain these differences, but the experiment does not identify
their allocation among the approximation effects above. Neither LSM nor
agreement among the deterministic methods is used as a continuous-model
accuracy certificate. The principal architectural comparison is with
directional splitting, an alternative approach to the joint continuation action.

Independent evaluation of a fitted stopping policy gives a lower bound in
expectation for the corresponding simulated stopping problem. The same-path
LSM estimate need not have this property: look-ahead and policy-suboptimality
effects can offset, while Euler discretization has no established bias sign
here. Replication standard errors capture sampling variability only; the
full interpretation is given in Appendix~\ref{subsec:lsm-two-sample}.

Two supplementary checks at the Scenario A $K=40$ production configuration
use 16 replications with 60,000 training paths and 60,000 independent
evaluation paths per replication. In the first, the paired in-minus-out
difference is $0.0153\pm0.0091$ (mean $\pm$ standard error); its 95\%
interval permits a materially sized same-path effect. In the second,
the same fitted policies are evaluated without refitting on a new stream
with coupled four-, eight-, and sixteen-substep simulations. The paired
sixteen-minus-four change is $-0.000524\pm0.000753$, supporting limited
substep sensitivity of the frozen policies in this configuration, not
vanishing simulation bias. The independently evaluated means are $6.0433$
in the first check and $6.0641$ at sixteen substeps in the second, compared
with FFT $6.073954$: the gaps are respectively 6.1 and 1.64 reported
standard errors. The two four-substep evaluation streams differ
by $0.021342\pm0.009385$ when paired by training replication, with a 95\%
interval $[0.001338,0.041345]$. Both checks are retained separately;
neither identifies how the FFT--policy difference divides between policy
suboptimality and the simulated versus CTMC pricing problems.
Appendix~\ref{subsec:lsm-two-sample} gives the design and
Table~\ref{tab:lsm-substep} reports the paired comparisons.

Second, the PROJ comparison illustrates the resolution requirements of its
intermediate transition-density representation. For a fixed positive variance
state with nondegenerate price diffusion, the local short-time diffusive
kernel has characteristic width proportional to $\sqrt{\Delta t}$ and peak
height proportional to $1/\sqrt{\Delta t}$. Thus reducing the propagation
interval at fixed spatial resolution can make the density increasingly
difficult to resolve. This local scaling describes concentration of a
probability density, not divergence of the exact transition operator; numerical
amplification concerns the resulting discrete projection/continuation action.

Under the stated common-grid configurations, several coarse settings produce
unusable continuation values, while increasing the complete price/projection
resolution can restore prices close to the deterministic cluster; for example,
PROJ-HiRes is close to the other deterministic methods in Scenario B at
$M=1000$. The observed behavior is consistent with an under-resolved density
representation. The supplementary diagnostics show finite-iteration
amplification on three tested coarse Scenario A grids and no net amplification
over the recorded run on the finest diagnostic grid. These results support
the resolution-based interpretation for the tested configurations without
identifying a unique cause for every unstable run.

The Fourier-block construction avoids this intermediate density-resolution
requirement by propagating the semigroup of the conservative joint generator
directly. Positivity and contractivity of the exact discrete semigroup,
together with the exercise obstacle, keep the on-grid CTMC recursion within
$[0,K]$. This structural stability does not eliminate spatial or boundary
approximation errors, and numerical propagation and off-grid interpolation
remain separate sources of error. The experiments demonstrate stable pricing
and favorable measured cost for the Fourier-block implementation under the
tested configurations, rather than a ranking of separately optimized methods
at equal continuous-model accuracy.

A same-spacing, nested-node domain check on the usable $700\times112$
Scenario A configuration yields an expanded-domain PROJ--FFT price difference
of $4.86\times10^{-5}$ (Table~\ref{tab:proj-domain-crosscheck}), providing a
price cross-check once the projection configuration is sufficiently resolved.

\paragraph{Repeated timings.}
The production-mesh timing experiment uses $T=5$ years and $M=1000$ exercise
intervals on the $700\times120$ grids in both scenarios. To reduce sensitivity
to transient machine load, each setup and online recursion is preceded by one
warm-up and then measured five times with BLAS/OpenMP restricted to one thread.
Table~\ref{tab:timing-M1000} reports the median wall-clock time together with the
observed minimum and maximum. ``Setup'' is reusable method-specific setup;
``online'' is the backward recursion for all three strikes using the cached
objects; and ``cold'' combines the median common model setup, median method setup,
and median online time. The components are benchmarked separately, so the cold
range is a derived diagnostic, not five end-to-end cold runs.

\begin{table}[htbp]\centering\scriptsize
\caption{Repeated $M=1000$ wall-clock timings in seconds. Setup and online entries are median
$[\min,\max]$ over five measured repetitions after one warm-up. Cold entries
are sums of separately measured component medians and ranges.}
\label{tab:timing-M1000}
\resizebox{\textwidth}{!}{%
\begin{tabular}{llrrr}\toprule
Scenario&Method&Reusable setup (s)&Online recursion (s)&Derived cold (s)\\\midrule
A&Poly&0.002 [0.001,0.004]&66.263 [64.235,145.875]&66.534 [64.466,146.178]\\
A&FFT&2.610 [2.564,2.623]&4.683 [4.507,4.920]&7.562 [7.301,7.843]\\
A&Layer split&18.892 [18.696,19.167]&17.684 [17.376,17.922]&36.845 [36.302,37.389]\\
A&PROJ&21.925 [21.802,22.082]&10.818 [10.728,10.859]&33.012 [32.761,33.241]\\
A&PROJ-HiRes&70.008 [69.464,70.386]&23.358 [22.983,23.465]&93.366 [92.448,93.851]\\
\addlinespace
B&Poly&0.001 [0.001,0.001]&49.337 [49.012,49.712]&49.440 [49.113,49.816]\\
B&FFT&2.244 [2.218,2.302]&4.983 [4.943,5.001]&7.329 [7.262,7.406]\\
B&Layer split&17.783 [17.749,17.972]&17.690 [17.115,18.476]&35.575 [34.964,36.552]\\
B&PROJ&16.354 [16.209,16.526]&10.963 [10.689,11.141]&27.419 [26.999,27.771]\\
B&PROJ-HiRes&56.151 [55.803,56.265]&23.373 [23.242,23.551]&79.524 [79.045,79.816]\\
\bottomrule
\end{tabular}}
\end{table}

Under the tested implementations and fixed grids, FFT has the smallest median
online and derived cold time. Scenario A Poly shows substantial variation,
which motivates reporting medians and ranges. These results demonstrate a
computational efficiency advantage for the Fourier-block method on the stated
pricing workloads. LSM remains a
price cross-check and is not included in these deterministic timings.

\begin{figure}[!htbp]
\centering
\safeincludegraphics[width=0.92\linewidth]{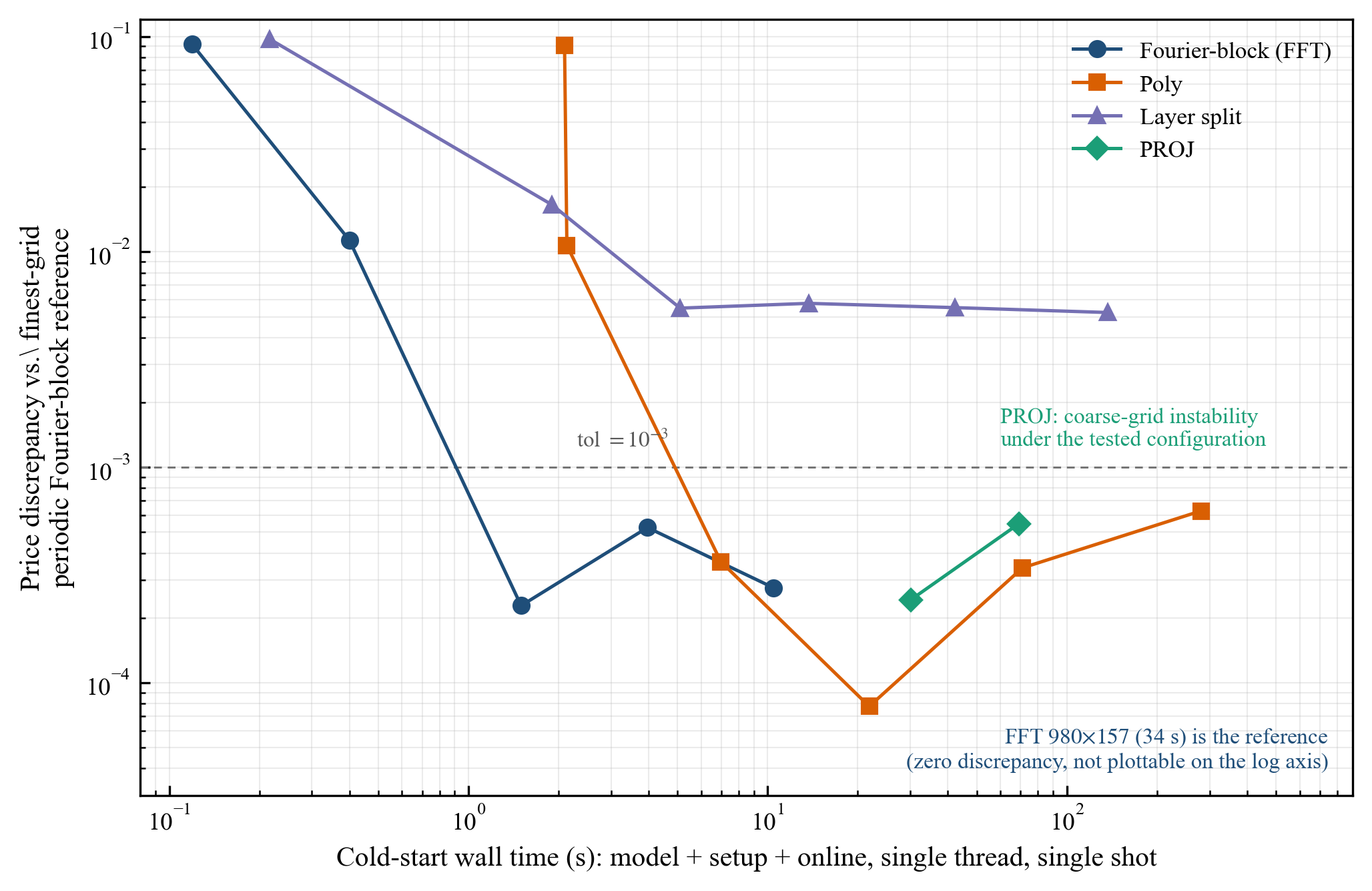}
\caption{Cost versus discrepancy from a finest-grid periodic
Fourier-block reference (Scenario~A, $K=40$, $T=5$ years, $M=1000$).
Each marker is one common-grid configuration from $150\times24$ to
$980\times157$ states; the horizontal axis is the single-shot
cold-start wall time (model + setup + online) under single-thread
BLAS. The reference is the FFT price on $980\times157$, not an
independent ground truth. Poly uses an endpoint-closed chain; the
layer split runs under the fixed splitting protocol; the coarser tested PROJ configurations do not yield usable prices and are not plotted.
Lines join measured configurations only.}
\label{fig:exp10-cost}
\end{figure}

\paragraph{Cost versus discrepancy from a finest-grid reference.}
\label{subsec:cost-discrepancy}
To complement the fixed-grid timings with a cost--discrepancy comparison,
Figure~\ref{fig:exp10-cost} and
Table~\ref{tab:exp10-cost-discrepancy} report a
six-level grid ladder from $150\times24$ to $980\times157$ states on the
Scenario A five-year configuration ($K=40$, $M=1000$): the same contract is
priced with all four propagators, and for each configuration we record the
single-shot cold-start wall time (model construction plus method setup plus
online recursion, single-thread BLAS) and the absolute price discrepancy from
a finest-grid periodic Fourier-block reference, namely the FFT price on
$980\times157$ (approximately $6.07375$). In particular, for method
$a$ and grid level $\ell$, the plotted quantity is
$d_{a,\ell}=|P_{a,\ell}-P_{\mathrm{FFT},\mathrm{ref}}|$, computed from
unrounded prices. The finest-grid value is a numerical reference for this
comparison, rather than an exact continuous-model price.

Among the non-reference configurations in this tested ladder, the
Fourier-block implementation has the smallest observed single-shot cold cost
at a discrepancy of at most $10^{-3}$ in price units: it first does so on the
$350\times60$ grid with a $1.5$-second cold start, whereas Poly first does so
on the same grid at $7.0$ seconds ($4.7\times$) and PROJ first does so from
$700\times112$ at $30.1$ seconds ($20\times$). The layer-split implementation
does not reach the $10^{-3}$ tolerance under the fixed splitting protocol (no
substep refinement): its discrepancy remains near $5\times10^{-3}$ on the
four finest tested grids. This residual discrepancy includes temporal
splitting error, which spatial refinement alone cannot remove, as well as
boundary effects. The exercise-mesh study in
Section~\ref{subsec:spatial-exercise-sensitivity} complements this fixed-step
comparison by reducing the propagation interval and recording the associated
price and cost changes. Poly uses an endpoint-closed price chain whereas the reference is
periodic, so Poly discrepancies include a boundary-closure component and
their non-monotonicity is not strict convergence to the periodic reference.
PROJ requires higher projection resolution on the coarser common-grid settings tested here;
those configurations are reported as unresolved/unstable in
Table~\ref{tab:exp10-cost-discrepancy} and are not plotted in
Figure~\ref{fig:exp10-cost}. Connecting lines in the figure join measured
configurations only and do not represent a continuous time--error relation.
The FFT reference configuration is excluded from tolerance-attainment
comparisons because its discrepancy is zero by definition. The quoted time
ratios summarize the measured single-shot cold-start costs at the stated
discrepancy threshold.

\begin{table}[htbp]\centering\footnotesize
\setlength{\tabcolsep}{3.5pt}
\caption{Cost versus discrepancy from a finest-grid periodic
Fourier-block reference (Scenario~A, $K=40$, $T=5$, $M=1000$). The
reference is the FFT price on $980\times157$ (6.07375); cold denotes a
single-shot model + setup + online time under single-thread BLAS.
\textbf{Bold} marks each method's first grid attaining a $10^{-3}$
discrepancy.}
\label{tab:exp10-cost-discrepancy}
\begin{tabular}{llrrrrrr}
\toprule
Method & Metric & $150\times24$ & $250\times40$ & $350\times60$ &
$500\times80$ & $700\times112$ & $980\times157$\\
\midrule
Fourier-block (FFT) & discrepancy & 9.20e-2 & 1.13e-2 & \textbf{2.28e-4} &
5.25e-4 & 2.76e-4 & ref\\
 & cold (s) & 0.1 & 0.4 & \textbf{1.5} & 4.0 & 10.4 & 34.2\\
\addlinespace
Poly & discrepancy & 9.12e-2 & 1.07e-2 & \textbf{3.64e-4} & 7.77e-5 &
3.42e-4 & 6.27e-4\\
 & cold (s) & 2.1 & 2.1 & \textbf{7.0} & 21.8 & 71.0 & 280.2\\
\addlinespace
Layer split & discrepancy & 9.71e-2 & 1.66e-2 & 5.48e-3 & 5.77e-3 &
5.51e-3 & 5.23e-3\\
 & cold (s) & 0.2 & 1.9 & 5.1 & 13.7 & 42.2 & 136.9\\
\addlinespace
PROJ & discrepancy & unstable & unstable & unstable & unstable & \textbf{2.43e-4} &
5.47e-4\\
 & cold (s) & -- & -- & -- & -- & \textbf{30.1} & 69.1\\
\bottomrule
\end{tabular}
\tablenote{PROJ cells marked \emph{unstable} do not yield usable prices below
$700\times112$ under the stated common-grid configuration; raw values are
not used in the price comparison. A dash denotes an omitted timing for an
unstable price configuration, not zero computational cost. Poly discrepancies include the
endpoint-versus-periodic boundary-closure component discussed in the text.
Layer split does not reach the $10^{-3}$ tolerance under the fixed splitting
protocol. Timings are single-shot cold runs, not repeated medians.}
\end{table}
\subsection{Spatial-grid and exercise-mesh sensitivity}
\label{subsec:spatial-exercise-sensitivity}
\paragraph{Spatial refinement.}
\label{subsec:spatial-self-convergence}

The five-year production benchmark above is complemented by a one-year
diagnostic designed to isolate the spatial behavior of the Fourier-block
scheme. We set $T=1$, $K=40$, and $M=1000$, retain the Scenario A and
Scenario B domains specified in Section~\ref{subsec:benchmark-settings}, and refine the
joint grid through
\[
\begin{gathered}
(50,8),\ (100,16),\ (200,32),\ (300,48),\ (400,64),\\
(500,80),\ (600,96),\ (720,120),\ (900,150),\ (1200,200).
\end{gathered}
\]
The $1200\times200$ calculation is used only as the finest-grid numerical
reference within the same fixed computational domain. It is not interpreted
as an exact solution and, because the price interval is not enlarged, this
experiment does not remove the periodic-boundary approximation.

\begin{figure}[!htbp]
\centering
\safeincludegraphics[width=\linewidth,keepaspectratio]{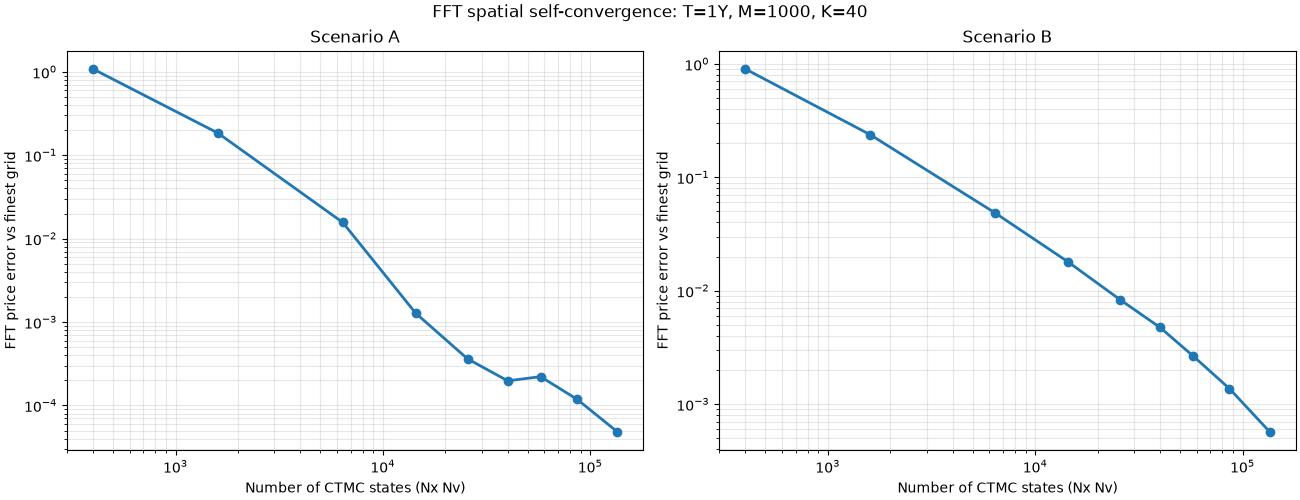}
\caption{FFT spatial self-convergence for Scenarios A and B with
$T=1$, $K=40$, and $M=1000$.
The vertical axis is the absolute price difference from the
$1200\times200$ FFT value on the same fixed domain.}
\label{fig:num-spatial}
\end{figure}

Figure~\ref{fig:num-spatial} shows clear stabilization in both scenarios, but
the difficult Scenario B requires a finer grid. In Scenario A, the FFT price
moves from $4.128070$ on the $50\times8$ grid to $3.046270$ on the
$1200\times200$ grid. The absolute difference from the finest-grid value is
$1.18\times10^{-4}$ at $720\times120$ and $4.83\times10^{-5}$ at
$900\times150$. The tail of the Scenario A sequence is not perfectly
monotone---the $500\times80$ and $600\times96$ points show a small oscillation---
so the figure is used as a self-convergence diagnostic, not as evidence
for a formal empirical order.

Scenario B converges more slowly on the same sequence. Its FFT price decreases
from $4.465184$ at $50\times8$ to $3.562900$ at $1200\times200$; the
corresponding finest-grid differences are $1.37\times10^{-3}$ at
$720\times120$ and $5.71\times10^{-4}$ at $900\times150$. This contrast is
consistent with Scenario B being numerically more demanding because it combines
larger volatility-of-volatility, stronger negative correlation, and more
intense jumps. The experiment therefore supports grid refinement as an
important part of the Fourier-block error budget, while leaving domain
expansion as a separate validation exercise.

\paragraph{Exercise-mesh refinement.}
\label{subsec:M-sensitivity}

We next fix the spatial grid at $700\times120$, set $T=1$ and $K=40$, and vary
the number of exercise intervals over
\[
M\in\{500,1000,2000,3000,5000\}.
\]
The FFT calculation uses the periodic price CTMC of
Algorithm~\ref{alg:fft}. The comparison method is the endpoint-closed
price--variance product implementation used as the literature splitting
benchmark. Their price difference includes both propagation-architecture and
boundary effects, so this $M$-sweep is not interpreted as a measurement of a
splitting convergence order. Its main purpose is to compare price behavior and
measured setup/online cost as the repeated-exercise workload increases.
The layer-split implementation uses one directional composition per exercise
interval, so increasing $M$ also refines its propagation step. This addresses
temporal refinement that is absent from the fixed-$M$ spatial ladder.
Because the exercise schedule changes at the same time, the experiment is
not a substep-only study at a fixed Bermudan schedule.

\begin{figure}[!htbp]
\centering
\safeincludegraphics[width=0.90\linewidth,keepaspectratio]{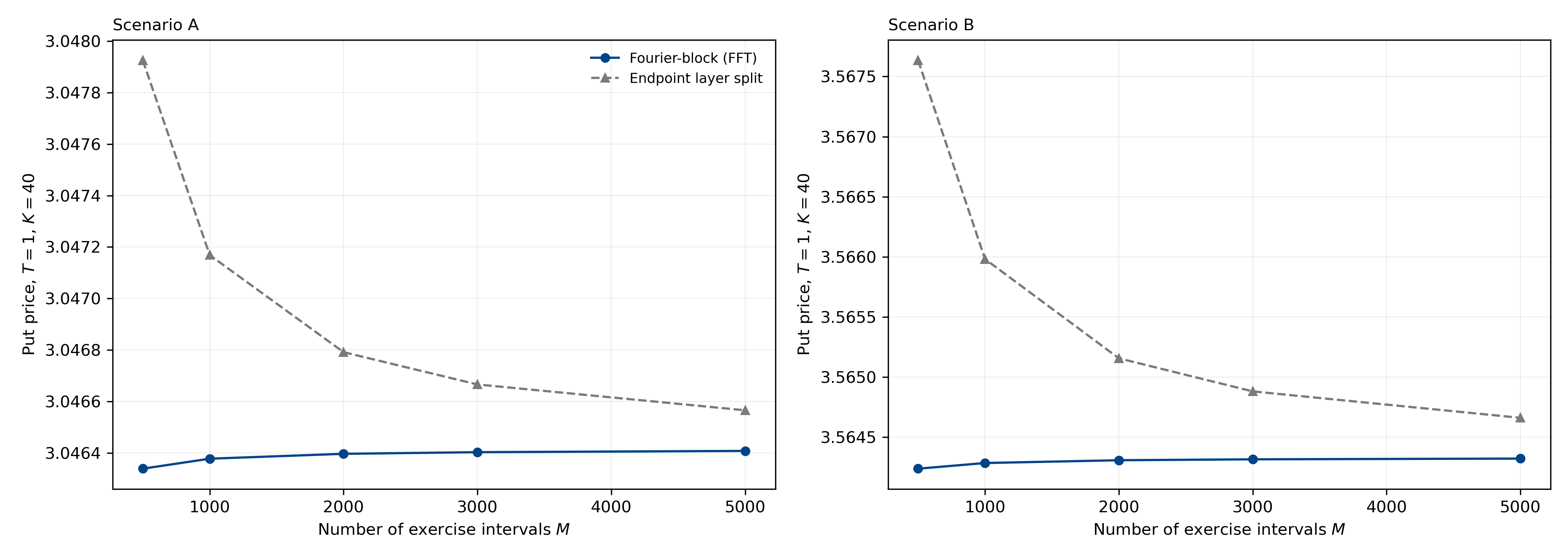}
\caption{Price sensitivity to the number of exercise intervals for the
Fourier-block (FFT) and Layer split methods on the common
$700\times120$ grid for $T=1$ and $K=40$.}
\label{fig:num-M-prices}
\end{figure}

\begin{figure}[!htbp]
\centering
\safeincludegraphics[width=0.90\linewidth,keepaspectratio]{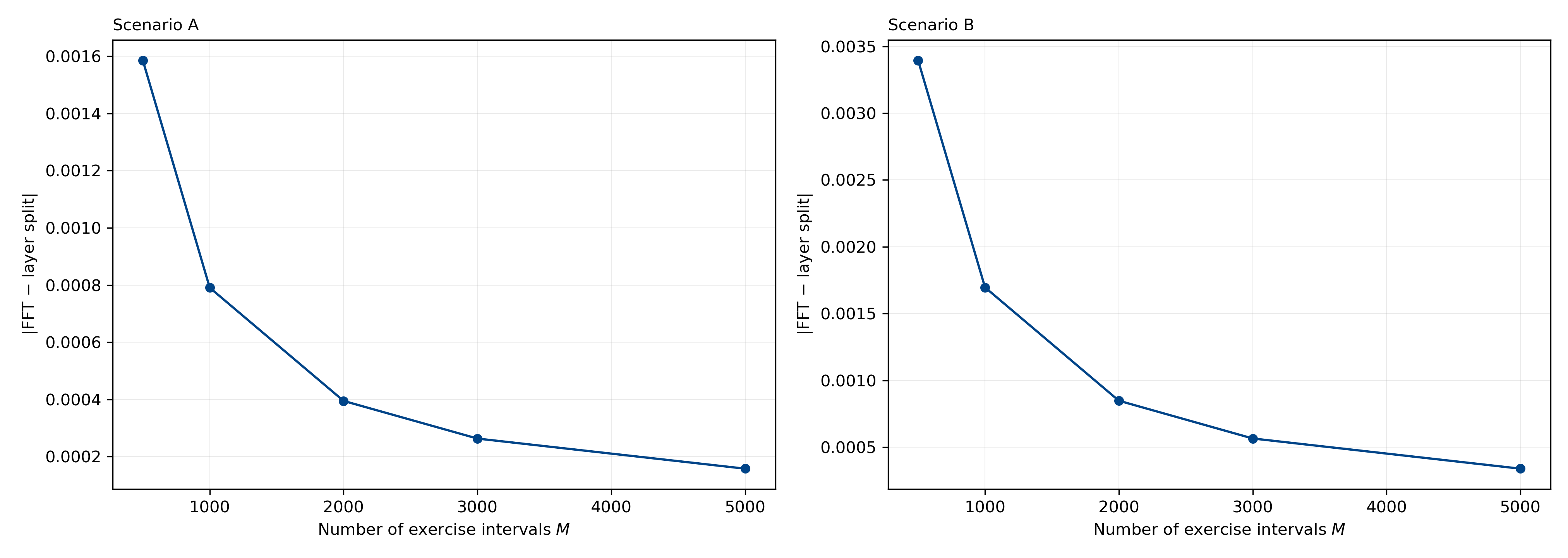}
\caption{Absolute discrepancy between the strict endpoint-closed layer split
and the periodic Fourier-block solver across the exercise meshes.}
\label{fig:num-M-discrepancy}
\end{figure}

The FFT values are already stable over this range. In Scenario A they change
from $3.046339$ at $M=500$ to $3.046408$ at $M=5000$; in Scenario B they move
from $3.564239$ to approximately $3.564323$. The layer-split values approach
the FFT values as the exercise mesh is refined. In Scenario A the absolute
difference decreases from $1.586\times10^{-3}$ at $M=500$ to
$1.577\times10^{-4}$ at $M=5000$. In Scenario B it falls from
$3.395\times10^{-3}$ to approximately $3.4\times10^{-4}$ over the same range.

The discrepancy decreases over the tested exercise meshes, but no
convergence order is inferred from this pattern because the two methods use
different price-boundary closures and different propagation architectures.
This sweep assesses exercise-mesh sensitivity; supplementary runtime curves
are reported in the auxiliary numerical appendix.
\subsection{Near-maturity exercise-boundary diagnostic}
\label{subsec:free-boundary}

The real-space obstacle also provides an economic diagnostic of the Fourier
propagation. On the $700\times120$ grid with $T=1$, $K=40$, and $M=1000$, we
extract the critical futures price at the initial variance level,
$F^*(\tau,V_0)$, from the exercise/continuation crossing on each variance layer
and then interpolate in variance to $V_0$.

\begin{figure}[!htbp]
\centering
\safeincludegraphics[width=0.90\linewidth,keepaspectratio]{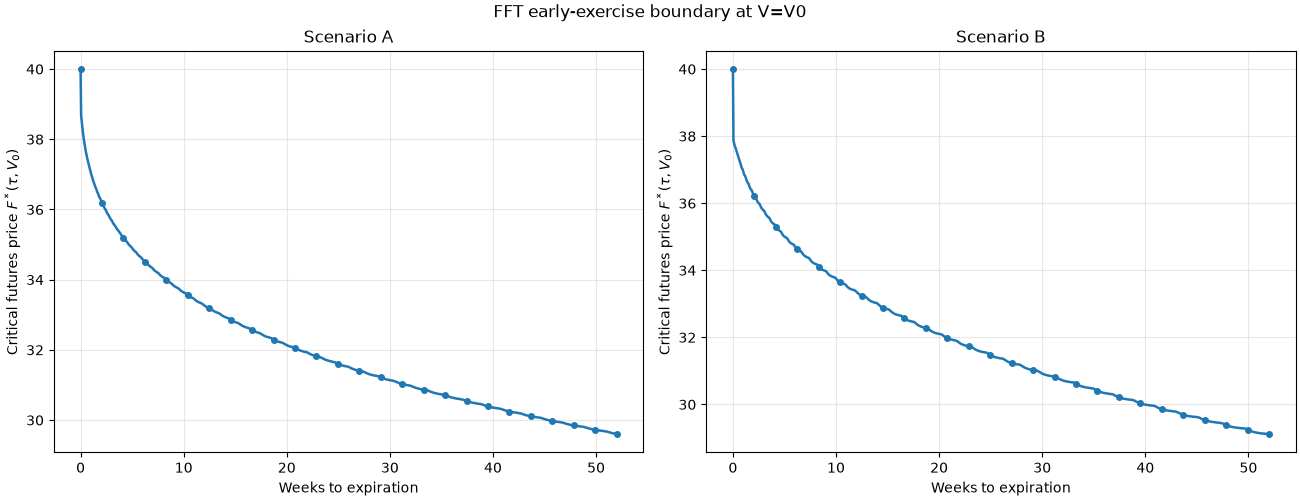}
\caption{FFT early-exercise boundary $F^*(\tau,V_0)$ for the two synthetic
scenarios on the $700\times120$ grid. The horizontal axis is weeks to
expiration, so $\tau=0$ is expiry. The terminal convention
$F^*(0,V_0)=K=40$ is conceptually distinct from the positive-time free
boundary in the jump-diffusion model.}
\label{fig:num-free-boundary}
\end{figure}

The terminal exercise rule gives $F^*(0,V_0)=K$. This terminal value should not,
however, be interpreted as the continuous limit of the positive-time free
boundary. Upward jumps create a strictly positive short-time continuation
benefit even when $F<K$. To see this directly, let
\[
 g(F)=(K-F)^+.
\]
For $F<K$, the futures martingale compensation implies
\begin{equation}
 (\mathcal L^{F,V}-r)g(F)
 =
 \lambda\mathbb E[(Fe^Z-K)^+]-r(K-F).
 \label{eq:maturity_boundary_generator}
\end{equation}
Hence any state for which the right-hand side is positive cannot belong to the
immediate-exercise region for every sufficiently small positive remaining
maturity.

For the double-exponential jump law used in the numerical scenarios,
\begin{equation}
 \mathbb E[(Fe^Z-K)^+]
 =
 \frac{pK}{\eta_+-1}
 \left(\frac{F}{K}\right)^{\eta_+},
 \qquad F<K.
 \label{eq:kou_short_maturity_gain}
\end{equation}
For $r>0$ and $\lambda p>0$, let $\bar F\in(0,K)$ denote the unique solution of
\begin{equation}
 r(K-\bar F)
 =
 \frac{\lambda pK}{\eta_+-1}
 \left(\frac{\bar F}{K}\right)^{\eta_+}.
 \label{eq:boundary_upper_threshold}
\end{equation}
The left-hand side is strictly decreasing and the right-hand side strictly
increasing on $(0,K)$, giving the stated unique root. At any
$F\in(\bar F,K)$ and fixed variance state, waiting for a sufficiently short
interval and then exercising has greater expected discounted payoff than
immediate exercise. Such an interval can be chosen within any positive
remaining maturity. Hence $(\bar F,K)$ is excluded from the
immediate-exercise region at every positive remaining maturity. The
interior-root statement is specific to $r>0$: when $r=0$ and
$\lambda p>0$, the generator gain is positive throughout $0<F<K$.
For $K=40$ and the parameters in
Table~\ref{tab:scenario-parameters},
\[
 \bar F_A=38.7126,\qquad
 \bar F_B=37.8775.
\]
These values are analytical necessary upper bounds for the positive-time
exercise boundary, not numerical estimates of the full free boundary.

Accordingly, the point $F^*(0,V_0)=40$ is a terminal convention and should be
interpreted separately from the positive-time boundary and not used to
infer a smooth connection at maturity. The positive-time curves decrease toward
the longer-horizon region and reach roughly $29$--$30$ at one year in both
scenarios. Scenario B lies modestly below Scenario A over much of the horizon,
consistent with stronger stochastic-volatility and jump activity increasing
continuation value. The plotted boundary remains a qualitative diagnostic:
raw exercise/continuation crossings and grid sensitivity, not visual
smoothness alone, determine numerical reliability. Discontinuity or loss of
boundary regularity at maturity is a known feature of American puts under jump
diffusions; see Bayraktar and Xing~\cite{BayraktarXing2009}.
\subsection{Same-generator verification and price-domain sensitivity}
\paragraph{Identical-generator verification.}
\label{subsec:same-generator}
To verify the implemented Fourier algebra independently of the continuous-model
approximation, we assemble a small $(N_x,N_v)=(17,7)$ periodic CTMC using
exactly the same local price rates, wrapped Kou jump coefficients, variance
generator, and discount rate as the Fourier-block code. For $\Delta=1/12$, we
compare direct dense propagation,
$U_{\rm dir}=\exp\!\left(\Delta(Q_{h,\mathrm{per}}^{[W]}-rI)\right)U_0$,
with the rFFT/block-exponential/irFFT action of Algorithm~\ref{alg:fft} on the
same fixed-seed test surface. The relative discrepancies
$\|U_{\rm FFT}-U_{\rm dir}\|_\infty/\|U_{\rm dir}\|_\infty$ are
$6.139\times10^{-16}$ in Scenario~A and $4.462\times10^{-16}$ in
Scenario~B. Thus the two implementations agree to double-precision roundoff,
validating the block construction, Fourier sign convention, vectorization, and
discount placement for the tested configuration. This is an implementation
check, not a test of continuous-model approximation error.

\paragraph{Price-domain expansion.}
\label{subsec:price-domain}
To isolate sensitivity to the periodic boundary, we enlarge only the
transformed-price domain while keeping the local price spacing essentially
unchanged. The test uses $T=1$, $M=1000$, $K=40$, $N_v=120$, the production
variance interval for each scenario, and the same interpolation and
jump-winding settings as the benchmark calculations. Table~\ref{tab:price-domain-expansion}
reports the resulting prices.

\begin{table}[htbp]
\centering\small
\caption{Price-domain expansion at approximately fixed local price spacing.}
\label{tab:price-domain-expansion}
\begin{tabular}{llrrr}
\toprule
Scenario & Price domain & $N_x$ & FFT price & $|\Delta P|$ from base\\
\midrule
A & $[-1.5,1.5]$ & 700 & 3.0463779881 & --\\
A & $[-1.875,1.875]$ & 875 & 3.0463853015 & $7.313\times10^{-6}$\\
A & $[-2.25,2.25]$ & 1049 & 3.0463855517 & $7.564\times10^{-6}$\\
\addlinespace
B & $[-2,2]$ & 700 & 3.5642860241 & --\\
B & $[-2.5,2.5]$ & 875 & 3.5642833198 & $2.704\times10^{-6}$\\
B & $[-3,3]$ & 1049 & 3.5642828821 & $3.142\times10^{-6}$\\
\bottomrule
\end{tabular}
\tablenote{The corresponding price spacings are approximately
$0.00429$ in Scenario~A and $0.00572$ in Scenario~B across all three
expansions, so the experiment primarily moves the periodic boundary rather
than changing local spatial resolution.}
\end{table}

Expanding the price domain by 50\% while preserving approximately the same
local spacing changes the price by less than $8\times10^{-6}$ in both
scenarios. The final 25\%-to-50\% expansion changes prices by only
$2.502\times10^{-7}$ in Scenario~A and $4.378\times10^{-7}$ in Scenario~B.
This provides direct numerical evidence that the reported one-year at-the-money
prices are insensitive to further enlargement of the tested periodic domains;
it is a domain-sensitivity diagnostic, not a bound on total continuous-model
error.
\section{Market data and calibration sample}\label{sec:data}
The SHFE gold option market provides an application setting for repeated
pricing across strikes and exercise calendars. The section proceeds from data
construction and contract conventions, to liquidity-based selection of the
underlying futures, to option-level calibration filters, and finally to the
smile structure of the retained sample. This empirical description is kept
separate from the controlled numerical validation of the pricing algorithm.
\subsection{Data sources and benchmark environment}

The Shanghai Futures Exchange (SHFE) operates an active gold derivatives market: over the option sample period, aggregate gold futures trading volume averages approximately 266,000 lots (about 266 tonnes; one lot corresponds to one kilogram) per day. Its gold option market, launched in December 2019, provides a broad cross-section of strikes and maturities with which to evaluate numerical option pricing methods. We therefore construct an empirical benchmark that combines gold spot prices, gold futures, gold futures options, and interest-rate information from official market sources.

The derivatives data are obtained from the Shanghai Futures Exchange (SHFE). The futures data span March 3, 2008--July 31, 2026 (gold futures trading itself began in January 2008), and the option data cover the complete history of the SHFE gold option market from its inception on December 20, 2019 to August 3, 2026, including 1,603 trading days; the two end dates reflect the respective data-retrieval cutoffs. The final option trading day in the file (August 3, 2026) has no matched futures closing prices; its 616 observations carry no underlying futures contract and are removed at the underlying-rank screen below, so the retained sample ends on July 31, 2026. The spot price series is based on Au99.99 gold prices from the Shanghai Gold Exchange (SGE). The risk-free interest rate is constructed from the three-month Chinese government bond yield-to-maturity (3-month YTM) provided by ChinaBond. The annualized continuously compounded rate used for discounting is obtained as

\[
r=\ln(1+y),
\]

where $y$ denotes the observed annualized three-month YTM, used as an
annual-effective-yield proxy in this conversion. This is a flat-rate
modeling convention rather than a fitted discount curve or an assertion
about the compounding convention of every source quotation. The effect of using a flat rate rather than a maturity-dependent
discount curve is not separately quantified in the empirical study.

The exchange announced the exercise-style revision on July~31, 2020. Its
implementation begins with the options on newly listed underlying futures
AU2212 on November~16, 2021: AU2212 and later series are American-style,
whereas earlier series retain European exercise until expiry
\cite{SHFE2020OptionRevision,SHFE2020OptionGuide}. Exercise style is
therefore assigned by underlying contract series, not solely by the
observation date. The market contracts are premium-paid options whose
exercise creates positions in the underlying futures
\cite{SHFE2020OptionGuide}. The discounted intrinsic payoff in
Assumption~\ref{ass:wellposed} represents the value obtained by immediate,
frictionless liquidation of that exercise position; it does not assert that
the exchange contract is cash-settled. Transaction costs and margin funding
are not modeled.

The binomial diagnostics and American CRR baselines use futures-price nodes
with $u=\exp(\sigma_B\sqrt{\Delta t})$, $d=u^{-1}$ and
\[
 p_F=\frac{1-d}{u-d},\qquad
 p_Fu+(1-p_F)d=1.
\]
Here $\sigma_B$ is the constant volatility used for that diagnostic or
baseline. Continuation is discounted separately by $e^{-r\Delta t}$;
the stock-tree probability $(e^{r\Delta t}-d)/(u-d)$ is not used.
An independent Black--76 comparison of this convention, with all numerical
inputs stated, is reported in Appendix~\ref{subsec:futures-tree-check}.

To assess the effect of applying the European Black--76 inversion to the American-style observations, we compare American and European values on a common 120-step binomial lattice. In a diagnostic sample of 2,000 deep in-the-money observations, where early exercise is most relevant, the premium has a median of 0.12\% of the European lattice value, a 95th percentile of 0.31\%, and a maximum of 0.72\%; it is smaller near the money. On the out-of-the-money reference subset itself, we isolate the early-exercise component of the inversion difference by inverting both the American and the European value of each quoted option on the same 120-step lattice, holding lattice discretization fixed across the two inversions; in a random subsample of 5,000 American-style observations, the resulting volatility difference has a median of 0.001, a 95th percentile of 0.006, and a maximum of 0.016 volatility points, economically negligible relative to the cross-sectional and time-series variation in observed implied volatilities. The lattice's own discretization relative to the analytic Black--76 inversion is likewise immaterial, with a median of 0.02 and a maximum of 0.28 volatility points. Black--76 implied volatilities are used as descriptive coordinates---for sample screening and surface characterization---and this lattice diagnostic does not establish the early-exercise approximation error under the stochastic-volatility/jump model of the pricing application; any subsequent price-based calibration should account explicitly for the exercise style of each contract. The same-day three-month yield is applied to all contracts, and discounting uses $e^{-r\tau}$ with each contract's own time to maturity $\tau$.

All option observations are constructed at the daily closing-price frequency: the futures closing price, option closing price, and corresponding interest rate are aligned by trading date and used consistently in the Black--76 implied volatility inversion \cite{Black1976}. Because the analysis is conducted at the daily frequency, bid--ask quotes are not available; closing prices represent the last transaction of the day, and the volume screen described below partly controls for their representativeness. All price and volume fields in the final calibration sample were verified against official SHFE data records.
\subsection{Liquidity-based selection of underlying futures}

A reliable calibration benchmark requires liquid underlying assets. Therefore, before selecting option contracts, we first construct the relevant futures universe.

On each trading day, all listed gold futures contracts are ranked according to their delivery dates. The label $T+k$ denotes the $(k+1)$-th nearest futures contract on that day, with $T+0$ the front (expiring) contract. This maturity-rank label describes the underlying futures contract and is distinct from the option's own time to maturity, which is denoted $\tau$ in calendar years (or $\tilde{\tau}$ in trading days). Calendar time is computed as days to expiration divided by 365; $\tilde{\tau}$ counts trading days.

The four nearest non-expiring ranks, $T+1$--$T+4$, account for
92.3\% of futures trading volume over the option sample period, and the daily
volume leader falls in this set on 96.4\% of trading days. The expiring rank
$T+0$ has a median daily volume of only 135 lots and is excluded. These facts
motivate the retained underlying universe. Supplementary
Figures~\ref{fig:f1}--\ref{fig:f2} document spot--futures co-movement and liquidity.

Before option-level screening, calls and puts account for 56.9\% and 43.1\%
of option volume on the retained universe; out-of-the-money contracts account
for 82.0\% of call volume and 85.1\% of put volume. Detailed rank--maturity
and call--put breakdowns are reported in Tables~\ref{tab:liq}--\ref{tab:liqcp}
in Appendix~\ref{app:market-data}. Futures rank and available option maturity
are linked by the listing structure and are not independent sorting variables.
\subsection{Option sample construction and application interface}\label{subsec:sample-filters}

The screened dataset provides cross-sectional option information for market description and price-based calibration applications. Sample construction defines a complete-chain universe and an out-of-the-money reference subset; the contracts actually used in a particular pricing application are selected separately. Guided by the concentration patterns documented in Table~\ref{tab:liq}, we apply a two-layer liquidity-based sample construction procedure.

At the underlying level, we restrict attention to futures ranks $T+1$--$T+4$. We define an option chain as the cross-section of contracts observed for a given trading date and underlying futures contract. At the option level, observations are retained only when they satisfy the following requirements:

\begin{enumerate}
\item option trading volume exceeds 100 lots;
\item option closing price exceeds CNY 0.5 per gram;
\item Black--76 implied volatility inversion is feasible;
\item each retained option chain contains at least six out-of-the-money options, including at least two calls and two puts.
\end{enumerate}

Requirement (iv) follows recent practice of constructing option curves from actively traded strikes subject to minimum completeness conditions \cite{AlexiouGoyalKostakisRompolis2025}. These restrictions select actively traded option contracts on chains satisfying the stated minimum out-of-the-money coverage requirements. The final complete-chain universe contains 51,583 option-day observations across 2,699 complete option chains. Of these, the 43,852 out-of-the-money observations---the high-strike calls and low-strike puts of each chain---form an OTM-only reference subset for calibration designs that impose this additional restriction. The in-the-money observations remain in the complete-chain universe for internal-consistency diagnostics and application-specific selection; they are not excluded by the chain-completeness requirement. Chain availability is thinner for $T+4$ (204 complete chains), reflecting lower far-maturity option activity. Within the $T+4$ segment, relaxing the 100-lot volume requirement to positive volume while retaining the price, IV-feasibility, and chain-completeness requirements increases the number of complete chains from 204 to 1,506. This relaxed $T+4$ sample supports far-maturity robustness analysis; the 2,699 chains reported above cover all retained ranks $T+1$--$T+4$ under the original screen.

\begin{figure}[!htbp]
\centering
\safeincludegraphics[width=\linewidth,keepaspectratio]{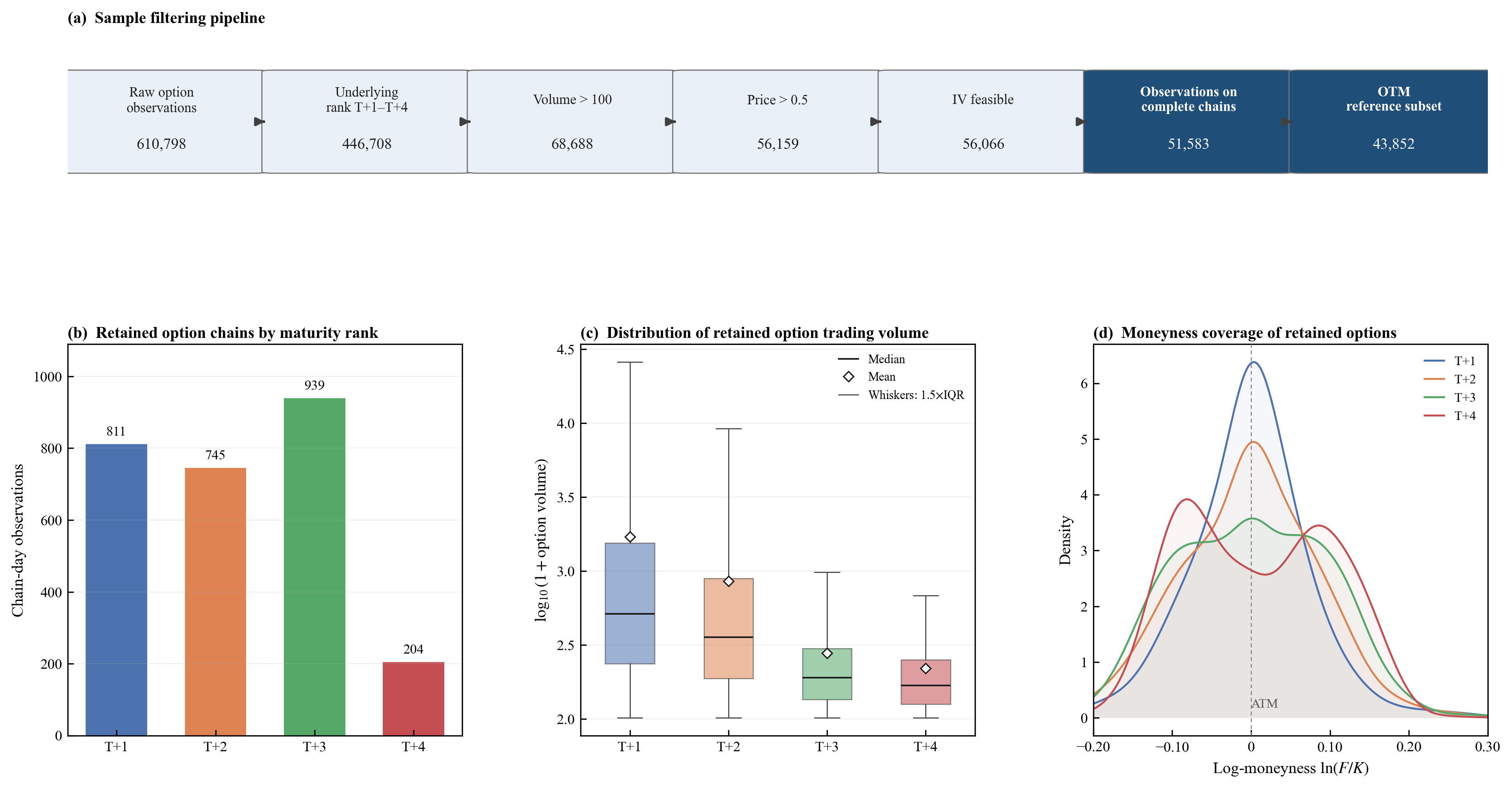}
\caption{Construction of the screened option universe. The figure reports the filtering procedure, retained option chains by futures maturity rank, trading activity, and moneyness coverage. The final two funnel stages distinguish all retained observations on complete chains (51,583) from the out-of-the-money reference subset (43,852). The reference subset is the out-of-the-money layer used by OTM-only calibration designs; it is distinct from the mixed-moneyness American-put cross-section used in the empirical application.}\label{fig:f3}
\end{figure}

We also examine the internal consistency of the retained quotes, computing both diagnostics on all 51,583 observations. Within each chain and contract type, 11.6\% of the 40,787 consecutive strike triples violate convexity, in that the middle price exceeds the linear interpolation of its two neighbors---computed on the actual, possibly non-uniform strike grid---by more than a numerical tolerance of $10^{-9}$; the median excess among violating triples is CNY 0.09 per gram. In addition, 14.7\% of the 7,098 matched same-strike call--put pairs---each pairing an out-of-the-money option with its in-the-money counterpart---violate the American futures-option parity bounds $e^{-r\tau}F-K\le C-P\le F-e^{-r\tau}K$ of \cite{RamaswamySundaresan1985}, with a median violation of CNY 0.46 per gram among violating pairs; this band is applied uniformly across the sample as a conservative diagnostic, since for the European-era contracts it nests the exact parity relation $C-P=e^{-r\tau}(F-K)$. Because the data record closing transactions rather than synchronized bid--ask quotes, these diagnostics may reflect differences in the timing of the last trades across option legs, as well as market frictions associated with futures-side arbitrage. We therefore treat them as measures of quote noise rather than impose an ex ante economic interpretation. The empirical summaries below use cross-sectional medians across chains and days to reduce the influence of individual inconsistent quotes. Table~\ref{tab:sample} and Figure~\ref{fig:f3} summarize the construction procedure and the properties of the retained sample.

The data description distinguishes four sample layers, each with a distinct role. The pre-screen panel of Table~\ref{tab:liq} describes the market as a whole and underlies the weekly-median surfaces and the factor analysis in Appendix~\ref{app:surface-data}; the volume-screened out-of-the-money universe (volume above 100 lots with feasible inversion, restricted to $|\ln(F/K)|\le 0.08$ and $10\le\tilde{\tau}\le 120$ trading days) underlies the skew estimates in Appendix~\ref{app:surface-data}; the complete-chain universe supports the retained-sample summaries and internal-consistency diagnostics, with its 43,852 out-of-the-money observations separately identified as an OTM-only reference subset; and the relaxed screen described above provides an alternative sample for assessing the far-maturity ($T+4$) segment.

Contract type, exercise style, and moneyness are separate selection dimensions. Restricting an application to American-style puts does not by itself impose an out-of-the-money restriction. The same-date American-put cross-section used in the empirical application includes both out-of-the-money and in-the-money puts and is therefore distinct from the OTM-only reference subset. Its exact selection rules and pricing results belong to the empirical application; the descriptive statistics reported here do not establish its calibration accuracy or the numerical robustness of its fitted parameters.

\begin{table}[htbp]
\centering\small
\caption{Construction and summary statistics of the screened SHFE gold option universe.}\label{tab:sample}

\medskip
\textit{Panel A: Sample construction}

\medskip
\begin{tabular}{lr}
\toprule
Raw option observations & 610,798\\
Underlying futures rank $T+1$--$T+4$ & 446,708\\
Option volume $>100$ & 68,688\\
Option close price $>0.5$ & 56,159\\
Black--76 IV inversion feasible & 56,066\\
Observations on complete chains & 51,583\\
\quad of which out-of-the-money (reference subset) & 43,852\\
\bottomrule
\end{tabular}

\medskip
\textit{Panel B: Summary statistics}

\medskip
\begin{tabular}{lrrrrr}
\toprule
Variable & Mean & Std.\ dev. & 25\% & Median & 75\%\\
\midrule
Option volume (lots) & 961 & 2,254 & 167 & 298 & 744\\
Open interest (lots) & 951 & 1,214 & 309 & 569 & 1,104\\
Implied volatility (\%) & 20.6 & 8.2 & 15.1 & 18.6 & 24.1\\
Log-moneyness $\ln(F/K)$ & $-$0.001 & 0.090 & $-$0.060 & 0.000 & 0.056\\
Time to maturity $\tilde{\tau}$ (trading days) & 38.9 & 23.1 & 21 & 35 & 52\\
Maturity $\tau$ (years) & 0.159 & 0.095 & 0.085 & 0.148 & 0.211\\
Futures price (CNY/g) & 626.2 & 239.1 & 402.5 & 563.2 & 790.3\\
Option price (CNY/g) & 9.72 & 15.97 & 1.66 & 4.34 & 10.98\\
\bottomrule
\end{tabular}

\tablenote{Panel~A filters are applied sequentially; counts refer to remaining observations. Panel~B reports summary statistics of the complete-chain sample ($N=51{,}583$), which comprises 43,852 out-of-the-money observations and 7,731 in-the-money observations. The out-of-the-money subset is a reference for OTM-only calibration designs, not a restriction on every empirical application.}
\end{table}

\subsection{Implied-volatility smile structure in the retained sample}

\begin{figure}[!htbp]
\centering
\safeincludegraphics[width=0.80\linewidth,keepaspectratio]{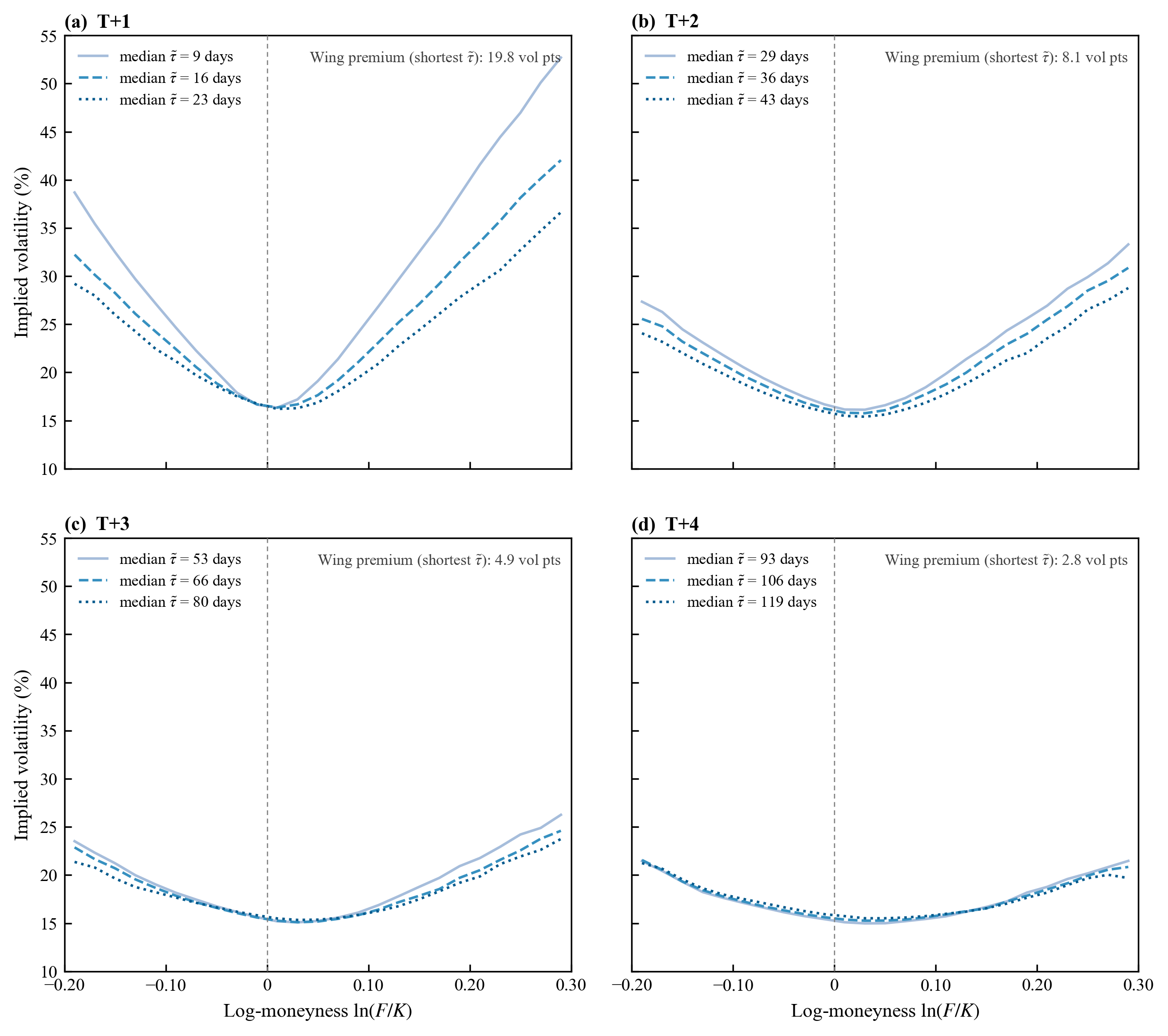}
\caption{Observed implied volatility smiles and wing premiums across futures maturity ranks. Within each rank, options are grouped into maturity terciles; each curve is the cross-chain-day median of implied volatilities demeaned by the same-day at-the-money benchmark and recentered at the median at-the-money level, and the legend reports the median maturity $\tilde{\tau}$ (trading days) of the corresponding tercile.}
\label{fig:f4}
\end{figure}

Against the backdrop of these surface dynamics, the retained sample provides a market setting for multi-strike pricing and calibration applications. Figure~\ref{fig:f4} presents the observed implied volatility smiles across futures maturity ranks. Within each rank, options are grouped into maturity terciles, and each legend reports the median maturity $\tilde{\tau}$ of the corresponding group.

Each curve is constructed in three steps. First, for every trading day and
underlying, the at-the-money implied volatility is obtained by linear
interpolation between the nearest out-of-the-money call and put, and each
option's implied volatility is demeaned by this same-day benchmark. Second,
the resulting deviations are aggregated into daily medians within
moneyness bins, so that each date--underlying pair contributes at most one
observation to each occupied bin within its maturity group. Third, the
curves report the cross-chain-day median deviation added back to the
median at-the-money level. Subtracting the same-day benchmark removes its
additive contribution before aggregation, and chain-day medians prevent
heavily quoted chains from contributing multiple observations to the same
bin. This reduces the influence of daily volatility-level variation and
uneven quote counts; it does not eliminate changes in the identities of
the chains or dates represented in different bins.

The option surface exhibits substantial nonlinear dependence across strike and maturity dimensions. Shorter maturity contracts display stronger wing effects, while longer maturities exhibit smoother volatility structures; the observed rank--maturity segments display progressively weaker smile intensity at longer horizons, consistent with the maturity-dependent patterns documented for equity index options \cite{BakshiCaoChen1997}, and the use of futures-option cross sections as a pricing benchmark follows \cite{BroadieChernovJohannes2007}. Because futures rank and option maturity are mechanically linked by the exchange's listing structure, this comparison is descriptive and does not identify a separate rank effect. The wing premium---the mean of the chain-day-by-moneyness-bin median implied-volatility deviations for $|\ln(F/K)|\ge 0.10$ within the shortest maturity group---declines from 19.8 volatility points for $T+1$ to 2.8 volatility points for $T+4$, while at-the-money implied volatilities remain close to 16\%. Such persistent nonlinear surface features imply repeated model evaluation across strikes and maturities: the Fourier-block propagators of the pricing method can be reused across strikes for a fixed parameter vector and maturity calendar, while changes in model parameters require rebuilding the affected generators and propagators. Evidence on the accuracy and computational cost of these evaluations is supplied by the controlled synthetic benchmarks of the pricing method, not by the market smile itself.

\section{Empirical calibration and pricing results}
\label{sec:empirical}

This section applies the Fourier-block pricing architecture to SHFE gold
options through a multi-day alternating calibration with all structural
parameters freely estimated. Section~7.1 sets out the protocol; the
estimation window is ten trading days on a three-chain maturity ladder
(September--October 2023), with the extreme-volatility February 2026
window retained as a stress check (Appendix~\ref{app:stress}). The
daily variance states are initialized from the underlying futures'
realized variance---the futures-first design of \cite{Bates2000}---and
then re-estimated freely from option quotes; Section~7.2 reports the
fit. Section~7.3 examines parameter identification through
boundary diagnostics and loss profiles, examines futures-side
realized variance descriptively, and discusses the economic interpretation.
Section~7.4 presents the two-measure state-space estimation on a
120-day panel, with parameter estimates within their search bounds,
day-clustered standard-error diagnostics, a nested-model ladder, and
a one-step state-transition diagnostic. Section~7.5 evaluates
cross-date pricing on 15 dates after the estimation window. The multi-day calibrations use American puts; the state-space panel
also includes American calls.

\subsection{Multi-day alternating calibration}
\label{subsec:multiday}

\paragraph{Design.}
The estimation pools option quotes across a window of consecutive
trading days, with all eight structural parameters
$(\kappa,\theta,\sigma,\rho,\lambda,p,\eta_+,\eta_-)$ estimated freely
and a separate variance state $V_0$ per trading day. No parameter is
fixed or set from auxiliary data. The calibration alternates between
two stages, following the design of
\cite{ChristoffersenHestonJacobs2009}: Stage~A holds the daily variance
states fixed and jointly estimates the shared structural vector on
the pooled relative-price residuals of all days; Stage~B holds the
structural vector fixed and recovers each day's $V_0$ from that day's
quotes alone. Rounds alternate until the relative improvement of the
joint objective falls below 0.1\%.

The volatility of volatility is parameterized to enforce the strict
Feller condition endogenously: $\sigma = z\,\bar\sigma(\kappa,\theta)$
with $\bar\sigma=\min\{2,\sqrt{2\kappa\theta/1.01}\}$ and
$z\in[0.04,0.999]$, so that $\chi_F=2\kappa\theta/\sigma^2>1.01$ at
every evaluation point and no ad hoc rejection of candidate vectors is
needed; the optimizer operates in this scaled coordinate.

Each day's $V_0$ starts from the futures side, following
\cite{Bates2000}. The initial value is the squared 60-day realized
volatility of the day's volume-leader underlying futures---the
own-contract rolling close-log-return volatility constructed from the
same futures panel as the universe of Section~\ref{sec:data}---and
Stage~B
then re-optimizes $V_0$ freely on $[0.005,0.35]$; the realized value is
a data-based starting point, never a constraint. The 20-day realized
variance is also recorded; in the February stress window it exceeds
the $V_0$ box (Appendix~\ref{app:stress}). This return-based statistic
may include jump variation as well as diffusion variance.

Global search uses twelve starting points---four informed seeds and
eight Latin-hypercube draws---each taken through a short Stage~A/B
iteration; among the starts whose short-run solutions keep every
parameter interior to its box, the lowest-RMSPE candidate is then
iterated to convergence.

The calibration uses relative pricing residuals pooled across days.
The reported pooled RMSPE weights each option quote equally. The search box is
\begin{center}
\begin{tabular}{lrr}
\toprule
Parameter & Lower & Upper\\
\midrule
$\kappa$ & 0.1 & 100\\
$\theta$ & 0.001 & 0.5\\
$\sigma=z\,\bar\sigma(\kappa,\theta)$ & $z=0.04$ & $z=0.999$\\
$\rho$ & $-0.99$ & 0.99\\
$\lambda$ & 0.1 & 8\\
$p$ & 0.01 & 0.99\\
$\eta_+$ & 1.1 & 50\\
$\eta_-$ & 0.5 & 50\\
$V_0$ (daily) & 0.005 & 0.35\\
\bottomrule
\end{tabular}
\end{center}
The estimation window is September~25--October~16, 2023 (ten
consecutive trading days across the National Day recess). Each (date,
chain) pair is priced on a $500\times60$ price--variance grid with the
periodic price coordinate of Section~3 and an exercise mesh of 100
dates per year ($\lceil100T\rceil$ dates for maturity $T$); the price domain is frozen per chain before
optimization at the eligible log-moneyness range widened by
$\max\{0.8,0.8\sqrt{T/(30/365)}\}$; the variance grid is the CIR
moment interval (mean $\pm5.5$ standard deviations at $T/2$, floored
at $10^{-4}$ and widened to contain $V_0$); and the normalized
surface is interpolated bicubically at $(V_0,\ln(F_0/K))$ with prices
floored at intrinsic value. Solver settings and the precise definition
of a round are given in Appendix~\ref{app:calibration-settings}.

\paragraph{Window and sample.}
The estimation window is September~25 to October~16, 2023---ten
consecutive trading days across the National Day recess. It sits in a calm volatility regime (the 20-day realized
volatility of the volume-leader futures is 6--9\%, near the 10th
percentile of the full sample) during a cumulative price advance, so
the option surface prices an accumulation rally rather than a
volatility episode. The sample consists of the American-put quotes
retained by the validity screen and the one-pass strike-convexity
screen within each futures chain: 178 quotes over the AU2312, AU2402,
and AU2404 chains, with window-start maturities of 0.16, 0.33, and
0.50 years (0.11, 0.28, and 0.44 on the last window day)---
a three-tier maturity ladder. The AU2404 chain is absent on three
days of the window (September~26--27 and October~13), on which the
two nearer chains remain.

\subsection{Multi-day results}
\label{subsec:multiday-results}

\begin{table}[htbp]
\centering\small
\setlength{\tabcolsep}{3.6pt}
\caption{Multi-day alternating calibration: final structural parameters.}
\label{tab:multiday}
\begin{tabular}{lrrrrrrrr}
\toprule
Window & $\kappa$ & $\theta$ & $\sigma$ & $\rho$ & $\lambda$ & $p$
& $\eta_+$ & $\eta_-$\\
\midrule
Sep--Oct 2023 (options only) & 9.797 & 0.0026 & 0.225 & $-0.829$ & 2.031 & 0.989 & 19.85 & 4.38\\
\bottomrule
\end{tabular}
\tablenote{The option-only vector estimates all eight structural parameters
freely under the alternating protocol and Feller parameterization;
boundary and identification diagnostics are reported in Section~7.3.}
\end{table}

\begin{table}[htbp]
\centering\small
\setlength{\tabcolsep}{3.4pt}
\caption{Multi-day calibration fit quality: four error metrics.}
\label{tab:multiday-daily}
\begin{tabular}{lrrrr}
\toprule
Window & Quotes & RMSPE (\%) & MAPE (\%) & RMSE / MAE (CNY/g)\\
\midrule
Sep--Oct 2023 (options only) & 178 & 8.56 & 6.39 & 0.352 / 0.248\\
\bottomrule
\end{tabular}
\tablenote{RMSPE is the root-mean-square percentage error, MAPE the
mean absolute percentage error, RMSE the root-mean-square price error
and MAE the mean absolute price error, both in CNY per gram. The relative errors are largest among the lower-priced,
deep out-of-the-money quotes (Table~\ref{tab:buckets}).}
\end{table}

\begin{table}[htbp]
\centering\small
\setlength{\tabcolsep}{3.8pt}
\caption{Per-day fit quality (RMSPE, \%).}
\label{tab:multiday-perday}
\begin{tabular}{lrrrrr}
\toprule
\multicolumn{3}{c}{September--October 2023 window} &
\multicolumn{3}{c}{}\\
\cmidrule(lr){1-3}\cmidrule(lr){4-6}
Date & Quotes & RMSPE & Date & Quotes & RMSPE\\
\midrule
Sep~25 & 14 & 3.4 & Oct~10 & 20 & 7.6\\
Sep~26 & 13 & 4.0 & Oct~11 & 18 & 9.2\\
Sep~27 & 16 & 10.0 & Oct~12 & 13 & 9.0\\
Sep~28 & 26 & 11.8 & Oct~13 & 10 & 3.6\\
Oct~9 & 25 & 10.3 & Oct~16 & 23 & 6.3\\
\bottomrule
\end{tabular}
\tablenote{Per-day RMSPE in percent under the equal-weight relative
price loss. October~13 is the thinnest day (10 quotes); the AU2404 chain
returns at full breadth on October~16.}
\end{table}

On the September--October window the alternating optimization runs
sixteen rounds with decreasing RMSPE, reaching the stopping threshold; the complete
campaign, including the twelve multi-start short iterations, completes
in under seven hours of wall-clock time.
Table~\ref{tab:multiday-daily} reports four complementary error
metrics for the option-only calibration: 8.56\% pooled RMSPE, 6.39\% MAPE,
0.352 CNY per gram RMSE, and 0.248 CNY per gram MAE. The numerical
benchmarks in Section~\ref{sec:numerics} assess the solver under controlled
configurations; the market residuals also reflect model and quote discrepancies.

\begin{table}[htbp]
\centering\small
\setlength{\tabcolsep}{4.5pt}
\caption{Option-only fit quality by moneyness bucket.}
\label{tab:buckets}
\begin{tabular}{lrrr}
\toprule
$|\ln(F/K)|$ & $N$ & RMSPE (\%) & MAPE (\%)\\
\midrule
$<5\%$ & 99 & 6.82 & 5.15\\
$5$--$10\%$ & 54 & 8.60 & 6.45\\
$10$--$15\%$ & 25 & 13.35 & 11.22\\
\bottomrule
\end{tabular}
\tablenote{Relative errors increase in the lower-priced wings, where
relative tick noise is largest. The near-the-money region includes
99 of 178 quotes. Across the three maturity chains, option-only RMSPE
is 8.81/8.17/8.84\% at window-start maturities of
$T=0.16/0.33/0.50$ years.}
\end{table}

Table~\ref{tab:buckets} decomposes the fit by moneyness and maturity.
Relative errors are larger in the wings, where low prices can amplify the
impact of quote ticks. The near-the-money RMSPE is 6.82\%, and the three
maturity chains exhibit similar RMSPEs. These magnitudes do not establish
that signed pricing errors are unbiased across maturities.

Table~\ref{tab:multiday} reports the
fitted option-only structural vector. Which parameters press against their box
bounds and what
that means for identification is examined in Section~7.3; the fitted
vector satisfies the strict Feller condition by construction
(the unrounded fit is constrained by the stated admissibility floor; Section~7.3).

\begin{figure}[!htbp]
\centering
\safeincludegraphics[width=0.7\linewidth,keepaspectratio]{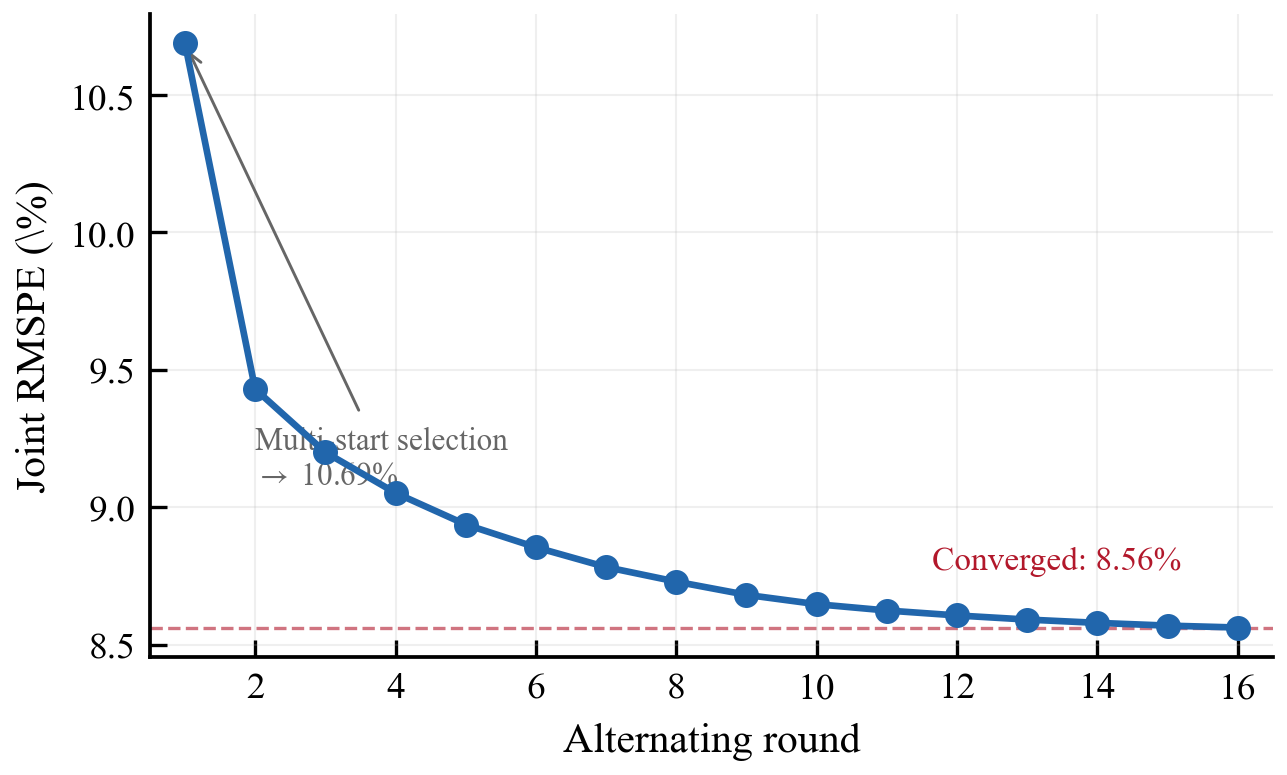}
\caption{Convergence of the alternating optimization: option-only RMSPE across
calibration days by round after the multi-start selection. The figure
label ``Joint RMSPE'' refers to the multi-day option fit.}
\label{fig:convergence}
\end{figure}

\paragraph{Multi-start selection.}
The twelve starts include four informed seeds and eight Latin-hypercube draws. After the short iteration the starts rank between 12.5\% and
184\% pooled RMSPE, with seven of the eight Latin-hypercube draws above
30\%. The selection screen then retains a start whose short-run score
is 28.6\%---interior, but ranked mid-table---and iterates it to
convergence: 10.7\% after the first full round, already below every
short-iteration score, decreasing monotonically to 8.56\% over sixteen
rounds (Figure~\ref{fig:convergence}). The dispersion suggests sensitivity to initialization and motivates
the multi-start search.

\subsection{Parameter identification and economic interpretation}
\label{subsec:identification}

\paragraph{Maturity coverage and parameter identification.}
Parameter identification depends on the maturity--strike region
covered by the calibration sample. Short-dated options help constrain
the current variance state and smile shape, whereas the long-run variance
level and mean-reversion speed are expressed through the maturity dimension.
In the Deutsche-Mark option study of \cite{Bates1996}, jump and
stochastic-volatility effects are difficult to separate at short
maturities. The September--October 2023 window spans 0.11--0.50 years;
the February 2026 stress window (Appendix~\ref{app:stress}) spans
0.03--0.14 years. These windows also differ in market regime and
quote count (178 versus 287), so they do not isolate maturity effects.

\begin{figure}[!htbp]
\centering
\safeincludegraphics[width=\linewidth,keepaspectratio]{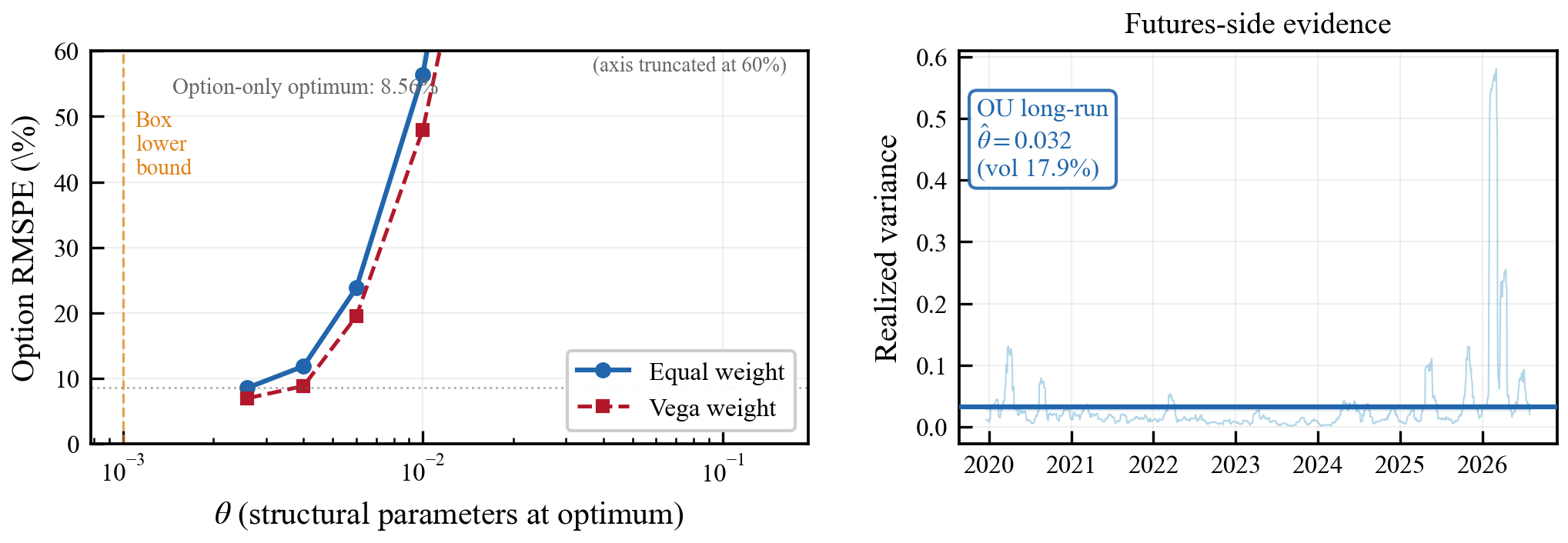}
\caption{Left: conditional $\theta$ loss profiles for the
September--October 2023 option-only window, holding the other
structural parameters fixed and re-estimating daily $V_0$ states.
Right: the historical volume-leader futures realized-variance series
through 2026; the horizontal line and label $0.032$ are from a
\emph{full-sample} descriptive OU fit, displayed only as background.
They are not moments used in the 2023 option-only calibration.
For the futures sample ending October~16, 2023, the corresponding
20-day rolling statistic has mean $0.019$. The non-overlapping-block
persistence calculation discussed in Section~\ref{subsec:identification}
is a separate diagnostic restricted to data through the window end,
not the full-sample OU fit.}
\label{fig:profiles}
\end{figure}

\paragraph{Boundary solutions and conditional loss profiles.}
Under the wide Feller box, the option-only fit has $p=0.989$ near its
0.99 upper bound, the volatility-of-volatility fraction $z=0.999$
near the Feller ceiling, and $\theta=0.0026$ near the lower part of
its search box (lower bound 0.001). Boundary estimates may signal
one-sided or weak identification, not necessarily optimization failure.
Figure~\ref{fig:profiles} (left) plots conditional loss profiles
for $\theta$: the remaining structural parameters are fixed at their
option-only optimum while daily states are re-estimated. Both the
equal-weight and vega-weighted losses increase as $\theta$ moves away
from the leftmost admissible value. The shape of these conditional
profiles does not establish that the full jointly reoptimized objective
has comparable curvature. Boundary solutions have also been reported
in longer option panels \cite{BakshiCaoChen1997,ChristoffersenHestonJacobs2009,Bates2000,HurnLindsayMcClelland2015}.

\begin{remark}[Short-maturity sensitivity of the long-run variance level]
For the CIR variance process, the integrated expected variance is
\[
 \int_0^T\mathbb E[V_s] \,ds
 =\theta T+(V_0-\theta)\frac{1-e^{-\kappa T}}{\kappa}.
\]
Holding $\kappa$ and $V_0$ fixed, its sensitivity to $\theta$ is
\[
 \frac{\partial}{\partial\theta}\int_0^T\mathbb E[V_s] \,ds
 =T-\frac{1-e^{-\kappa T}}{\kappa}
 =\frac{\kappa T^2}{2}+O(\kappa^2T^3),
 \qquad \kappa T\ll1.
\]
When $\kappa T\ll1$, this local expansion of \emph{integrated
expected variance} indicates weak sensitivity to $\theta$, which,
together with variance--jump parameter trade-offs, may complicate
finite-sample identification. It is not an identification theorem for
option prices: additional quotes or longer maturities may improve
identification. Since $\kappa T$ is not uniformly small across the
fitted maturities, the expansion is only a short-horizon heuristic.
\end{remark}

\paragraph{The Feller ceiling.}
The volatility of volatility ends near its admissibility frontier:
$\chi_F=2\kappa\theta/\sigma^2$ is just above the imposed floor of
1.01 in the September--October fit. Unconstrained Heston-type fits
may violate the Feller condition \cite{Bates1996,ErakerJohannesPolson2003}.
Enforcing the condition by construction makes this binding constraint
explicit; it does not by itself establish parameter identification.

\paragraph{Parameter trade-offs and futures-side diagnostics.}
The conditional $\theta$ profiles favor a low long-run variance level,
but re-optimizing $\theta$ and $\kappa$ together within the option-only
objective produces an alternative fit (11.5\% RMSPE), revealing a
trade-off between mean-reversion speed and long-run level.

The futures-side statistic is a 20-day rolling realized variance
of own-contract close-to-close log returns for the volume-leader future.
Restricting this series to information available by the calibration
window end (October~16, 2023) gives 924 daily observations and a
sample mean of $0.019$, rather than the full-sample descriptive level
$0.032$ shown in Figure~\ref{fig:profiles}. Its lag-one rolling
correlation is $0.983$, which is affected by window overlap: even
independent daily squared returns give lag-one correlation
$(N-1)/N=0.95$ for a 20-day moving average.
Non-overlapping 20-day block means of the truncated rolling series
have lag-one correlation $0.46$. Although the blocks of rolling
observations do not overlap, adjacent blocks can still share underlying
return observations.

For comparison only, a stationary CIR process with exponential
covariance decay gives the correlation between consecutive
\emph{block averages} of length $\Delta$ as
\[
 \operatorname{Corr}(\bar V_j,\bar V_{j+1})
 =\frac{(1-e^{-a})^2}{2(a-1+e^{-a})},
 \qquad a=\kappa\Delta.
\]
Setting $\Delta=20/252$ and formally matching $0.46$ gives an
illustrative $\kappa\approx16.5$ per year (half-life approximately
11 trading days). This is not an estimate of the latent physical
mean-reversion speed: realized return variation is not a noiseless
block integral of the CIR diffusion and may contain jump variation
and measurement noise. A valid joint estimator must construct
model moments for the \emph{actual} return-based block statistic;
no futures-moment joint-calibration results are claimed here.

The present option-only estimation therefore supports a limited
identification conclusion: different variance and jump parameter
combinations can produce similar prices, and the reported
conditional profiles do not resolve that confounding.

\paragraph{Joint estimation with futures information.}
Jointly fitting option prices and futures-side moments is a natural
extension. Such an estimator requires model-implied moments that account
for the realized-variance statistic's construction, overlapping return
windows, jumps, and measurement noise. The option-only calibration does
not use these moment restrictions.

One source of information not imposed by the alternating protocol is
the daily variance-state transition law: each $V_0$ is fitted separately.
As a diagnostic, fixing the structural vector and recovering states
on a sixty-day panel (June--August 2023) yields an interior fitted
long-run level ($\hat\theta=0.006$, about 7.8\% volatility) and a
nine-trading-day half-life. These path-based descriptive quantities
vary by segment, however, and do not replace a joint state-space
estimation. Section~\ref{subsec:statespace} considers that extension.

\paragraph{Economic reading.}
The option-only fitted vector suggests a model-dependent economic interpretation. The
September--October window prices an accumulation rally in a calm
volatility regime: jump intensity $\lambda=2.03$ per year with
up-jump probability $p=0.99$, mean jump sizes $1/\eta_+=5.0\%$ (up)
and $1/\eta_-=22.8\%$ (down), a return--variance correlation
$\rho=-0.83$, and a variance half-life $\ln 2/\kappa\approx18$
trading days. The fitted jump contribution to the log-return second moment is
$\lambda\E[Z^2]\approx0.013$, compared with the long-run diffusion
variance $\theta\approx0.003$. This decomposition is specific to the
fitted risk-neutral model and is subject to the identification limitations
discussed above.

\paragraph{Jump asymmetry and skew.}
The SHFE gold option surface exhibits predominantly positive
short-maturity skew (83\% of trading days;
Appendix~\ref{app:surface-data}). In a Heston-type model,
stochastic-volatility skew is transmitted through the correlation
channel, which is commonly linked to the equity leverage effect
\cite{ChristoffersenHestonJacobs2009}; gold options display a different pattern in this sample, and the Kou
jump block can generate positive skew through its jump asymmetry
parameter $p$. The fitted negative correlation and asymmetric jump distribution are
consistent with the jump component contributing to positive skew. The
data do not uniquely identify the relative contributions of these channels;
related futures-option evidence is discussed in
\cite{BakshiCaoChen1997,BroadieChernovJohannes2007}.

\begin{figure}[!htbp]
\centering
\safeincludegraphics[width=\linewidth,keepaspectratio]{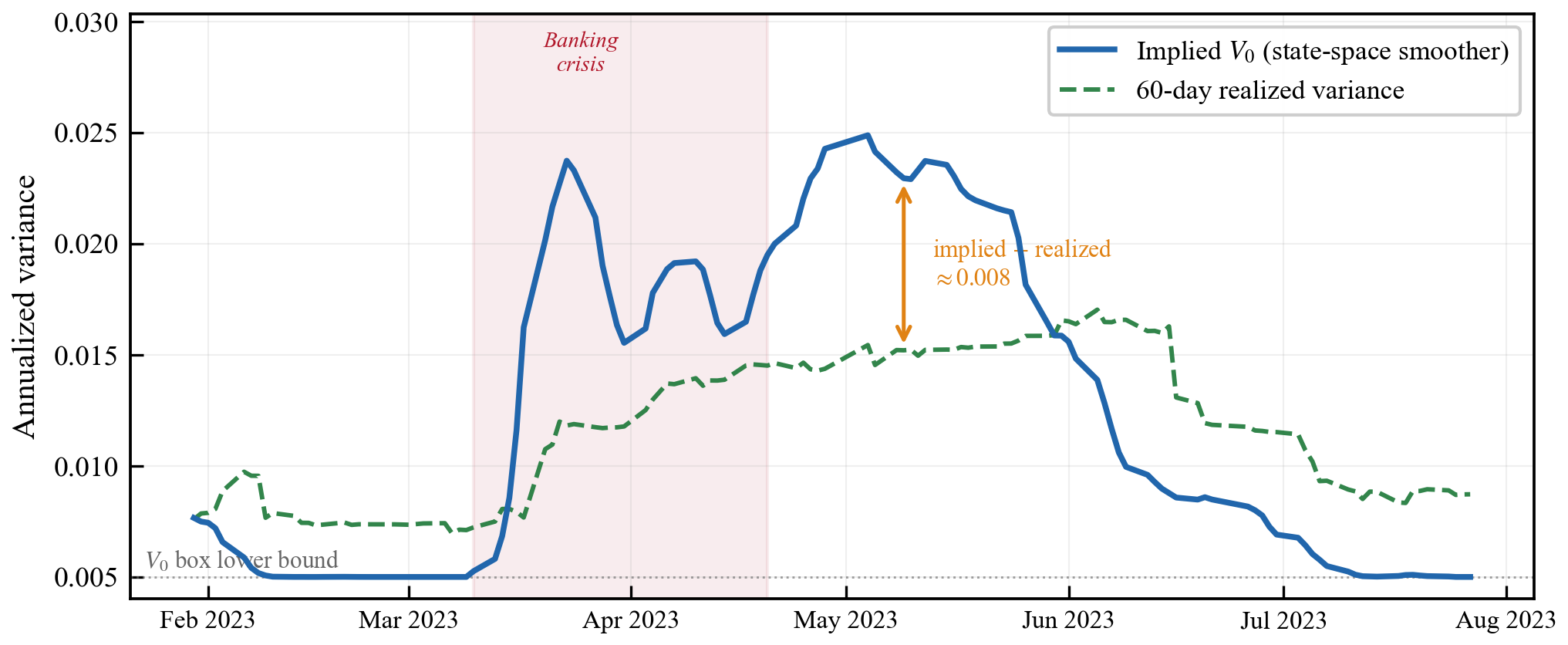}
\caption{Implied variance state from the two-measure state-space
smoother (solid) against the trailing 60-day realized variance of the
volume-leader futures (dashed), over the 120-trading-day estimation
panel. The smoothed state rises around the March 2023 banking crisis (shaded)
and declines toward the state-box floor in the final months. Because
the state is smoothed over the panel, the comparison with a trailing
statistic is descriptive and does not establish predictive lead time. The state-space level
premium is $\gamma=+0.81$ (Section~7.4); its approximate conditional
delta-method 95\% interval is $[-3.9,+5.5]$. The
dotted line marks the
$V_0$ search-box lower bound; the forward-looking variance risk
premium of \cite{CarrWu2009}, which relates the implied state to
subsequently realized variance over the premium horizon, is the
natural object of the rolling extension.}
\label{fig:v0-tracking}
\end{figure}

\paragraph{Implied versus realized variance.}
The daily $V_0$ series is the protocol's state-recovery output: on
the September--October window it sits above the 60-day realized
variance used as its Stage-B start;
Figure~\ref{fig:v0-tracking} (Section~7.4) shows the same comparison
over the 120-day state-space panel. This is the
implied-versus-realized comparison of \cite{Bates2000}, read off a
calibrated state; the forward-looking variance risk premium of
\cite{CarrWu2009} is the natural object of the rolling extension, and
a filter-based treatment along the lines of
\cite{BardgettGagliardiniOssola2019} provides a related estimation
route.

One alternating round exercises the strike-normalized surface reuse
of Section~3.3 throughout: each Stage~A pricing evaluation solves one
surface per (date, chain) pair and reuses it across all strikes of
that chain. On the live 22-put chain of October~18, 2023, pricing the
full chain takes 0.88 seconds on the reuse path against roughly seven
seconds when each strike is priced separately (medians of repeated
runs; prices identical to machine precision).

\subsection{Two-measure state-space estimation}
\label{subsec:statespace}

The identification analysis of Section~7.3 motivates a design that
links the daily states by the model's own dynamics and separates the
two measures: the variance state follows its physical dynamics
$(\kappa,\theta_P,\sigma)$ through a variance-standardized
CIR state-transition penalty, while pricing uses the
risk-neutral level $\theta_Q$, with $\kappa$ and $\sigma$ shared and
the premium reported as $\gamma=\theta_Q/\theta_P-1$. Both option
wings enter: American calls join the puts (the call obstacle mirrors
the put recursion; the European-limit put--call parity of the
implementation matches $e^{-rT}(F-K)$ to $0.06\%$ of the futures
price). The structural parameters and daily states are estimated jointly;
the transition residuals are standardized by the model-implied conditional
variance.

The estimation criterion is transition-penalized least squares. Let
$\psi=(\kappa,\theta_P,\sigma,\theta_Q,\rho,\lambda,p,\eta_+,\eta_-)$
be the nine structural parameters, and let
$V_0,\ldots,V_{119}$ be the \emph{120} daily variance states.
For an option quote $n$ on day $i$, denote its market price by
$P_{i,n}$ and its model price by $\widehat P_{i,n}(\psi,V_i)$.
Write $m_i(\psi,V_i)$ and $v_i(\psi,V_i)$ for the conditional mean
and variance of the one-trading-day CIR transition from $V_i$.
The objective is
\begin{equation}
\begin{aligned}
 Q(\psi,V_0,\ldots,V_{119})
 ={}&\sum_{i=0}^{119}\sum_{n\in\mathcal O_i}
 \frac{\bigl(\widehat P_{i,n}(\psi,V_i)-P_{i,n}\bigr)^2}
 {P_{i,n}^2}\\
 &+\sum_{i=0}^{118}
 \frac{\bigl(V_{i+1}-m_i(\psi,V_i)\bigr)^2}
 {v_i(\psi,V_i)}.
\end{aligned}
\label{eq:ss-objective}
\end{equation}
Here $\mathcal O_i$ indexes the option quotes observed on day $i$.
There are 3{,}198 pricing residuals and exactly 119 state-transition
residuals. Since the transition part includes only squared standardized
innovations and omits the parameter-dependent $\sum_{i=0}^{118}\log v_i$
term, the criterion differs from a Gaussian transition quasi-likelihood.
The two residual blocks have the relative weights displayed in
\eqref{eq:ss-objective}. The optimization alternates between a
Gauss--Newton update of the structural vector with the states fixed
and a backward--forward sweep updating the daily states conditional on
their neighbours, stopping when the relative improvement in the objective
falls below 0.1\%.

On the 120-trading-day panel January~30--July~27, 2023 (1{,}486 puts
and 1{,}712 calls), the estimator attains 13.40\% pooled RMSPE, with
all nine structural parameter estimates within their search bounds:
$\theta_P=0.0036$, $\theta_Q=0.0065$, $\kappa=2.9$ (half-life near
60 trading days), and $\gamma=+0.81$.

Table~\ref{tab:se-main} reports standard-error diagnostics.
Their calculation uses 3{,}198 pricing residuals and 119 standardized
state-transition residuals defined by the penalized
least-squares criterion \eqref{eq:ss-objective}. The residual
Jacobian with respect to the nine structural parameters is evaluated
numerically. The daily states are held at their smoothed estimates
when forming the reported covariance: these standard errors are
therefore conditional on the estimated state path, not full joint
estimator standard errors. The day-clustered sandwich calculation
allows within-day dependence between option residuals and an assigned
state-transition residual, but does not explicitly represent all
cross-day dependence created by adjacent transition links.

Under this conditional calculation the $t$-statistic for $\theta_P$
is only 0.8. The delta-method 95\% interval for
$\gamma=\theta_Q/\theta_P-1$ is $[-3.9,+5.5]$, including zero.
Because $\theta_P$ is imprecisely estimated and appears in a
denominator, a first-order ratio approximation may be unstable;
the interval is indicative rather than a guaranteed-coverage result.
The positive premium is therefore only a descriptive point estimate.
The estimation and pricing-fit results, rather than formal inference
on $\gamma$, are the focus of this panel exercise.

\begin{table}[htbp]
\centering\small
\setlength{\tabcolsep}{4.5pt}
\caption{Parameter estimates with conditional day-clustered sandwich
standard-error diagnostics (120-day panel).}
\label{tab:se-main}
\begin{tabular}{lrrrrrrrrr}
\toprule
& $\kappa$ & $\theta_P$ & $\sigma$ & $\rho$ & $\lambda$ & $p$
& $\eta_+$ & $\eta_-$ & $\theta_Q$\\
\midrule
Est. & 2.90 & 0.0036 & 0.022 & 0.39 & 2.23 & 0.85 & 26.3 & 14.2 & 0.0065\\
SE & 0.26 & 0.004 & 0.007 & 0.12 & 0.17 & 0.02 & 0.74 & 0.97 & 0.001\\
$t$ & 11.3 & 0.8 & 3.1 & 3.3 & 13.3 & 42.7 & 35.5 & 14.7 & 6.6\\
\bottomrule
\end{tabular}
\tablenote{Day-clustered sandwich standard errors from the full
penalized least-squares residual vector
(3{,}198 relative option-pricing residuals plus 119 standardized
state-transition residuals, conditional on the smoothed daily states).  Same-day option residuals and the transition residual
linking day $i$ to day $i+1$ share a cluster.  $\sigma$ is reported
in physical units.  These standard errors condition on the smoothed states and do not
incorporate full latent-state uncertainty or all cross-day dependence.
The $t$-statistics are based on unrounded estimates and standard errors.
The delta-method 95\% interval for $\gamma=\theta_Q/\theta_P-1$
is $[-3.9,+5.5]$; the ratio approximation may be unreliable when
$\theta_P$ is estimated imprecisely.}
\end{table}

A one-step state-transition diagnostic propagates the previous day's
fitted state through the model dynamics without re-solving the daily
state, then prices that day's full cross-section.  It yields
14.80\% RMSPE, compared with 13.40\% for the smoothed in-sample
fit---a difference of 1.4 percentage points.  Because this
diagnostic uses the fitted state sequence, it should not be
interpreted as a strictly real-time, out-of-sample forecast.
The monthly errors vary across market regimes: the banking-crisis months of March--April fit at
17--18\%, the remaining months at 10--13\%, with June lowest at
9.8\%. A grid-refinement diagnostic provides a check on numerical
sensitivity: refining the pricing grid from $250\times35$ to
$500\times60$ changes the pooled RMSPE by at most 0.11 percentage
points.

\begin{figure}[!htbp]
\centering
\safeincludegraphics[width=\linewidth,keepaspectratio]{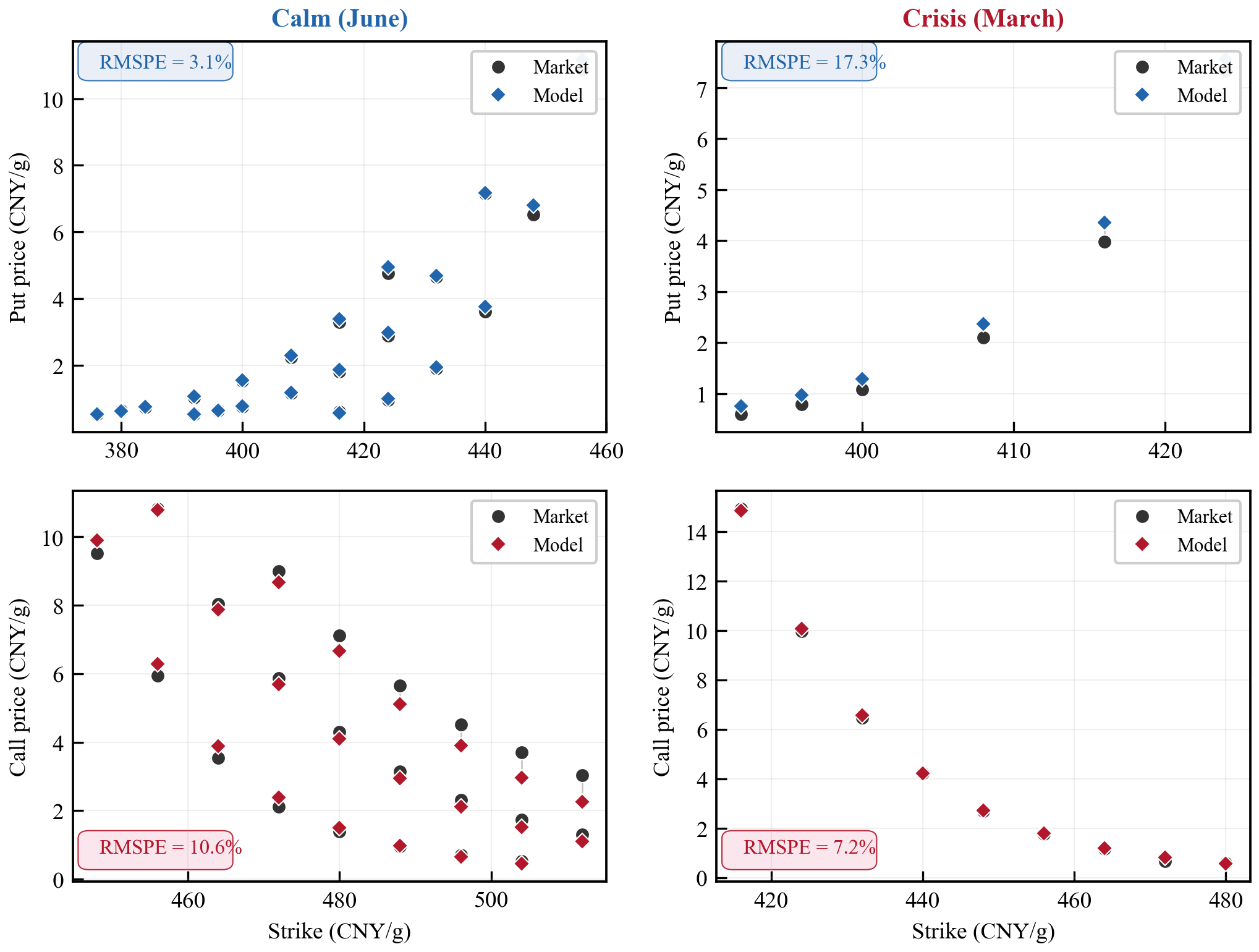}
\caption{Market vs.\ model prices across strikes on two
representative dates (left: calm June; right: crisis March),
separately for puts (top) and calls (bottom). The plotted RMSPEs
refer to the respective displayed cross-sections and should not be
compared directly with the separate single-date recalibrations in
Table~\ref{tab:ladder-main}.}
\label{fig:smile-main}
\end{figure}

\begin{table}[htbp]
\centering\small
\caption{Nested-model ladder on representative dates (single-date
calibration; \% RMSPE).}
\label{tab:ladder-main}
\begin{tabular}{lrrr}
\toprule
Model & $k$ & Calm (Jun 7) & Crisis (Mar 15)\\
\midrule
Black--76 (per-chain vol) & 1 & 25.50 & 14.40\\
Kou jump diffusion & 5 & 2.68 & 1.43\\
Heston stochastic vol. & 5 & 4.19 & 1.11\\
Full SVJ & 9 & \textbf{1.87} & \textbf{0.55}\\
\bottomrule
\end{tabular}
\tablenote{On the two representative dates, removing either the
jump or stochastic-volatility component increases the in-sample
pricing RMSPE relative to the combined model.}
\end{table}

Table~\ref{tab:ladder-main} shows that the combined nine-parameter
specification has the lowest in-sample RMSPE on these two dates;
removing either component increases the error, while the
Black--76 benchmark has substantially larger discrepancies.
These single-date comparisons do not by themselves establish
out-of-sample model superiority. Figure~\ref{fig:smile-main} provides a separate illustration of
pricing fit across strikes.

\subsection{Cross-date pricing evaluation}
\label{subsec:oos}

To assess the cross-date transfer of the fitted structural vector,
we evaluate the option-only estimate on 15 selected trading dates
between October~17 and November~10, 2023, after the estimation window,
using the three maturity chains as they roll forward.
The eight structural parameters are frozen at their calibration-window
values; only the evaluation day's $V_0$ is re-estimated from that
day's own option quotes, as in Stage~B. This tests cross-date
pricing with same-day state re-estimation, not a fully prospective
price forecast.

\begin{table}[htbp]
\centering\small
\caption{Cross-date option-only pricing evaluation over 15 selected
trading dates, with daily-state re-estimation.}
\label{tab:oos}
\begin{tabular}{lrrrrr}
\toprule
Vector & Days & Quotes & Pooled RMSPE (\%) & Mean daily (\%) & SD (\%)\\
\midrule
Option-only (frozen) & 15 & 293 & 13.59 & 12.95 & 3.54\\
\bottomrule
\end{tabular}
\tablenote{The structural vector is fixed at the option-only
window estimate of Table~\ref{tab:multiday}; each evaluation day's
$V_0$ is re-estimated from that day's own quotes. The in-sample
option-only result is reported in Table~\ref{tab:multiday-daily}.}
\end{table}

The held-out-date pooled RMSPE is 13.59\% across 293 quotes,
compared with 8.56\% on the calibration window. These are different
market dates and quote selections, so the change is descriptive
rather than a controlled decomposition of the error. 

\section{Conclusion}
This paper develops a Fourier-block CTMC method for American futures-option
pricing under stochastic volatility and jumps, addressing the repeated
application of a large joint two-dimensional CTMC semigroup. The construction
integrates joint-state CTMC discretization, the finite-state volatility
treatment used in CTMC--PROJ, and the frequency-domain continuation principle
of FST at the level of a conservative discrete generator. Its periodic price
structure reduces the full joint action $e^{\Delta tQ}$ to variance-sized
frequency blocks while retaining the complete discrete variance coupling.
The resulting method provides an unsplit alternative to directional
price--variance propagation, with stability inherited from the Markov
semigroup and no intermediate transition-density projection.

The theoretical results establish conservation and stopping stability for
the finite-state recursion, together with sequential convergence to
the American value under the stated Feller, mesh, expanding-domain,
jump-winding, and exercise-mesh conditions. On the identical
periodic generator, direct dense matrix-exponential propagation and the
Fourier-block implementation agree to machine precision. In the common-grid
cost--discrepancy experiment, the Fourier-block method first reaches the
$10^{-3}$ reference-price threshold at $350\times60$ with a 1.5-second cold
cost, compared with 7.0 seconds for the endpoint CTMC on the same grid. Under
the fixed no-substep protocol, the layer-split calculation retains a
price discrepancy of approximately $5\times10^{-3}$ on the finer spatial
grids; spatial refinement alone does not remove its temporal splitting error.
The dedicated exercise-mesh study shows decreasing FFT--splitting
differences as the time step is reduced and reports the associated
computational cost. The tested PROJ implementation reaches the same discrepancy threshold
after increasing the complete price/projection resolution, at 30.1 seconds on
the $700\times112$ configuration. These comparisons support stable and
efficient Fourier-block propagation under the tested configurations. LSM
serves as an auxiliary simulation-based reference, with differences reflecting
the distinct numerical approximations to the valuation problem rather than
errors measured against an exact price.

The market application estimates the structural parameters of the
SHFE gold American-option model on a ten-day, three-chain window
(September--October 2023: 178 puts, option-only RMSPE 8.56\%).
The analysis documents sensitivity of the long-run variance level to
the available maturity coverage and a trade-off with mean-reversion
speed. It does not claim that the short-maturity information alone
uniquely determines these parameters. On a separate 120-day two-measure
state-space panel (January--July 2023: 1{,}486 puts and 1{,}712 calls),
the pooled RMSPE is 13.40\%. Although the nine structural estimates
lie within their search bounds, the physical long-run variance level
remains weakly identified and the sign of the estimated variance
premium ($\gamma=+0.81$) is not statistically established. A one-step
state-transition diagnostic gives an RMSPE 1.4 percentage points above
the smoothed in-sample fit. In a 15-date cross-date option-only pricing
evaluation with same-day variance-state re-estimation, the pooled RMSPE
is 13.59\%. A nested-model comparison gives lower in-sample error for
the combined jump--stochastic-volatility model on two representative
dates. For a live chain of 22 American puts, strike-normalized surface
reuse reduces pricing time from roughly seven seconds to 0.88 with
identical unrounded prices.

The empirical analysis is ongoing. Further work will extend the cross-date
pricing evaluation, develop model-consistent futures-side moment conditions,
and examine uncertainty in the latent variance states.

The present block structure relies on conditional translation invariance of the
transformed price coordinate and on a uniform periodic price grid, so it does
not apply directly to general local--stochastic-volatility models with explicit
price dependence. A natural extension is a three-factor state $(X,V,r)$ with a
stochastic short rate. When conditional translation invariance in $X$ is
retained, the Fourier transform would continue to act only in the price
coordinate, while variance and short-rate states remain jointly coupled within
each frequency block. This suggests a higher-dimensional variance--rate block
structure with the same joint-semigroup representation and reuse principle.

\appendix
\section{Proofs of local CTMC consistency}
\label{app:consistency}
\subsection{Local diffusion-part consistency}
\label{app:local-proof}

Fix a compact $K\subset\R\times(0,\infty)$. By grid exhaustion, all grid
points in $K$ are interior for sufficiently small $h$. Taylor expansion in
the variance coordinate gives, for $\ell=j\pm1$,
\begin{equation}
f(x_i,v_\ell)-f(x_i,v_j)
=f_v(x_i,v_j)(v_\ell-v_j)
+\frac12f_{vv}(x_i,v_j)(v_\ell-v_j)^2+R^v_{j\ell},
\label{eq:app_taylor_v}
\end{equation}
where $|R^v_{j\ell}|\le C_K|v_\ell-v_j|^3$. Hence, using
\eqref{eq:moment_matching},
\begin{align}
&\sum_{\ell}q^v_{j\ell}
   [f(x_i,v_\ell)-f(x_i,v_j)]
 -\mu(v_j)f_v(x_i,v_j)-\frac12\eta^2(v_j)f_{vv}(x_i,v_j)
 \nonumber\\
&\hspace{10mm}=\frac12
 \bigl(\widehat\eta_{h,j}^{\,2}-\eta^2(v_j)\bigr)f_{vv}(x_i,v_j)
 +\sum_{\ell=j\pm1}q^v_{j\ell}R^v_{j\ell}.
\label{eq:app_v_error}
\end{align}
If $h_K^v$ is the maximum variance mesh width in a fixed compact
neighbourhood of $K$, then
\begin{equation}
\left|\sum_{\ell=j\pm1}q^v_{j\ell}R^v_{j\ell}\right|
\le C_Kh_K^v\sum_{\ell=j\pm1}q^v_{j\ell}(v_\ell-v_j)^2
 =C_Kh_K^v\widehat\eta_{h,j}^{\,2}.
\label{eq:app_v_bound}
\end{equation}
The effective variance is uniformly bounded on $K$, and
\eqref{eq:rebalancing_error} bounds the first term on the right-hand side of
\eqref{eq:app_v_error} by
$\tfrac12\|f_{vv}\|_{\infty,K}|\mu(v_j)|h_K^v$. Both terms therefore tend to
zero uniformly on $K$.

The corresponding expansion in $x$ yields
\begin{align}
&\sum_{m}q^{x,j}_{im}[f(x_m,v_j)-f(x_i,v_j)]
 -b(v_j)f_x(x_i,v_j)-\frac12a^2(v_j)f_{xx}(x_i,v_j)
 \nonumber\\
&\hspace{10mm}=\frac12
 \bigl(\widehat a_{h,ij}^{\,2}-a^2(v_j)\bigr)f_{xx}(x_i,v_j)
 +\sum_{m=i\pm1}q^{x,j}_{im}R^x_{im},
\label{eq:app_x_error}
\end{align}
where, with $h_K^x$ defined analogously,
\begin{equation}
\left|\sum_{m=i\pm1}q^{x,j}_{im}R^x_{im}\right|
 \le C_Kh_K^x\widehat a_{h,ij}^{\,2}\longrightarrow0.
\label{eq:app_x_bound}
\end{equation}
The additional second-moment term in \eqref{eq:app_x_error} is bounded by
$\tfrac12\|f_{xx}\|_{\infty,K}|b(v_j)|h_K^x$ by
\eqref{eq:rebalancing_error}. Since the coefficient functions are bounded on
$K$, adding these estimates proves locally uniform convergence of the
rebalanced reference generator to $\cL_D$. This is the time-homogeneous form
of the generator error $\epsilon^{(n_1,n_2)}(f)$ used by Ding and Ning
\cite[Eq.~(2.17), Proposition~3, and Appendix~A]{DingNing2021}.
\subsection{Compound-Poisson jump consistency}
\label{app:jump-proof}

The rates \eqref{eq:jump_rates_new} imply the exact identity
\begin{equation}
(Q_{h,J}f_h)(x_i,v_j)
=\lambda\int_{\R}
 [f(\pi_h(x_i+z),v_j)-f(x_i,v_j)]\nu_Z(dz).
\label{eq:app_projected_jump}
\end{equation}
Consequently,
\begin{equation}
|(Q_{h,J}f_h)(x_i,v_j)-(\cJ f)(x_i,v_j)|
\le\lambda\int_{\R}
|f(\pi_h(x_i+z),v_j)-f(x_i+z,v_j)|\nu_Z(dz).
\label{eq:app_jump_error}
\end{equation}
Fix $\varepsilon>0$ and select $R$ so that
\begin{equation}
2\lambda\|f\|_\infty \nu_Z(\{z:|z|>R\})<\varepsilon/2.
\label{eq:app_tail}
\end{equation}
If $(x_i,v_j)\in K$ and $|z|\le R$, then both $(x_i+z,v_j)$ and
$(\pi_h(x_i+z),v_j)$ belong to a fixed compact set $K_R$ for all sufficiently
small $h$. Let
\begin{equation}
\omega_{f,K_R}(\delta)
=\sup\{|f(x,v)-f(y,v)|:(x,v),(y,v)\in K_R,\ |x-y|\le\delta\}.
\end{equation}
Grid refinement gives
$\sup_{y\in K_R}|\pi_h(y)-y|\le h_{K_R}^x\to0$, and uniform continuity
gives $\omega_{f,K_R}(h_{K_R}^x)\to0$. Therefore the integral in
\eqref{eq:app_jump_error} over $|z|\le R$ is at most
$\lambda\omega_{f,K_R}(h_{K_R}^x)<\varepsilon/2$ for all sufficiently small
$h$, uniformly over grid points in $K$. Equation~\eqref{eq:app_tail} controls
the complement, proving \eqref{eq:jump_consistency_new}.

This is the finite-measure specialization of the discretized jump generator
$\Lambda_J^{(n)}$ constructed by Mijatovi\'c and Pistorius
\cite[Appendix~A.1]{MijatovicPistorius2009}. Because the compound-Poisson
measure $\lambda\nu_Z$ is finite, no small-jump compensation or singular-kernel
estimate is required.

\section{Proofs of semigroup and stopping-value convergence}
\subsection{Uniform generator consistency on a Feller core}
\label{app:proof-global_core_consistency}
\begin{proof}
On a fixed rectangle $[-R,R]\times[0,H]$, the coefficient functions and the
derivatives of $f$ are bounded. At every interior variance node, including
nodes tending to zero, Taylor expansion and the moment identities give an
error bounded by
\[
 C_f\left[|\mu(v_j)|h_{H}^v
       +h_{H}^v\widehat\eta_{h,j}^{\,2}\right].
\]
Since $\widehat\eta_{h,j}^{\,2}\le\sigma^2H+
\sup_{[0,H]}|\mu|h_H^v$, the bound tends to zero uniformly on the entire
lower layer. The corresponding price error is bounded by
$C_f[|b(v_j)|h_R^x+h_R^x\widehat a_{h,ij}^{\,2}]$ and also tends to zero
uniformly, including at $v=0$. The rebalanced price stencil is applicable
there even though $a^2(0)=0$.

At the lower variance endpoint, a one-sided Taylor expansion gives
\[
 q^v_{1,2}[f(x,v_2^h)-f(x,v_1^h)]
 =q^v_{1,2}\Delta_h f_v(x,v_1^h)
   +O(q^v_{1,2}\Delta_h^2).
\]
Equations~\eqref{eq:cir_entrance_closure}--\eqref{eq:lower_boundary_moments}
show uniform convergence to $\kappa\theta f_v(x,0)$ on bounded price sets.
The continuous variance operator at $v_1^h$ has the same limit.

Choose a rectangle containing the support of $f$ and a fixed neighbourhood.
At sufficiently remote local boundaries, both endpoints of every local
transition are outside this support by the escape condition. At other nodes
outside the rectangle, nearest-neighbour stencils cannot reach the support
for small $h$, by the mesh condition on stencils meeting compact sets.
Thus the local contribution is identically zero there; increasing coefficients
at remote nodes cannot change this conclusion.

For jumps, fix $R_0>0$ and enlarge the price projection of the support by
$R_0+1$. Within that set, cell-projection consistency follows from uniform
continuity and the refined mesh, for all variance nodes including zero.
Outside it, a projected jump into the support requires $|z|>R_0$ for small
$h$. Both jump actions are bounded by
$2\lambda\|f\|_\infty\nu_Z(|z|>R_0)$. Outside the variance projection of
the support they vanish identically. Taking the grid limit and then
$R_0\to\infty$ proves the global result.
\end{proof}
\subsection{Kurtz semigroup convergence}
\label{app:proof-kurtz_semigroup}
\begin{proof}
The admissible affine representation checked in Section~\ref{subsec:model}
identifies the transformed jump--Heston process with the conservative
regular affine process on $D$. Thus $T(t)$ is a strongly continuous
contraction semigroup on $C_0(D)$, and $C_c^\infty(D)$ is a core of its
generator by \cite[Theorem~2.7]{DuffieFilipovicSchachermayer2003}.
Each $T_h(t)$ is a
contraction in the grid sup norm by Proposition~\ref{prop:generator}, and the
restriction operators satisfy $\|\pi_h\|\le1$ and
$\|\pi_hf\|_{\infty,E_h}\to\|f\|_{\infty,D}$ for every $f\in C_0(D)$:
grid density captures the supremum on a sufficiently large compact set,
while vanishing at infinity controls its complement. Lemma~\ref{lem:global_core_consistency}
therefore verifies the generator condition of the varying-space
semigroup approximation theorem, \cite[Chapter~1, Theorem~6.1]{EthierKurtz1986}.
Here the limiting Banach space is $C_0(D)$, the approximating spaces are
$\mathbb R^{E_h}$ with the sup norm, and the approximating generators are
$Q_h^J$. For every core function $f$, the choice $f_h=\pi_hf$ is in the
domain of the finite-matrix generator and satisfies both approximation
conditions: $f_h-\pi_hf=0$ and
$Q_h^Jf_h-\pi_h\cL f\to0$ in the grid sup norm.
Equation~\eqref{eq:kurtz_semigroup} follows.

For $y_h\to y$, the Feller property makes $T(t)f$ continuous, so
\eqref{eq:kurtz_semigroup} implies convergence of
$P_t^h(y_h,\cdot)$ against every $f\in C_0(D)$. Both the approximating and
limiting measures have total mass one. On the locally compact state space
$D$, convergence against $C_0(D)$ together with preservation of total mass is
weak convergence of probability measures. This proves
\eqref{eq:transition_weak}, and the $C_b(D)$ statement follows from the
Portmanteau theorem.
\end{proof}
\subsection{Fixed-schedule Bermudan convergence}
\label{app:proof-values}
\begin{proof}
All exact value functions take values in $[0,K]$, and the limiting value
functions are continuous by backward induction. A conservative Feller
transition is weakly continuous in its initial state: convergence on $C_0(D)$
and preservation of mass imply weak convergence, hence continuity on
$C_b(D)$. Suppose, as the backward-induction hypothesis, that the next-date
grid value $g_h$ converges to $g$ along every converging sequence of grid
states. This is stronger than pointwise convergence. It implies
$\sup_{z\in C\cap E_h}|g_h(z)-g(z)|\to0$ for every compact $C\subset D$:
otherwise there are $h_n\to0$ and $z_n\in C\cap E_{h_n}$ for which the
difference stays above a fixed positive number. A subsequence has
$z_n\to z\in C$; the induction hypothesis gives $g_{h_n}(z_n)\to g(z)$,
while continuity gives $g(z_n)\to g(z)$, a contradiction. No separate
equicontinuity assumption is used in this step.
For $y_h\to y$, the measures $P_{\Delta t}^h(y_h,\cdot)$ converge weakly,
and hence are tight. Split the integral of $g_h-g$ over a common compact
set and its complement. The first part tends to zero; the second is bounded
by $2K$ times an arbitrarily small tail probability. The integral of $g$
converges by the assumption. Continuation values therefore converge, and
taking the maximum with the continuous payoff preserves convergence.
Starting at the terminal payoff proves the assertion.
\end{proof}
\subsection{Exercise-mesh limit to the American value}
\label{app:proof-american}
\begin{proof}
For each fixed $m$, apply Corollary~\ref{cor:values}. For any stopping time
$\tau\le T$, round it upward to $\tau_m=\min\{t\in\mathcal E_m:t\ge\tau\}$.
These are stopping times with $\tau_m\downarrow\tau$. Right continuity of
$Y$, continuity of $\Phi_K$, and $0\le\Phi_K\le K$ give convergence of the
discounted expectations by dominated convergence. The schedule values
increase and are bounded above by $V_K^A$. Taking the supremum over $\tau$
proves the result. This argument gives no joint spatial/exercise rate and
requires no convergence of optimal stopping boundaries.
\end{proof}
\subsection{Convergence of the periodic chain on expanding domains}
\label{app:proof-periodic-convergence}
\begin{proof}
The variance estimates are those of Lemma~\ref{lem:global_core_consistency}.
On a fixed price compact set the uniform local stencil is eventually
unwrapped. At the seam, its source and targets are outside the support of
$f$, so its action is zero; other stencils outside a neighbourhood of the
support also vanish.

For the infinite-winding jump operator, write its action as an integral over
the translation-equivariant nearest ring-cell map. For $|z|\le R_0$, any
source whose target lies in a fixed compact support must itself lie within
an $R_0+\delta_h$ neighbourhood of a translated copy of that support.
Nonzero periodic copies are eventually more than $R_0+\delta_h$ from the
fundamental price interval, because the distances from the support to both
endpoints diverge. Consequently no small jump wraps into the support, and
ordinary nearest-cell consistency applies uniformly to the remaining sources.
The tail $|z|>R_0$ contributes at most
$2\lambda\|f\|_\infty\nu_Z(|z|>R_0)$. Let $h\to0$ and then $R_0\to\infty$.

Finite winding changes the action by at most
$2\lambda\varepsilon_{W_h}\|f\|_\infty$ by
\eqref{eq:winding_generator_error}. The generators are conservative, so the
same contraction-semigroup approximation theorem applies. Weak transition
convergence, bounded-payoff backward induction and the nested exercise-limit
argument then give the stated stopping conclusions.
\end{proof}
\section{Additional semigroup and operator bounds}
\subsection{Error bound for numerical block propagation}
For interval $\Delta t_n$, write
\[
 P_{n,k}=e^{\Delta t_n(\widehat Q_k^{[W]}-rI_m)}
\]
for the exact cached block propagator and $\widetilde P_{n,k}$ for its numerical
approximation. If
$\|\widetilde P_{n,k}-P_{n,k}\|_2\le\delta_{n,k}$, use the unitary DFT
$\widehat u_k^{\mathrm u}$ to obtain the sufficient exact-arithmetic bound
\begin{align}
 \| (\widetilde P_n-P_n)\widetilde u^{n+1}\|_\infty
 &\le\left(\sum_{k=0}^{N-1}\delta_{n,k}^2
       \|\widehat u_k^{\mathrm u}\|_2^2\right)^{1/2}\nonumber\\
 &\le\max_k\delta_{n,k}\,\|\widetilde u^{n+1}\|_2.
 \label{eq:block_error_bound}
\end{align}
The full-spectrum sum includes the conjugate partners. A rigorous numerical
certificate requires justified block-error and arithmetic bounds; differences
against a high-accuracy reference are diagnostics, not automatically such a
certificate. Positivity and conservation hold for the exact real-space
semigroup; small floating-point violations are measured separately.
\subsection{Difference between endpoint-closed and periodic semigroups}
For a common uniform grid and the same variance closure, set
$R_h=Q_h^J-Q_{h,\mathrm{per}}$. For any vector $z$,
\begin{align}
 &(e^{t(Q_h^J-rI)}-e^{t(Q_{h,\mathrm{per}}-rI)})z\nonumber\\
 &\quad=\int_0^t e^{(t-s)(Q_h^J-rI)}R_h
                     e^{s(Q_{h,\mathrm{per}}-rI)}z\,ds.
 \label{eq:boundary_duhamel}
\end{align}
For $\|z\|_\infty\le K$, the one-step bound satisfies
\[
 \int_0^t e^{-r(t-s)}
 \|R_he^{s(Q_{h,\mathrm{per}}-rI)}z\|_\infty\,ds
 \le Kt e^{-rt}\|R_h\|_\infty.
\]
This yields a whole-recursion bound $KT\|R_h\|_\infty$, but it may be
uninformative because seam rates increase as the grid is refined.
No small-norm, low-rank, or vanishing boundary correction is assumed.
Domain-expansion experiments at interior initial states are therefore essential.
Equation~\eqref{eq:boundary_duhamel} is an analysis identity;
Algorithm~\ref{alg:fft} does not implement a boundary correction.
The put payoff is bounded but not periodic as a function on the line, and
its seam discontinuity must not be confused with a frequency cutoff.
\section{Supplementary numerical diagnostics}
\subsection{Two-sample LSM policy evaluation and substep sensitivity}
\label{subsec:lsm-two-sample}

The supplementary experiment exp12 uses Scenario A with $K=40$, $T=5$
and $M=1000$, retaining the production six-term regression basis and four
simulation substeps per exercise interval. Each of 16 replications fits
the stopping rule on 60,000 training paths and evaluates the frozen rule
on 60,000 independent paths. It also records the in-sample estimate on
the training panel. Replication streams use
\texttt{SeedSequence(20261005)}, distinct from the production seed 23.
There is no reduction of the training sample to fund the evaluation sample.
The recorded evaluation performs no regression refit. At time zero it
retains the production rule of taking the maximum of immediate intrinsic
value and the panel mean of discounted subsequent cash flows. In this
particular test $F_0=K=40$, so immediate intrinsic value is zero and all
discounted put cash flows are nonnegative. The final maximum therefore
does not change the continuation estimate. For positive initial intrinsic
value, choosing that decision from the evaluation-panel mean would be an
additional sample-dependent selection, not evaluation of a fully frozen
time-zero decision; that case is not tested here.

The second check, exp15, regenerates the same four-substep training
policies using the same training seeds and fitting procedure. A new
independent evaluation stream, \texttt{SeedSequence(20261006)}, supplies
60,000 paths per replication at each of four, eight, and sixteen substeps.
Policies are not refitted. For a coarse step containing $k$ fine steps,
the fine standard-normal drivers are summed and divided by $\sqrt{k}$;
with the coarse step's $\sqrt{h}$ factor this gives exactly the sum of
the fine Brownian increments. Positive and negative jump sizes are summed
without rescaling. Each variance path evolves at its own substep size,
and the fitted policy acts only at the common exercise dates. The three
values are therefore paired within each replication, while the exp12
and exp15 evaluation streams differ.

\begin{table}[htbp]
\centering\small
\caption{LSM policy evaluation and coupled substep sensitivity at the
Scenario A $K=40$, $T=5$, $M=1000$ configuration.}
\label{tab:lsm-two-sample}
\label{tab:lsm-substep}
\begin{tabular}{lrr}
\toprule
Quantity & Mean $\pm$ standard error & 95\% interval\\
\midrule
\multicolumn{3}{l}{First evaluation stream (exp12)}\\
In-sample policy value & $6.058595\pm0.005913$ & $[6.045991,6.071199]$\\
Independent policy value & $6.043313\pm0.005026$ & $[6.032601,6.054024]$\\
Paired in-minus-out & $0.015282\pm0.009146$ & $[-0.004213,0.034777]$\\
\addlinespace
\multicolumn{3}{l}{Second evaluation stream (exp15; frozen policies)}\\
Four-substep value & $6.064654\pm0.006050$ & $[6.051759,6.077550]$\\
Eight-substep value & $6.064242\pm0.006024$ & $[6.051402,6.077082]$\\
Sixteen-substep value & $6.064130\pm0.005984$ & $[6.051375,6.076886]$\\
Paired eight-minus-four & $-0.000413\pm0.000569$ & $[-0.001625,0.000800]$\\
Paired sixteen-minus-eight & $-0.000111\pm0.000484$ & $[-0.001142,0.000920]$\\
Paired sixteen-minus-four & $-0.000524\pm0.000753$ & $[-0.002129,0.001081]$\\
\addlinespace
Second-minus-first, four substeps & $0.021342\pm0.009385$ & $[0.001338,0.041345]$\\
\bottomrule
\end{tabular}
\tablenote{Each stream uses 16 replications and 60,000 independent
training and evaluation paths per replication. Standard errors are
computed across replication values or replication-level paired
differences, as appropriate. Intervals use the $t_{15}$ approximation.
The cross-stream comparison is paired by the shared training replication.
FFT $6.073954$ is a CTMC reference, not a continuous-model truth value.
The two streams are not pooled.}
\end{table}

The first stream's in-minus-out interval permits effects comparable to its
FFT--policy discrepancy; failure to reject zero is not an equivalence
result. Its in-sample mean is within approximately 0.9 combined standard
errors of the original 300-replication estimate $6.053149$, a consistency
check, not an independent audit of path simulation. In the second
stream, the paired substep changes are small relative to its FFT--policy
gap, but their intervals do not prove that sixteen substeps have converged
or that all simulation bias is negligible. Its sixteen-substep policy
interval includes FFT, whereas the first stream's independent-policy
interval does not. The cross-stream four-substep interval excludes zero
under the same $t_{15}$ approximation; this difference is reported rather
than dismissed as immaterial or used to pool the two estimates.

A supplementary one-step consistency check compares the update formulas under
identical state inputs and drivers in 18 configurations, covering zero
and two positive variance levels, no jump and positive or negative jumps,
and two step sizes. It finds zero price and variance discrepancy in those
tests, with Brownian increments standardized by $1/\sqrt{k}$. These checks find no update-formula
inconsistency, but are not an end-to-end simulator equivalence proof and
do not identify the cause of the cross-stream difference.

Conditional on a frozen non-anticipating stopping rule, independent
evaluation is unbiased for that rule's expected payoff under the simulated
discrete process. This expected payoff is a lower bound on the optimal
stopping value of that same process, not automatically on the continuous
jump--Heston value or on the value of a different CTMC approximation.
Writing $V_{\mathrm{sim}}^*$ for the optimal value of the simulated
discrete problem and $V_{\mathrm{sim}}^{\widehat\pi}$ for the expected
payoff of the fitted rule, the comparison decomposes as
\[
 V_{\mathrm{FFT}}-V_{\mathrm{sim}}^{\widehat\pi}
 =\bigl(V_{\mathrm{FFT}}-V_{\mathrm{sim}}^*\bigr)
  +\bigl(V_{\mathrm{sim}}^*-V_{\mathrm{sim}}^{\widehat\pi}\bigr).
\]
Only the second term is necessarily nonnegative. The first includes
differences between the simulated and CTMC pricing problems and has no
established sign here. The first independently evaluated mean is 6.1 of
its standard errors below FFT, whereas the second stream's sixteen-substep
mean is 1.64 standard errors below it. Neither comparison separates these
terms or certifies continuous-model accuracy; the substep experiment
measures only the sensitivity of the frozen policies over the tested
resolutions.
\subsection{PROJ kernel and continuation-action diagnostics}
\label{subsec:proj-operator-diagnostic}

The supplementary experiment exp13 uses Scenario A at $\Delta t=0.005$.
For each tested grid, the inverse FFT of the discounted convolution symbol
is compared, transition by transition, with the corresponding entry of
$e^{\Delta tQ_V}e^{-r\Delta t}$. The reported maximum absolute mass
discrepancies are approximately $10^{-14}$--$10^{-11}$. Signed kernel
mass is therefore consistent with the intended variance-chain transition
mass for these tests. This check concerns the convolution kernel; it does
not by itself verify constant preservation in the complete online action.
The complete-action diagnostic applies coefficient reconstruction, zero-padding,
FFT/symbol propagation, inverse FFT and retention of the first $N_x$
columns in the same order as that action. Its constant-input test finds
nonzero defects $P\mathbf1-e^{-r\Delta t}\mathbf1$ on all four grids.
The reported variance-averaged endpoint defects are approximately
$-0.595$ and $-0.405$. On the index range recorded as
$5,\ldots,N_x-40$, maximum defects are approximately $0.029$ for
$150\times24$ and $0.47$ for $700\times112$. This range is not a
uniformly defined deep-interior region: it retains part of the left-edge
band, and a fixed node count represents different physical distances
on different grids. These measurements do not establish that the
boundary treatment is harmless or identify the cause of amplification.

Zero-padding is a representation choice, not evidence that the option
value vanishes at both price boundaries. For each fixed variance state,
the normalized put payoff $\phi_1$ defined in
Section~\ref{subsec:pricing-algorithm}
tends to one as the transformed price tends to $-\infty$ and to zero
as it tends to $+\infty$. Thus a boundary explanation based on decay to
zero at both ends is not applicable to this put problem. The reported
near-strike one-step test and the constant-input test do not replace
a workload-specific domain-sensitivity check for the PROJ comparator.

The tested kernels have negative weights, with reported minimum near
$-0.3$ and negative-entry fractions of 22--31\%. The reported
negative-mass ratios decrease across the four grids in
Table~\ref{tab:proj-operator-diagnostic}. Each ratio is the summed
absolute mass of negative kernel entries divided by the summed absolute
mass of all kernel entries. One-step continuation of the
normalized put payoff $(1-e^x)^+$ produces small negative ripples near
the strike, of approximately $10^{-4}$--$6\times10^{-4}$ in magnitude.
Negative projection coefficients or kernel entries alone do not establish
instability of a complete pricing method.

The second diagnostic iterates the implemented linear continuation action
on a random positive vector for up to 200 steps, stopping early when the
norm exceeds $10^{30}$ (or falls below $10^{-30}$); the recorded
trajectories have 162, 187, 200, and 200 steps on the four grids,
respectively. The iteration retains coefficient
reprojection and omits only the exercise obstacle. Table~\ref{tab:proj-operator-diagnostic}
records the reported per-step geometric growth factors $g$. These are
finite-iteration measurements, not certified spectral-radius estimates.
With $a_n=\|P^nU_0\|_\infty$ and $T$ the recorded length of a trajectory,
exp13b specifies the statistic over the last 100 recorded values (99
log-increments),
\[
 g=\exp\!\left(\frac1{99}\sum_{n=T-98}^{T}
       (\log a_n-\log a_{n-1})\right)
   =\left(\frac{a_T}{a_{T-99}}\right)^{1/99}.
\]
No stepwise normalization is applied. The averaging windows are defined
per trajectory and are therefore not aligned across grids with different
recorded lengths.

\begin{table}[htbp]
\centering\small
\caption{Reported PROJ diagnostics on four Scenario A grids at $\Delta t=0.005$.}
\label{tab:proj-operator-diagnostic}
\begin{tabular}{lrrl}
\toprule
Grid & Negative-mass ratio & Growth factor $g$ & Price configuration\\
\midrule
$150\times24$ & $2.88\times10^{-2}$ & 1.5406 & unstable\\
$250\times40$ & $1.60\times10^{-2}$ & 1.4562 & unstable\\
$500\times80$ & $4.97\times10^{-3}$ & 1.1424 & unstable\\
$700\times112$ & $2.93\times10^{-3}$ & 0.9995 & usable\\
\bottomrule
\end{tabular}

\par\smallskip\footnotesize Negative-mass ratio: summed absolute mass of negative kernel entries relative to summed absolute mass of all entries. The growth factor $g$ is a finite-iteration measurement over the last 100 recorded norms, not a certified spectral radius; see the appendix text.
\end{table}

The correspondence between large amplification and unstable prices supports
a resolution-sensitive continuation mechanism on these tested grids.
Finite-iteration growth can nevertheless include nonnormal transients;
$g>1$ alone is not proof of spectral radius greater than one, and a
single run with $g<1$ does not establish power boundedness. Extrapolating
$g^{1000}$ additionally assumes that the measured growth persists and
does not quantitatively predict the nonlinear obstacle recursion.
Transition-mass agreement and repeated-action diagnostics address different
properties of the numerical operator. Together they support the interpretation
of resolution-sensitive continuation, but do not by themselves isolate its
cause. The short-time density scaling discussed in Section~\ref{sec:numerics}
explains why spatial and projection resolution must be considered jointly with
the propagation interval. The
$350\times60$ exp10 configuration, the $700\times120$ tests at $M=2000$, and
the unstable Scenario B configurations have not been covered by these four diagnostics.
In particular, intrinsic-value collapse is not attributed to overflow or
NaN handling without a corresponding execution trace.

\paragraph{Same-spacing price-domain cross-check.}
Experiments exp14 and exp14b retain Scenario A, $K=40$, $T=5$, $M=1000$
and the 112-node variance grid on $[10^{-4},0.60]$. The baseline price
grid has 700 nodes on $[-1.5,1.5]$ and spacing $\delta_x=3/699=1/233$.
The original expansion to 933 nodes on $[-2,2]$ has the same spacing,
but shifts the grid phase by half a cell. The nested-node check instead
uses 934 nodes with half-width
$1.5+117(3/699)=2.0021459227\ldots$, preserving the original node
locations up to floating-point rounding by adding 117 intervals on each
side. This configuration is distinct from the $700\times120$ LSM
reference comparison in Table~\ref{tab:lsm-substep}.

\begin{table}[htbp]
\centering\small
\caption{Scenario A at-the-money PROJ--FFT price-domain cross-check at
fixed price spacing and variance grid.}
\label{tab:proj-domain-crosscheck}
\begin{tabular}{lrrr}
\toprule
Price grid & PROJ price & FFT price & PROJ $-$ FFT\\
\midrule
700 nodes, half-width 1.5 & 6.07350986 & 6.07402878 & $-5.19\times10^{-4}$\\
933 nodes, half-width 2 & 6.07350972 & 6.07346110 & $+4.86\times10^{-5}$\\
934 nodes, nested expansion & 6.07350989 & 6.07346126 & $+4.86\times10^{-5}$\\
\bottomrule
\end{tabular}
\tablenote{Parameters, variance domain and exercise dates are fixed;
$\delta_x=1/233$ in all rows. Prices are rounded only for display.
No expanded-domain runtime comparison is reported.}
\end{table}

Relative to the baseline, the nested expansion changes PROJ by approximately
$3.0\times10^{-8}$ and FFT by $-5.68\times10^{-4}$; the expanded
PROJ--FFT difference is $4.8625\times10^{-5}$. The original and nested
expansions give very similar price changes, supporting the interpretation
that the observed domain sensitivity is not primarily driven by this
particular grid-phase change. This limited-configuration check neither
decomposes the boundary error nor establishes continuous-model accuracy.
\subsection{Futures-tree convention check}
\label{subsec:futures-tree-check}

An independent constant-volatility calculation checks the zero-drift
futures-tree convention against the analytic Black--76 European put price.
For each row of Table~\ref{tab:futures-tree-check}, we use $r=0.02$,
$\sigma_B=0.20$, 800 equally spaced steps and $\Delta t=T/800$.
The lattice uses $p_F=(1-d)/(u-d)$ and step discount $e^{-r\Delta t}$;
no early-exercise obstacle is imposed in this European check.
The analytic value is
\[
 P_{76}=e^{-rT}\{K\Phi(-d_2)-F\Phi(-d_1)\},\qquad
 d_1=\frac{\log(F/K)+\tfrac12\sigma_B^2T}{\sigma_B\sqrt T},\quad
 d_2=d_1-\sigma_B\sqrt T.
\]
The script \texttt{futures\_tree\_check.py} records these inputs and also
computes the result of deliberately substituting the stock-tree probability.

\begin{table}[htbp]
\centering\small
\caption{Independent European futures-tree check against Black--76.}
\label{tab:futures-tree-check}
\begin{tabular}{rrrrrr}
\toprule
$F$ & $K$ & $T$ & $P_{76}$ & Tree value & Relative difference (\%)\\
\midrule
760 & 820 & 0.150 & 65.083894 & 65.081925 & $-0.003025$\\
760 & 896 & 0.085 & 135.802791 & 135.802448 & $-0.000252$\\
600 & 650 & 0.300 & 58.913973 & 58.917791 & $+0.006482$\\
\bottomrule
\end{tabular}
\end{table}

Here relative difference is $100(P_{\mathrm{tree}}/P_{76}-1)$, using
unrounded prices. The small, mixed-sign discrepancies are consistent with
lattice discretization for these configurations. With the stock-tree
probability substituted and every other input unchanged, the relative
differences are instead $-2.8785\%$, $-0.9483\%$ and $-4.5119\%$.
Their sizes depend on option moneyness and cannot be inferred from $rT$
alone. This is a check of the stated convention with an independent
implementation, not a rerun of the production market-tree routine or a
validation of American values under stochastic volatility and jumps.
\subsection{Exercise-mesh runtime diagnostic}
\label{subsec:M-runtime}

\begin{figure}[!htbp]
\centering
\safeincludegraphics[width=0.90\linewidth,keepaspectratio]{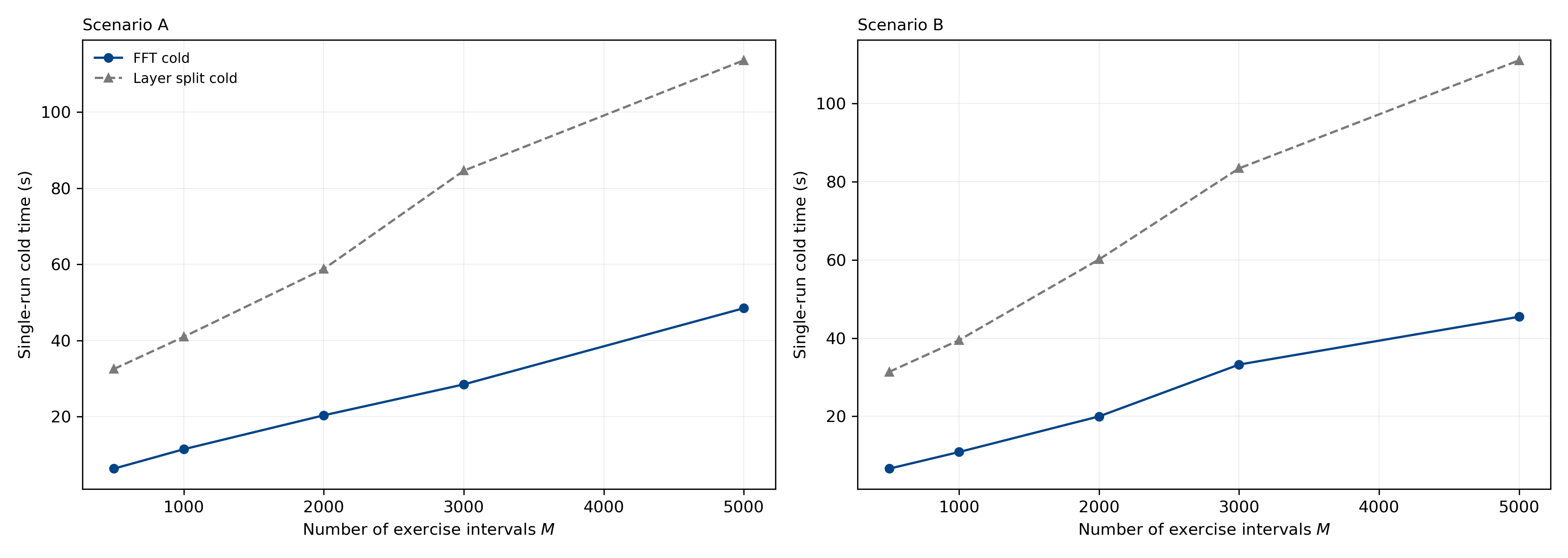}
\caption{Single-run runtime diagnostic across exercise meshes on the common
$700\times120$ grid. ``Cold'' denotes setup plus online recursion. These
single-run timings are retained as a workload diagnostic; the repeated
$M=1000$ measurements in Table~\ref{tab:timing-M1000} remain the primary timing
evidence.}
\label{fig:num-M-runtime}
\end{figure}

Figure~\ref{fig:num-M-runtime} reports single-run setup-plus-online and online
times for the same one-year $M$-sweep. At every tested mesh the strict
endpoint-closed layer split is materially more expensive than the Fourier-block
recursion. For example, in Scenario A at $M=5000$ the measured online times are
approximately $21.8$ seconds for FFT and $81.0$ seconds for the layer split,
with cold times of approximately $23.3$ and $93.2$ seconds, respectively.

The single-run sweep also reveals substantial machine-load variability. In
Scenario A, for instance, the FFT online time at $M=3000$ exceeds the
$M=5000$ time, and analogous irregularities appear in Scenario B. These
nonmonotonic wall-clock observations cannot represent algorithmic complexity,
since the online FFT recursion performs the same operations at each exercise
step and therefore has work proportional to $M$ on a fixed grid. We therefore
do not estimate scaling exponents from Figure~\ref{fig:num-M-runtime}. The
controlled comparison is Table~\ref{tab:timing-M1000}, which uses a warm-up and
five single-thread repetitions and reports medians and ranges. The $M$-sweep
is retained as a workload diagnostic for these implementations, without
estimating a scaling exponent or attributing every timing difference to the
Fourier-block representation.
\subsection{Differences from the LSM benchmark at the production mesh}

For the production exercise mesh, the LSM estimate provides an independent
Monte Carlo price benchmark. Table~\ref{tab:performance} shows absolute price
differences from the $M=1000$ LSM benchmark. These numbers should be interpreted
as benchmark differences, not errors relative to an exact solution,
because LSM retains regression and time-discretization error even when its
sampling uncertainty is small.

\begin{table}[htbp]\centering\small
\caption{Absolute price differences from the LSM benchmark at $M=1000$.}
\label{tab:performance}
\begin{tabular}{llrrrr}\toprule
Scenario&Method&$K=35$&$K=40$&$K=45$&Mean absolute difference\\\midrule
A&Poly&0.011352&0.020190&0.029175&0.020239\\
A&FFT&0.014055&0.020805&0.029362&0.021407\\
A&Layer split&0.015468&0.026043&0.036827&0.026113\\
A&PROJ&0.011422&0.020286&0.029289&0.020332\\
A&PROJ-HiRes&0.011423&0.020287&0.029291&0.020334\\\addlinespace
B&Poly&0.016528&0.027499&0.036956&0.026994\\
B&FFT&0.016631&0.027559&0.037016&0.027069\\
B&Layer split&0.024915&0.038537&0.050627&0.038026\\
B&PROJ&2.309592&1.942714&1.383553&1.878620\\
B&PROJ-HiRes&0.015946&0.026793&0.036171&0.026303\\
\bottomrule
\end{tabular}
\tablenote{The corresponding LSM replication standard errors for $K=35,40,45$ are
$(1.124,1.450,1.734)\times10^{-3}$ in Scenario A and
$(1.455,1.814,2.073)\times10^{-3}$ in Scenario B.}
\end{table}
\section{Supplementary market-data diagnostics}\label{app:market-data}
\begin{figure}[!htbp]
\centering
\safeincludegraphics[width=0.82\linewidth,keepaspectratio]{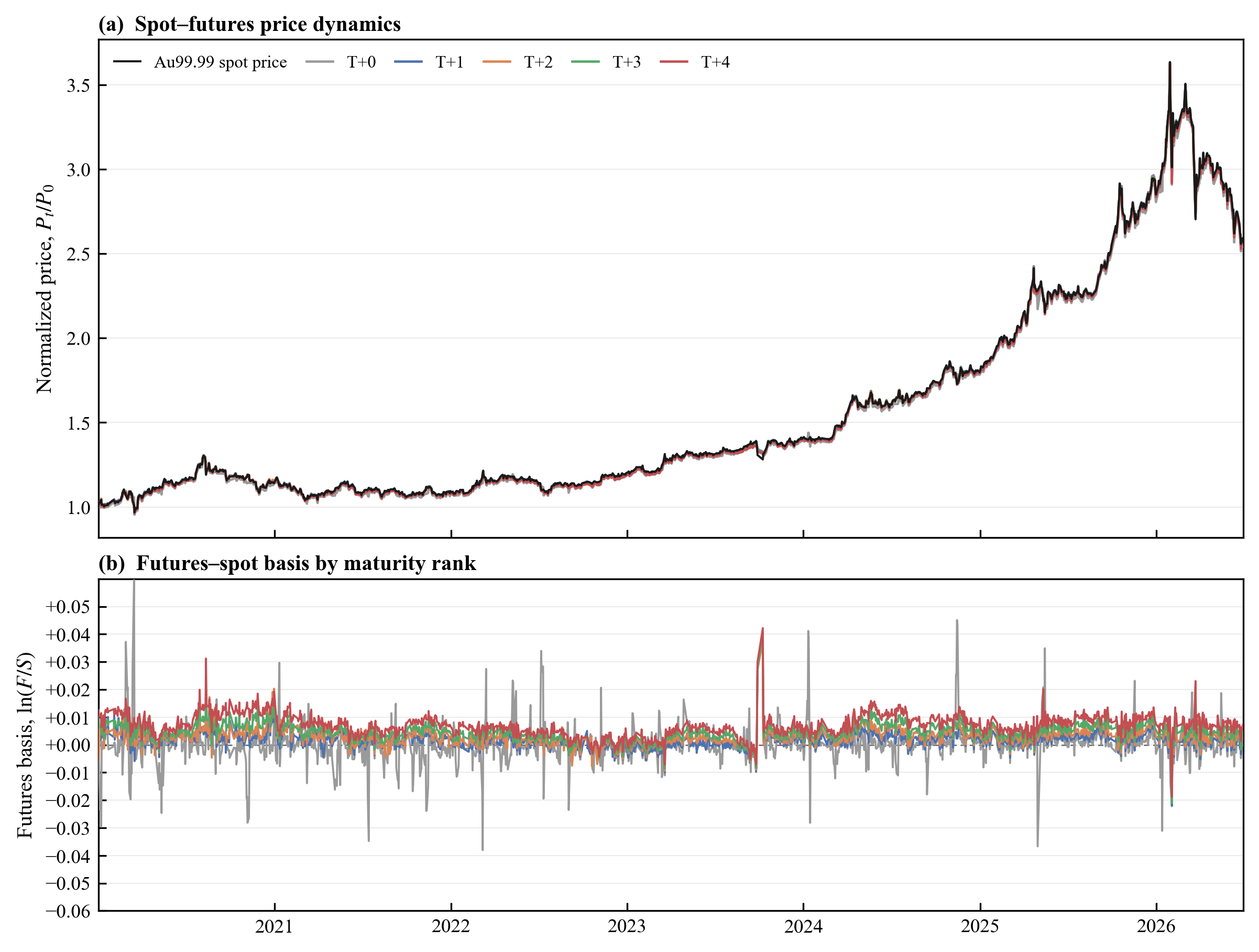}
\caption{Gold spot and futures price dynamics. Panel A compares Au99.99 spot prices with the constructed maturity-rank series $T+0$--$T+4$; all series track spot closely in levels. Panel B reports the logarithmic futures--spot basis for each maturity rank $T+0$--$T+4$; the band is tight and ordered by maturity, while the expiring contract $T+0$ drifts more widely. The figure period is restricted by SGE spot availability (January 2020--June 2026).}
\label{fig:f1}
\end{figure}

\begin{figure}[!htbp]
\centering
\safeincludegraphics[width=0.95\linewidth,keepaspectratio]{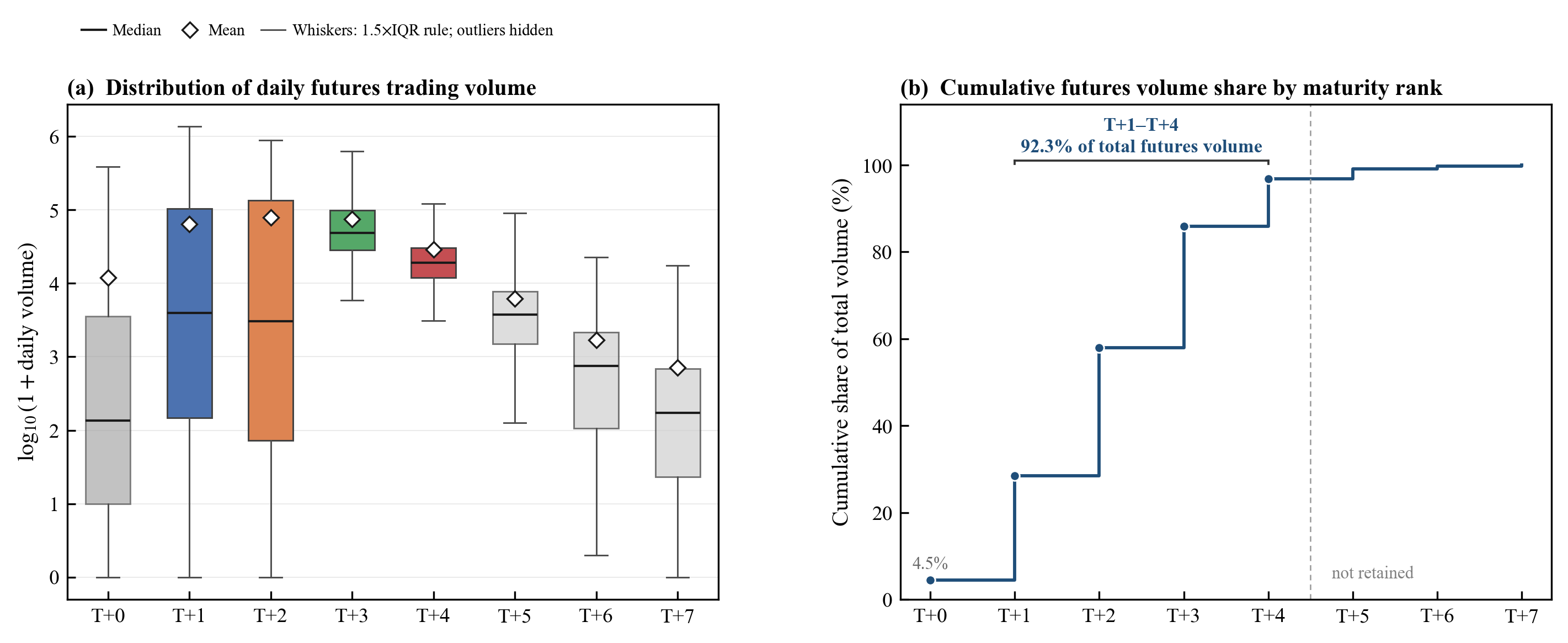}
\caption{Futures liquidity concentration across maturity ranks. Panel A reports daily futures trading volume distributions. Panel B reports cumulative futures volume shares across maturity ranks.}
\label{fig:f2}
\end{figure}

\begin{table}[htbp]
\centering
\caption{Option trading volume on the $T+1$--$T+4$ underlying universe.}\label{tab:liq}
\small
\begin{tabular}{llrrrr}
\toprule
Maturity & Moneyness & $T+1$ & $T+2$ & $T+3$ & $T+4$\\
\midrule
\multirow{3}{*}{\shortstack[l]{Short\\($\tilde{\tau}\le 21$)}} & OTM & 68.4 (42,764) & 0.2 (248) & --- & ---\\
 & ATM & 11.4 (4,036) & 0.0 (22) & --- & ---\\
 & ITM & 3.6 (42,764) & 0.0 (248) & --- & ---\\
\addlinespace
\multirow{3}{*}{\shortstack[l]{Medium\\($22\le\tilde{\tau}\le 42$)}} & OTM & 14.0 (12,793) & 76.2 (40,495) & --- & ---\\
 & ATM & 1.8 (1,280) & 8.9 (4,046) & --- & ---\\
 & ITM & 0.8 (12,793) & 4.9 (40,495) & --- & ---\\
\addlinespace
\multirow{3}{*}{\shortstack[l]{Long\\($43\le\tilde{\tau}\le 84$)}} & OTM & --- & 8.4 (7,481) & 81.7 (53,360) & 1.5 (663)\\
 & ATM & --- & 0.7 (712) & 6.5 (5,046) & 0.1 (74)\\
 & ITM & --- & 0.6 (7,481) & 8.2 (53,360) & 0.2 (663)\\
\addlinespace
\multirow{3}{*}{\shortstack[l]{Ultra-long\\($\tilde{\tau}> 84$)}} & OTM & --- & --- & 3.0 (4,217) & 80.3 (50,953)\\
 & ATM & --- & --- & 0.2 (370) & 6.2 (5,174)\\
 & ITM & --- & --- & 0.4 (4,217) & 11.6 (50,953)\\
\midrule
\multicolumn{2}{l}{Aggregate option volume (million lots)} & 37.7 & 16.9 & 6.4 & 1.9\\
\bottomrule
\end{tabular}
\tablenote{The table uses all 446,708 option-day observations written on $T+1$--$T+4$ underlyings before option-level screening. Cells report the share of the column's aggregate option volume (\%) and, in parentheses, the number of option-day observations; ``---'' indicates absent contract types. Options are grouped by time to maturity $\tilde{\tau}$ in trading days (21 trading days $\approx$ one calendar month) and by moneyness: ATM if $|\ln(F/K)|\le 0.01$, with OTM and ITM assigned according to contract type. The banded structure reflects exchange listing mechanics: an option expires shortly before its underlying's delivery month, so the underlying's maturity rank determines the maturities at which its options exist. The OTM and ITM counts coincide because calls and puts are listed at identical strikes, so every out-of-the-money contract has an in-the-money counterpart of the opposite type; the corresponding volume shares nonetheless differ, reflecting where trading takes place.}
\end{table}

\begin{table}[htbp]
\centering
\caption{Option trading volume by contract type on the $T+1$--$T+4$ underlying universe.}\label{tab:liqcp}
\small
\textit{Panel A: Calls}\\
\medskip
\begin{tabular}{llrrrr}
\toprule
Maturity & Moneyness & $T+1$ & $T+2$ & $T+3$ & $T+4$\\
\midrule
\multirow{3}{*}{\shortstack[l]{Short\\($\tilde{\tau}\le 21$)}} & OTM & 66.8 (18,685) & 0.2 (103) & --- & ---\\
 & ATM & 11.7 (2,018) & 0.0 (11) & --- & ---\\
 & ITM & 3.9 (24,079) & 0.0 (145) & --- & ---\\
\addlinespace
\multirow{3}{*}{\shortstack[l]{Medium\\($22\le\tilde{\tau}\le 42$)}} & OTM & 14.6 (5,675) & 74.2 (17,622) & --- & ---\\
 & ATM & 2.0 (640) & 9.5 (2,023) & --- & ---\\
 & ITM & 1.0 (7,118) & 5.9 (22,873) & --- & ---\\
\addlinespace
\multirow{3}{*}{\shortstack[l]{Long\\($43\le\tilde{\tau}\le 84$)}} & OTM & --- & 8.7 (3,421) & 79.9 (23,055) & 1.6 (268)\\
 & ATM & --- & 0.7 (356) & 7.3 (2,523) & 0.2 (37)\\
 & ITM & --- & 0.6 (4,060) & 9.5 (30,305) & 0.2 (395)\\
\addlinespace
\multirow{3}{*}{\shortstack[l]{Ultra-long\\($\tilde{\tau}> 84$)}} & OTM & --- & --- & 2.8 (2,028) & 77.2 (22,352)\\
 & ATM & --- & --- & 0.2 (185) & 7.0 (2,587)\\
 & ITM & --- & --- & 0.4 (2,189) & 13.7 (28,601)\\
\midrule
\multicolumn{2}{l}{Aggregate call volume (million lots)} & 21.5 & 9.9 & 3.4 & 0.9\\
\bottomrule
\end{tabular}

\medskip
\textit{Panel B: Puts}\\
\medskip
\begin{tabular}{llrrrr}
\toprule
Maturity & Moneyness & $T+1$ & $T+2$ & $T+3$ & $T+4$\\
\midrule
\multirow{3}{*}{\shortstack[l]{Short\\($\tilde{\tau}\le 21$)}} & OTM & 70.6 (24,079) & 0.2 (145) & --- & ---\\
 & ATM & 11.0 (2,018) & 0.0 (11) & --- & ---\\
 & ITM & 3.2 (18,685) & 0.0 (103) & --- & ---\\
\addlinespace
\multirow{3}{*}{\shortstack[l]{Medium\\($22\le\tilde{\tau}\le 42$)}} & OTM & 13.3 (7,118) & 79.0 (22,873) & --- & ---\\
 & ATM & 1.4 (640) & 8.0 (2,023) & --- & ---\\
 & ITM & 0.5 (5,675) & 3.5 (17,622) & --- & ---\\
\addlinespace
\multirow{3}{*}{\shortstack[l]{Long\\($43\le\tilde{\tau}\le 84$)}} & OTM & --- & 8.1 (4,060) & 83.9 (30,305) & 1.5 (395)\\
 & ATM & --- & 0.8 (356) & 5.7 (2,523) & 0.1 (37)\\
 & ITM & --- & 0.5 (3,421) & 6.7 (23,055) & 0.2 (268)\\
\addlinespace
\multirow{3}{*}{\shortstack[l]{Ultra-long\\($\tilde{\tau}> 84$)}} & OTM & --- & --- & 3.3 (2,189) & 83.2 (28,601)\\
 & ATM & --- & --- & 0.2 (185) & 5.4 (2,587)\\
 & ITM & --- & --- & 0.3 (2,028) & 9.6 (22,352)\\
\midrule
\multicolumn{2}{l}{Aggregate put volume (million lots)} & 16.2 & 7.0 & 3.0 & 1.0\\
\bottomrule
\end{tabular}
\tablenote{Same sample and conventions as Table~\ref{tab:liq}. Within each panel, cells report the share of the column's aggregate volume of that contract type (\%) and, in parentheses, the number of option-day observations. Because calls and puts are listed at identical strikes, the out-of-the-money count of one contract type equals the in-the-money count of the other in the same cell; the two panels' volume shares nonetheless differ, reflecting where trading takes place.}
\end{table}
\subsection{Implied-volatility surface dynamics and skew diagnostics}\label{app:surface-data}

\begin{figure}[!htbp]
\centering
\safeincludegraphics[width=\linewidth,keepaspectratio]{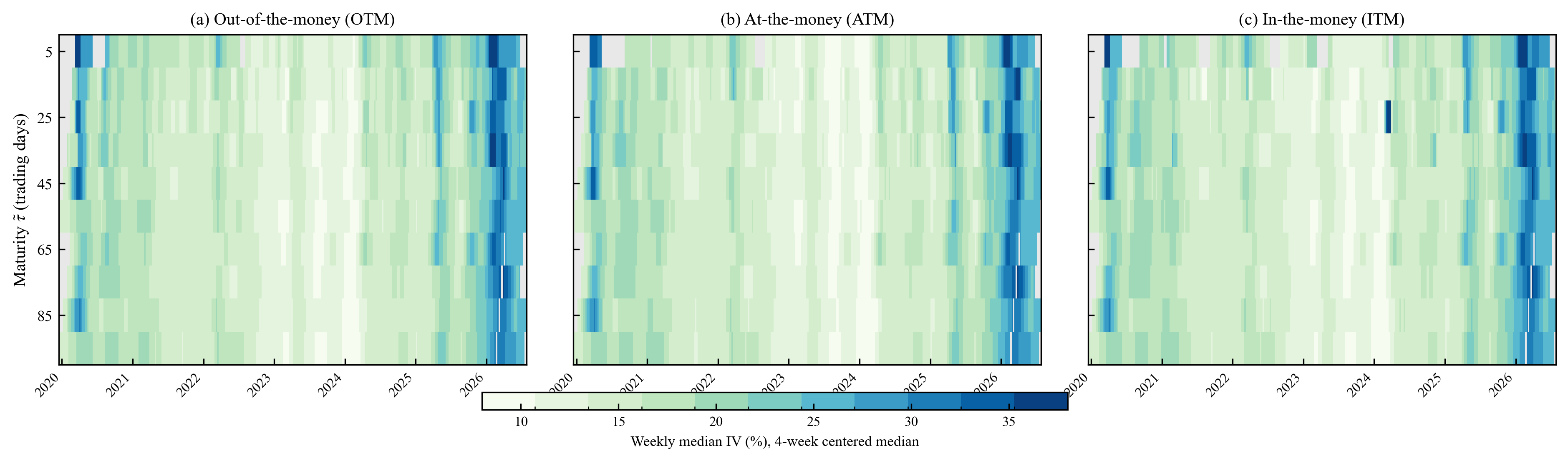}
\caption{Implied volatility surfaces by moneyness class. Panels (a)--(c) report the out-of-the-money, at-the-money, and in-the-money surfaces, restricted to the standardized band $|\ln(F/K)|\le 0.05$ (at-the-money: $|\ln(F/K)|\le 0.01$). Each cell is the weekly median implied volatility (\%) over all option-day observations with feasible close-price inversion on the $T+1$--$T+4$ underlying universe (before option-level screening), smoothed with a four-week centered rolling median; maturities are binned at ten-trading-day intervals, and gaps of up to six weeks are linearly interpolated for display; light-grey cells denote the missing values remaining after this display interpolation.}
\label{fig:f5s}
\end{figure}

We now characterize the empirical object that numerical pricing methods must confront: the implied volatility surface and its dynamics over the sample period (Figure~\ref{fig:f5s}). For every trading week and maturity bin, each panel reports the median implied volatility over all option-day observations with feasible close-price inversion on the $T+1$--$T+4$ underlying universe, restricted to a standardized moneyness band $|\ln(F/K)|\le 0.05$ around the at-the-money point. The weekly median reduces the influence of individual stale quotes; because in-the-money options trade thinly (Table~\ref{tab:liq}), the corresponding panels rest on fewer active quotes and rely more heavily on this aggregation. This description deliberately uses the pre-screen panel to maximize cross-sectional coverage; where individual-quote noise matters more---the skew estimates below---the volume screen is imposed instead. Implied volatilities peak in early 2020, decline through 2021--2023, and rise again over 2024--2026, coinciding with the increase in gold prices. Readings are highest at short maturities: the median out-of-the-money implied volatility declines from 17.9\% at $\tilde{\tau}\le 21$ trading days to 16.1\% beyond 84 days.

\begin{figure}[!htbp]
\centering
\safeincludegraphics[width=0.85\linewidth,keepaspectratio]{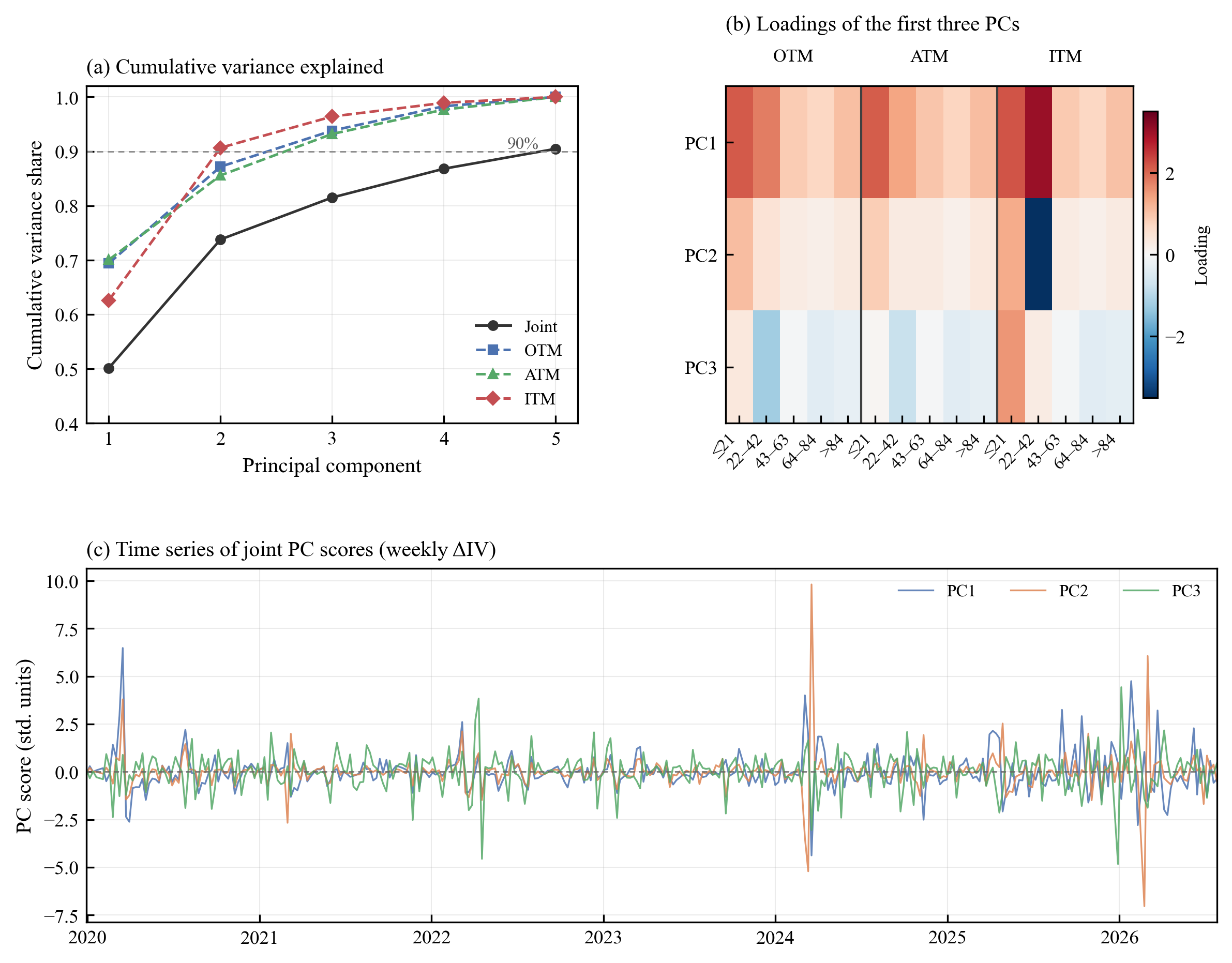}
\caption{Principal components of weekly implied volatility changes. Panel (a): cumulative variance explained by the first five components for the joint grid and within each moneyness class. Panel (b): loadings of the first three joint components across the moneyness-by-maturity grid. Panel (c): standardized time series of the joint component scores.}
\label{fig:f6pca}
\end{figure}

To characterize the factor structure of these dynamics, we follow the principal-component analysis of implied volatility surface changes in \cite{ContDaFonseca2002} and \cite{GoncalvesGuidolin2006}. We form a weekly panel (339 weeks) of median implied volatilities on the moneyness classes of Figure~\ref{fig:f5s}, with maturities aggregated into five bins; all fifteen grid columns are observed on at least 60\% of weeks, and remaining gaps of at most eight weeks are linearly interpolated. Covariance PCA is then applied to the resulting 337 weekly first differences. Table~\ref{tab:pca} and Figure~\ref{fig:f6pca} report the results. The first three factors explain 81.5\% of the joint variation. The leading factor loads with the same sign on every column and its scores track the weekly average change in implied volatility with a correlation of 0.98, identifying a level factor (50.1\% of variation); the second factor is a moneyness contrast concentrated in in-the-money cells, and the third is curvature-like. Within each moneyness class the structure is even more concentrated, with the first three components explaining 93.8\% (OTM), 93.2\% (ATM), and 96.4\% (ITM) of weekly variation. Restricting the panel to observations that pass the option-level volume screen leaves the qualitative structure intact---the leading factor remains a level factor, and the first three components explain 88.2\% of the reduced ten-column panel's variation---indicating that the volume-screened sample preserves the main qualitative factor structure of the broader market. The market is therefore rich in cross-sectional shape but relatively low-dimensional in its weekly dynamics: a candidate pricing model must reproduce the dominant level movements while also matching the smile features documented below.

\begin{table}[htbp]
\centering\small
\caption{Principal component analysis of weekly implied volatility changes.}\label{tab:pca}
\begin{tabular}{lrrrrr}
\toprule
Panel & Grid columns & PC1 & PC2 & PC3 & PC1--3\\
\midrule
Joint (OTM+ATM+ITM) & 15 & 50.1 & 23.7 & 7.7 & 81.5\\
OTM & 5 & 69.4 & 17.8 & 6.6 & 93.8\\
ATM & 5 & 70.0 & 15.5 & 7.6 & 93.2\\
ITM & 5 & 62.6 & 28.1 & 5.8 & 96.4\\
\midrule
Joint, volume $>$ 100 screen & 10 & 67.4 & 13.6 & 7.2 & 88.2\\
\bottomrule
\end{tabular}
\tablenote{Cells report variance shares (\%) of principal components of weekly first differences of weekly median implied volatilities. The main rows use all option-day observations with feasible close-price inversion on the $T+1$--$T+4$ universe, restricted to the moneyness band of Figure~\ref{fig:f5s}; after interpolation and removal of weeks with remaining gaps, this panel contains 337 first differences. The last row restricts the joint panel to observations passing the volume screen, under which five grid columns fall below 40\% weekly coverage and are dropped; the resulting panel contains 336 first differences. Remaining gaps of at most eight weeks are linearly interpolated; the PCA uses the covariance matrix without standardization.}
\end{table}

\begin{figure}[!htbp]
\centering
\safeincludegraphics[width=0.85\linewidth,keepaspectratio]{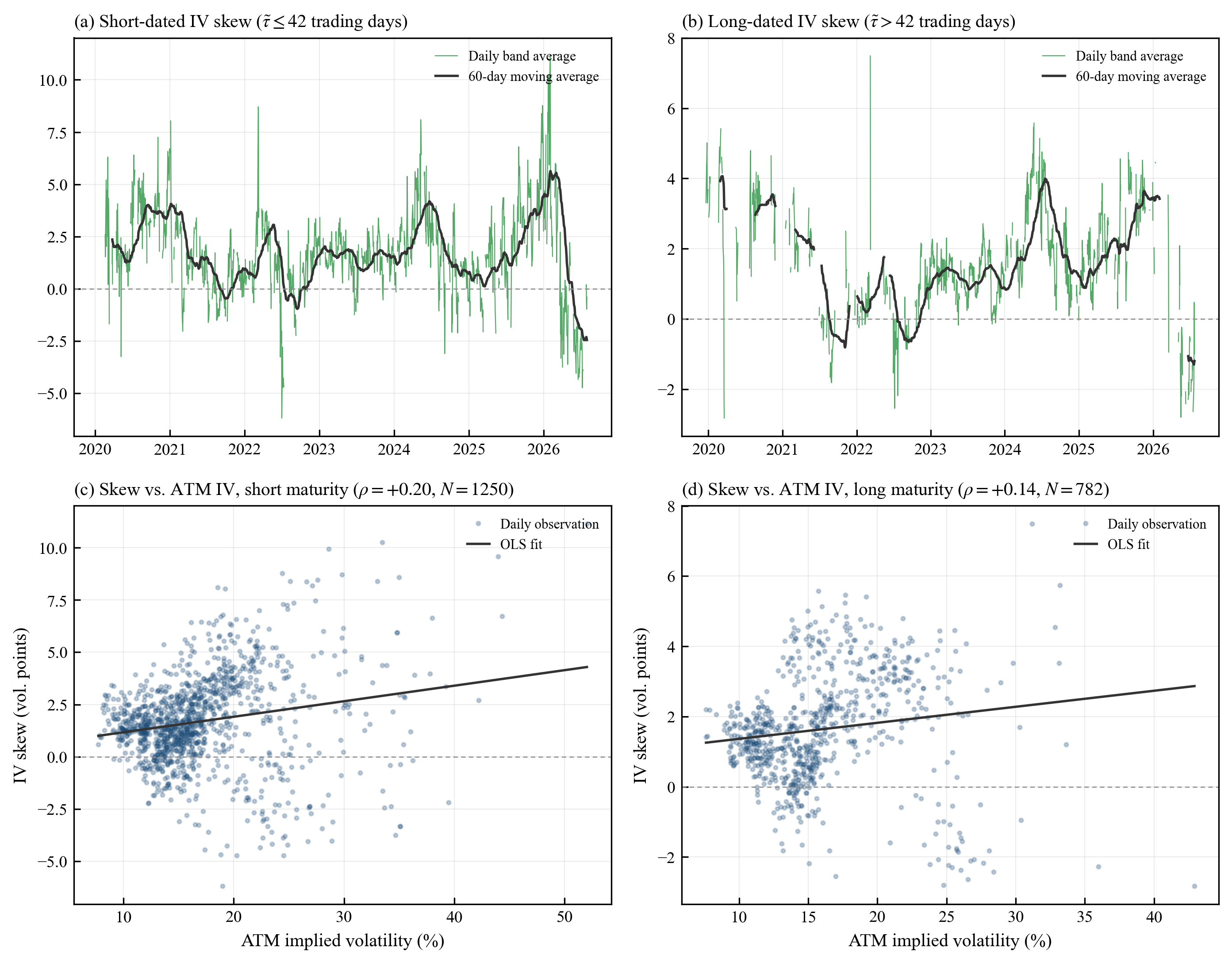}
\caption{Implied volatility skew and its relation to the volatility level. Following the difference-based skew constructions of \cite{XingZhangZhao2010,Yuan2022}, the skew is the call-wing minus the put-wing band average of the volume-screened OTM universe, with maturities split at the two-month line ($\tilde{\tau}\le 42$ versus $> 42$ trading days), the ATM band as $|\ln(F/K)|\le 0.025$, and the wings as the bands $0.025<|\ln(F/K)|\le 0.08$ on either side. Panels (a)--(b): daily skew (thin) and 60-trading-day moving averages (thick). Panels (c)--(d): daily skew against the ATM band average, with the OLS fit and the correlation $\rho$.}
\label{fig:f7skew}
\end{figure}

To characterize the asymmetry of the implied volatility surface, we follow the difference-based skew constructions of the option literature \cite{XingZhangZhao2010,Yuan2022} and construct daily series of implied volatility characteristics as averages over fixed moneyness and maturity bands of the volume-screened OTM universe, splitting maturities at the two-month line of Table~\ref{tab:liq}: short (long) maturities are $\tilde{\tau}\le 42$ ($> 42$) trading days, the at-the-money band is $|\ln(F/K)|\le 0.025$, and the out-of-the-money call and put wings are the bands $0.025<|\ln(F/K)|\le 0.08$ on either side. Figure~\ref{fig:f7skew} plots the implied volatility skew---the difference between the call-wing and put-wing band averages, defined on days when both wings are quoted---for both maturity buckets, together with its scatter against the ATM level. Three features stand out. First, the skew is positive on $83\%$ of days at short maturities and $89\%$ at long maturities, averaging $+1.7$ and $+1.5$ volatility points, respectively: out-of-the-money calls trade at higher implied volatilities than equally distant out-of-the-money puts, in contrast to the negative skew commonly documented for equity-index options. This pattern is consistent with relatively strong pricing of upside gold exposure, although the option data alone do not identify its economic source. Second, the skew is only weakly related to the volatility level: the correlation between the skew and ATM implied volatility is $0.20$ in levels at short maturities and $0.14$ at long maturities, while the corresponding correlations in daily changes are $+0.28$ and $+0.07$. The distinct variation in level and tilt would be difficult to reproduce using a mechanism that moves the entire surface only through a common volatility-level factor. The term structure is on average slightly downward sloping: the spread between the long- and short-maturity ATM implied volatilities averages $-0.3$ points and is negative on $55\%$ of days. Third, the implied volatility series are strongly persistent: the daily node series underlying Table~\ref{tab:skew} have first-order autocorrelations between $0.87$ and $0.97$, decaying to roughly $0.4$--$0.6$ at lag 30 and somewhat lower for the thinnest long-maturity put-wing series. These patterns motivate stochastic volatility and asymmetric jumps. Both components can affect the smile, and the descriptive statistics do not separately identify their parameters.

\begin{table}[htbp]
\centering
\caption{Summary statistics of daily implied volatilities at fixed moneyness and maturity nodes.}\label{tab:skew}
\small
\setlength{\tabcolsep}{4.5pt}
\begin{tabular}{llrrrrrrrr}
\toprule
$\tilde{\tau}$ & $\ln(F/K)$ & $N$ & Mean & Std.\ dev. & Skewness & Kurtosis & AC(1) & AC(15) & AC(30)\\
\midrule
21 & $-0.05$ (call wing) & 513 & 20.01 & 6.02 & 1.30 & 2.49 & 0.95 & 0.56 & 0.35\\
 & $-0.025$ & 527 & 18.48 & 6.01 & 1.27 & 2.46 & 0.95 & 0.57 & 0.37\\
 & $0$ (ATM) & 543 & 17.57 & 5.91 & 1.23 & 2.12 & 0.95 & 0.57 & 0.40\\
 & $+0.025$ & 551 & 17.58 & 5.64 & 1.18 & 1.86 & 0.95 & 0.57 & 0.42\\
 & $+0.05$ (put wing) & 538 & 18.52 & 5.43 & 1.23 & 1.86 & 0.95 & 0.57 & 0.43\\
\addlinespace
42 & $-0.05$ (call wing) & 940 & 18.12 & 5.36 & 1.36 & 3.05 & 0.96 & 0.70 & 0.54\\
 & $-0.025$ & 1008 & 17.33 & 5.49 & 1.42 & 3.20 & 0.97 & 0.72 & 0.56\\
 & $0$ (ATM) & 997 & 16.54 & 5.36 & 1.40 & 3.11 & 0.96 & 0.72 & 0.56\\
 & $+0.025$ & 950 & 16.00 & 5.02 & 1.36 & 2.80 & 0.96 & 0.70 & 0.56\\
 & $+0.05$ (put wing) & 841 & 15.94 & 4.66 & 1.54 & 3.79 & 0.96 & 0.65 & 0.57\\
\addlinespace
63 & $-0.05$ (call wing) & 663 & 16.64 & 4.76 & 1.23 & 1.89 & 0.97 & 0.70 & 0.55\\
 & $-0.025$ & 753 & 16.26 & 4.91 & 1.36 & 2.71 & 0.96 & 0.70 & 0.59\\
 & $0$ (ATM) & 736 & 15.48 & 4.70 & 1.42 & 2.91 & 0.95 & 0.69 & 0.53\\
 & $+0.025$ & 697 & 14.85 & 4.36 & 1.40 & 2.75 & 0.95 & 0.68 & 0.44\\
 & $+0.05$ (put wing) & 599 & 14.29 & 3.89 & 1.58 & 3.82 & 0.92 & 0.59 & 0.50\\
\addlinespace
84 & $-0.05$ (call wing) & 316 & 15.40 & 4.31 & 1.05 & 0.38 & 0.95 & 0.61 & 0.48\\
 & $-0.025$ & 341 & 15.26 & 4.52 & 1.01 & 0.37 & 0.97 & 0.64 & 0.56\\
 & $0$ (ATM) & 346 & 14.66 & 4.35 & 1.22 & 1.13 & 0.97 & 0.62 & 0.57\\
 & $+0.025$ & 293 & 13.58 & 3.81 & 1.69 & 3.74 & 0.91 & 0.56 & 0.22\\
 & $+0.05$ (put wing) & 250 & 12.77 & 3.06 & 2.19 & 8.49 & 0.87 & 0.41 & 0.00\\
\bottomrule
\end{tabular}
\tablenote{Statistics (in \%) of daily implied volatilities at fixed moneyness and maturity nodes, obtained by linear interpolation of the volume-screened OTM quotes in the (log-moneyness, maturity) plane at $\ln(F/K)\in\{\pm 0.05,\pm 0.025,0\}$ and $\tilde{\tau}\in\{21,42,63,84\}$ trading days. $N$ is the number of trading days on which the node is defined; kurtosis is in excess of the Gaussian value; AC($k$) is the autocorrelation at lag $k$, computed over the ordered sequence of trading days on which the node is defined. The construction follows Table~3 of \cite{Yuan2022}.}
\end{table}

\section{Calibration search box and solver settings}
\label{app:calibration-settings}

\subsection{Sample screening}
\label{app:screening-stages}

The multi-day calibration of Section~\ref{sec:empirical} uses the
validity screen and the one-pass within-chain strike-convexity screen
described there, applied uniformly across all (date, chain) pairs of
the estimation window. The September--October window retains 178
quotes; the stress window of
Appendix~\ref{app:stress} retains 287. No other exclusion is applied
at the calibration stage, and the underlying universe is the
complete-chain sample of Section~\ref{sec:data} without modification.

\subsection{An extreme-volatility stress window}
\label{app:stress}

The February~2--13, 2026 window estimates the model at the most
volatile point of the sample: the 20-day realized volatility of the
underlying futures reaches 0.65 (against a full-sample
90th percentile of 0.279), and the 20-day realized variance exceeds
the $V_0$ box. The window is estimated under the protocol of
Section~\ref{subsec:multiday} without modification. The fit is
RMSPE~16.74\% with six structural parameters interior, the daily $V_0$ states track the decay of the episode, and
its boundary parameters are the same ones diagnosed in
Section~7.3. The window is reported as a stress check on the
protocol, not as an estimation window: its maturities
($0.03$--$0.14$ years) may provide limited information about the
long-run variance level.

\subsection{Multi-day alternating search}
\label{app:multiday-search}

The multi-day calibration of Section~\ref{subsec:multiday} uses the
following round structure. Each \emph{round} alternates two stages. In
Stage~A the daily $V_0$ states are held fixed and the eight shared
structural parameters are estimated jointly on the pooled relative
price residuals of all days by bounded nonlinear least squares
(trust-region reflective, box-scaled) in the Feller-scaled coordinates,
with a budget of 20 pricing evaluations, retaining the best candidate
encountered. In Stage~B the structural vector is held fixed and each
day's $V_0$ is minimized separately by bounded scalar Brent search on
the $V_0$ box, with each day's update accepted only if it improves that
day's objective.

Global search runs twelve starting points (four informed seeds and
eight Latin-hypercube draws in the scaled coordinates, with the Feller
fraction drawn from $[0.1,0.9]$); each start receives one Stage~A
iteration with a budget of five pricing evaluations followed by one
Stage~B pass. Among the starts whose short-run solutions keep every
parameter interior to its box, the best by joint RMSPE is then
iterated by full
rounds until the relative improvement of the joint mean-squared
objective falls below 0.1\% (at most twenty rounds). Every daily
$V_0$---including the initial states of every start---is initialized
at the squared 60-day realized volatility of the day's volume-leader
underlying futures, and remains free on $[0.005,0.35]$ throughout.

The cross-date pricing evaluation of Section~7.5 re-solves each evaluation
day's $V_0$ by the same Stage-B search, with one addition guarding
against local minima of the scalar objective: the search is started
from the futures-side realized-variance initialization and from a
fixed grid of four interior points spanning the box, and the best
resulting $V_0$ is retained. No structural parameter is touched on the
evaluation day.

\end{document}